\documentclass[fleqn,usenatbib]{mnras}

\usepackage{newtxtext,newtxmath}
\usepackage{longtable}
\usepackage{lipsum}
\usepackage[utf8]{inputenc}
\usepackage[autostyle]{csquotes}
\usepackage{ltablex}
\usepackage{rotating}
\usepackage{multicol}
\usepackage{epstopdf}
\usepackage{graphicx}
\usepackage{mathrsfs}
\usepackage{adjustbox}
\usepackage{subfigure}
\usepackage{lscape}
\usepackage{longtable}
\usepackage[table]{xcolor}
\usepackage[T1]{fontenc}

\usepackage{pdflscape}

\DeclareRobustCommand{\VAN}[3]{#2}
\let\VANthebibliography\thebibliography
\def\thebibliography{\DeclareRobustCommand{\VAN}[3]{##3}\VANthebibliography}

\usepackage{graphicx}	% Including figure files
\title[GX 339$-$4 and H 1743$-$322 : X-ray view]{Long term wide-band spectro-temporal studies of  outbursting black hole candidate sources GX 339$-$4 and H 1743$-$322: {\it AstroSat} and {\it NuSTAR} results}

\author[Aneesha et al.]{
U. Aneesha,$^{1}$\thanks{E-mail: aneesha@iitg.ac.in}
Santabrata Das,$^{1}$\thanks{E-mail: sbdas@iitg.ac.in}
Tilak B. Katoch$^{2}$
and Anuj Nandi$^{3}$\thanks{E-mail: anuj@ursc.gov.in}
\\
$^{1}$Indian Institute of Technology Guwahati, Guwahati, 781039, India\\
$^{2}$Department of Astronomy and Astrophysics, Tata Institute of Fundamental Research, Homi Bhabha Road, Colaba, Mumbai, 400005, India \\
$^{3}$Space Astronomy Group, ISITE Campus, U. R. Rao Satellite centre, Outer Ring Road, Marathahali, Banglore, 560037, India
}

\date{Accepted XXX. Received YYY; in original form ZZZ}

\pubyear{2015}

\begin{document}
\label{firstpage}
\pagerange{\pageref{firstpage}--\pageref{lastpage}}
\maketitle

\begin{abstract}

We present a comprehensive spectro-temporal analyses of recurrent outbursting black hole sources GX 339$-$4 and H 1743$-$322 using available \textit{AstroSat} and \textit{NuSTAR} archival observations during $2016-2024$. The nature of the outburst profiles of both sources are examined using long-term {\it MAXI/GSC} and {\it Swift/BAT} lightcurves, and {\it failed} as well as {\it successful} outbursts are classified. Wide-band ($0.5-60$ keV) spectral modelling with disc (\texttt{diskbb}) and Comptonized (\texttt{Nthcomp}) components indicates that GX 339$-$4 transits from hard ($kT_{\rm bb}=0.12-0.77$ keV, $\Gamma_{\rm nth}=1.54-1.74$ and $L_{\rm bol}=0.91-11.56$\% $L_{\rm Edd}$) to soft state ($kT_{\rm in}~[\approx{kT}_{\rm bb}]=0.82-0.88$ keV, $\Gamma_{\rm nth}=1.46-3.26$, $L_{\rm {bol}}=19.59-30.06\%L_{\rm Edd}$) via intermediate state ($kT_{\rm in}~[\approx{kT}_{\rm{bb}}]=0.56-0.88$ keV, $\Gamma_{\rm nth}=1.76-2.66$, $L_{\rm {bol}}=2.90-16.09\%L_{\rm Edd}$), whereas H 1743$-$322 transits from quiescent to hard state ($\Gamma_{\rm nth}=1.57-1.71$, $L_{\rm {bol}}=2.08-3.48\%L_{\rm Edd}$). We observe type-B and type-C Quasi-periodic Oscillations (QPOs) in GX 339$-$4 with increasing frequencies ($0.10 - 5.37$ Hz) along with harmonics. For H 1743$-$322, prominent type-C QPOs are observed in frequency range $0.22-1.01$ Hz along with distinct harmonics. Energy-dependent power spectral studies reveal that fundamental QPO and harmonics disappear beyond $20$ keV in GX 339$-$4, whereas fundamental QPO in H 1743$-$322 persists up to $40$ keV. We also observe that type-C ${\rm QPO}_{\rm rms}\%$ decreases with energy for both sources although such variations appear marginal for type-B QPOs. Additionally, we report non-monotonic behavior of photon index with plasma temperature and detection of annihilation line. Finally, we discuss the relevance of the observational findings in the context of accretion dynamics around black hole binaries.

\end{abstract}

\begin{keywords}
	accretion, accretion discs – black hole physics – radiation mechanisms:general – X-rays: binaries – stars: individual: 
	GX 339$-$4  – stars: individual: H 1743$-$322.
\end{keywords}

%%%%%%%%%%%%%%%%%%%%%%%%%%%%%%%%%%%%%%%%%%%%%%%%%%

%%%%%%%%%%%%%%%%% BODY OF PAPER %%%%%%%%%%%%%%%%%%

\section{Introduction}

In black hole X-ray binaries (BH-XRBs), the black hole accretes matter from the companion star and the temperature of the accreted matter becomes significantly high to produce X-ray emission. Considering the long term source  variability, XRBs are broadly classified as persistent or transient sources \citep{Chen-etal1997,Corral-etal2016,Tetarenko-etal2016}. In the case of persistent sources, such as Cygnus X$-$1 \cite[]{Ankur-etal2021}, LMC X$-$1 and LMC X$-$3 \cite[and references therein]{Bhuvana-etal2021,Bhuvana-etal2022}, the central BH is steadily fed by the companion winds resulting the stable bolometric luminosity. On the other hand, black hole X-ray transients (BH-XRTs) are the low-mass X-ray binaries (LMXBs) harbouring an accreting BH that orbits with a low-mass ($\lesssim 2 \, {\rm M}_\odot$) companion star. BH-XRTs are faint objects that mostly remain in the quiescent state ($\rm L_{x} \sim 10^{31-33}$ erg s$^{-1}$) and accrete matter at very low rate \citep{Tanaka-Lewin1995,McClintock-Remillard2006}. However, bright outbursts are occasionally triggered in BH-XRTs (GX 339$-$4 and H 1743$-$322), which usually last for hundreds of days and the source luminosity increases to $L_{\rm x} \sim 10^{36-39}$ erg s$^{-1}$ \citep{Corral-etal2016,Tetarenko-etal2016}. Note that the transient LMXBs like GRS 1915$+$105 sustains outburst for an extended period (over $25$ years) and hence are classified as persistent-cum-aperiodic variable source \cite[and references therein]{Belloni-etal2000,Nandi-etal2001,Sreehari-etal2020,Athulya-etal2022,Seshadri-etal2022,Athulya-Nandi2023}.

The energy spectra of BH-XRTs often consist of a thermal emission component originating from the accretion disc \cite[]{Makishima-etal1986,Mitsuda-etal1984}, and a non-thermal component resulted due to the inverse-Comptonisation \cite[]{Sunyaev-Titarchuk1980, Titarchuk1994,Chakrabarti-Lev1995} of soft photons by `hot' corona ($\sim 100$ keV). In addition, BH-XRBs display rapid X-ray timing variabilities that are commonly examined by analysing the power density spectrum (PDS) \cite[]{vanderKlis-1989}. The PDS often exhibits Quasi-periodic Oscillation \cite[QPO; see][for a recent review]{Ingram-Motta2019} features along with a broad-band noise continuum. In particular, Low-Frequency QPOs (LFQPOs) are observed over a wide range of frequencies ranging from mHz to few tens of Hz and are classified as type-A, type-B and type-C QPOs, respectively \cite[]{Wijnands-etal1999,Homan-etal2001,Remillard-etal2002,Casella-etal2005}. Indeed, type-B and type-C LFQPOs are characteristically different indicating their different physical origin. On the contrary, type-A LFQPO is rarely observed and is poorly understood.

During an outburst, the BH-XRTs generally evolve through distinct spectral states, namely Low/Hard State (LHS), Hard-Intermediate State (HIMS), Soft-Intermediate State (SIMS), and High/Soft State (HSS). These states occupy different positions in the hardness intensity diagram \cite[HID;][]{Homan-Belloni2005,McClintock-Remillard2006,Belloni-etal2016,Nandi-etal2012,Radhika-nandi2014,Sreehari-etal2019b,Sreehari-Nandi2021,Nandi-etal2024}) and form a hysteresis loop \cite[]{Miyamoto-etal1995,Belloni-etal2005,Nandi-etal2018}. For a {\it successful} outburst, a source transits through all the spectral states, otherwise it is categorized as a {\it failed} outburst \cite[]{Tetarenko-etal2016,Alabarta-etal2021}.
 
In a successful outburst, when the source accretes matter at very low rate, the emergent X-ray emissions yield quiescent state (QS). As the source transits to LHS, X-ray emissions are dominated by the high energy photons where the photon index ($\Gamma$) lies in the range of $1.5-1.6$. In LHS, the PDS is dominated by the broad-band noise component with a maximum fractional rms amplitude of $30-50 \%$ \cite[]{Homan-etal2001} and usually type-C QPO features are observed. Next, the source makes transition to intermediate states (HIMS and SIMS), where the soft photon contribution becomes significant with $\Gamma \sim 1.7-2.5$, and the rms noise in the PDS becomes weaker (typically rms $\sim 5-20 \%$). The intermediate states (HIMS and SIMS) are distinguished from each other due to their different timing properties \citep{Belloni-etal2016}. Note that the type-C QPOs are observed in the LHS and HIMS, whereas type-A and type-B QPOs are seen in SIMS \citep{Belloni-etal2005,Radhika-nandi2014,Radhika-etal2016,Belloni-etal2016}. Subsequently, the source transits to HSS \cite[]{Homan-etal2001,Belloni-2005,Remillard-McClintock2006,Nandi-etal2012,Radhika-etal2016,Nandi-etal2018}, where the source becomes bright and the radiation spectrum is dominated by the thermal disc photons with steep photon index $\Gamma \sim 3$. In HSS, PDS displays powerlaw signature with rms less than $2-3\%$. Finally, the source returns back to LHS via intermediate states and subsequently fades into quiescence. Indeed, these changes of the spectro-temporal properties evidently indicate that the spectral state transition during the evolution of an outburst seems to be associated with the accretion geometry.
	
Being motivated with the unique outburst features of BH-XRTs, we intend to study the long term wide-band spectro-temporal properties of two LMXB sources, namely GX 339$-$4 and H 1743$-$322. The source GX 339$-$4 was discovered in $1973$ by satellite OSO-7 \cite[]{Markert-etal1973} and the source undergoes regular outbursts in every $2-3$ years \cite[]{Sreehari-etal2018}. Since its discovery, the source has been studied extensively \cite[]{Mendez-etal1997,Corbel-etal2003,Hynes-etal2003,Zdziarski-etal2004,Belloni-etal2005,Shaposhnikov-Titarchuk2009,Nandi-etal2012,Titarchuk-Seifina2021}. GX 339$-$4 is located at the sky position ${\rm RA}=17^{h}02^{m}49^{s}.38$ and ${\rm Dec}=-48^{\circ}47^{\prime}23^{\prime\prime}.2$. Meanwhile, the absorption lines study in near-IR \cite[]{Heida-etal2017} indicates the presence of a giant K-type companion confirming previous finding \cite[]{Hynes-etal2004, Darias-etal2008}. The mass and distance of the source are found as $8.28-11.89 ~\rm{M}_{\odot}$ \cite[]{Sreehari-etal2019a} and $\rm d=8.4 \pm 0.9$ kpc \cite[]{Parker-etal2016}. During the outburst, GX 339$-$4 was found to demonstrate fast spectral state transitions, where it traces all the four spectral states \cite[]{Belloni-etal2005,Debnath-etal2010,Nandi-etal2012}. Further, the temporal studies of GX339$-$4 reveals the presence of type-A, type-B and type-C QPOs \cite[]{Motta-etal2011}.

The transient LMXB source H 1743$-$322 (IGR J17464$-$3213 or XTE J17464$-$3213) is located in the sky at ${\rm RA}=17^{h}46^{m}15^{s}.596$ and ${\rm Dec}=-32^{\circ}14^{\prime}00^{\prime\prime}.860$ at a distance of $8.5 \pm 0.8$ kpc \citep{Steiner-etal2012}. The companion star is believed to be a late-type main-sequence star \cite[]{Chaty-etal2015} and the mass of the BH primary is estimated as $11.21_{-1.96}^{+1.65} \, \rm M _{\odot}$ \cite[]{Molla-etal2017,Tursunov-Kolos2018}. The brightest outburst of H 1743$-$322 was observed in $2003$ ($\sim 100$ cts s$^{-1}$ in {\it RXTE/ASM}) by {\it RXTE} and {\it INTEGRAL} \cite[]{Steiner-etal2012,Homan-etal2005a,Remillard-etal2006,Kalemci-etal2006}. Note that the source underwent multiple outbursts with a recurrence period of $0.5-1$ year after the $2003$ outburst \cite[]{Aneesha-Mandal2020,Stiele-Yu2016}. H 1743$-$322 was found in all spectral states and the HID showed hysteresis pattern \cite[]{Aneesha-Mandal2020}. Furthermore, this source exhibited high-frequency QPO (HFQPO) at $240$ Hz during $2003$ outburst \cite[]{Homan-etal2003} and mHz QPO at $\sim 11$ mHz during $2010$ outburst \citep{Altamirano-Strohmayer2012}.

In order to understand the underlying accretion scenarios during the outburst phases, in this paper, we analyse the spectro-temporal properties of GX 339$-$4 and H 1743$-$322 using the available archival {\it AstroSat} and {\it NuSTAR} data during $2016-2024$. Broad-band spectral coverage along with excellent timing capabilities of {\it AstroSat} \cite[]{Agrawal-2001} and {\it NuSTAR} \cite[]{Harrison-etal2013} allow us to investigate the nature of the accreting systems. Towards this, we examine the long term temporal variability of both sources using {\it MAXI/GSC} and {\it SWIFT/BAT} lightcurve. Further, in order to understand the overall nature of the outbursts, we examine HID of both sources. We conduct the wide-band spectral modelling that explains spectral characteristics of GX 339$-$4 and H 1743$-$322. We also performed a PDS analysis to examine the QPO features and observed that both sources exhibit type-B and/or type-C QPOs. With this, we perceive the nature of the accreting system present in the LMXBs under considerations.

The plan of the paper is as follows. In \S\ref{sec:obs-red}, we describe the observations and data reduction techniques. We discuss outburst profile and HID of outbursts for both sources in \S\ref{sec:HID}. We present the results of the spectral analysis in \S\ref{sec:spec_ana}. In \S\ref{sec:timing}, we carry out the timing analysis. In \S\ref{sec:Dis-con}, we present discussion and finally conclude.

\begin{table*}
    \centering
    \caption{Details of \textit{AstroSat} and \textit{NuSTAR} observations of GX 339$-$4 and H 1743$-$322 during $2016-2024$. Columns $1-9$ denote Source name, Epoch, Observation ID, Mission Name, Obs. Date, \textit{SXT} Exposure Time, \textit{LAXPC} Exposure Time, \textit{NuSTAR} Exposure Time,  Hardness Ratio, respectively. Observations in quiescent phase are highlighted in light gray color. See the text for details.}
    \label{tab:obs_log}
    \normalsize
    \begin{adjustbox}{width=0.88\textwidth} 
    \begin{tabular}{l @{\hspace{0.3cm}} c @{\hspace{0.3cm}} c @{\hspace{0.3cm}}c @{\hspace{0.3cm}} c @{\hspace{0.3cm}} c @{\hspace{0.3cm}} c @{\hspace{0.3cm}} c @{\hspace{0.3cm}} c @{\hspace{0.3cm}} c @{\hspace{0.3cm}} r}	
    \hline	\hline
    Source&Epoch&Observation ID&Mission&Obs. Date (MJD) & \multicolumn{3}{c}{Effective Exposure (ksec)} & Hardness Ratio$^\dagger$ &References\\
    \cline{6-8}
    &  &  & &  & {\it SXT} & {\it LAXPC} & {\it NuSTAR}  \\ 
    \hline
    \rowcolor{lightgray}
    
    &AS1.01&T01\_031T01\_9000000368 & \textit{AstroSat}&$11/03/2016~(57458.54)$ &$\sim 6.3$ &$\sim 10.1$&-&$1.83^{+0.19}_{-0.05}$ \\
    \rowcolor{lightgray}
    
    &AS1.02 &G05\_253T01\_9000000560 &\textit{AstroSat}& $24/07/2016~(57593.40)$ &$\sim 39.1$ & $\sim 53.7$&-&$1.13^{+0.09}_{-0.13}$\\

    &NU1.03&80302304002 &\textit{NuSTAR}&$02/10/2017~(58028.15)$&- & -&$\sim 21.5$&$2.62^{+0.05}_{-0.98}$ &$1,2$\\

    &AS1.04& A04\_109T01\_9000001578&\textit{AstroSat}&$04/10/2017~(58030.88) $& $\sim 5.6$& $\sim  21.6$&-& $2.54^{+0.16}_{-0.23}$&$3,4$\\

    &NU1.05& 80302304004&\textit{NuSTAR}&$25/10/2017~(58051.57)$ & $-$& $-$&$\sim 17.9$&$2.69^{+0.02}_{-0.01}$&$1,2,3$ \\

    &NU1.06& 80302304005&\textit{NuSTAR}&$02/11/2017~(58059.89)$ & -& -&$\sim 18.9$&$2.80^{+0.24}_{-0.13}$& $1,2,3$\\
		
    &NU1.07& 80302304007&\textit{NuSTAR}&$30/01/2018~(58148.36)$ & -& -&$\sim 28.9$& $2.62^{+0.15}_{-0.02}$& $1,2$ \\

    &NU1.08& 90401369002&\textit{NuSTAR}&$19/12/2018~(58471.18)$ & -& -&$\sim 0.6$& $2.08^{+1.27}_{-1.39}$ \\
    
    &NU1.09& 90401369004&\textit{NuSTAR}&$05/01/2019~(58488.63)$& -& -&$\sim $3.6&$2.83^{+0.46}_{-0.83}$ &$1$\\
		
    &AS1.10& A05\_166T01\_9000003192&\textit{AstroSat}& $22/09/2019~(58748.70)$ &$\sim 11.5$&$\sim 34.5 $ &-&$2.73^{-0.01}_{+0.57}$&$4$\\
		
    &NU1.11& 90502356004&\textit{NuSTAR}& $29/12/2019~(58846.58)$ &- &- &$\sim 0.6$&$0.35^{+0.05}_{-0.09}$\\
			
    &NU1.12& 80502325002&\textit{NuSTAR}& $20/02/2020~(58899.13)$ &- &- &$\sim 18.0$&$0.13^{+0.02}_{-0.07}$\\
			
    &NU1.13& 80502325004&\textit{NuSTAR}&$29/02/2020~(58908.65)$ &- &- &$\sim 20.3$&$0.13^{+0.05}_{-0.06}$\\
			
    &NU1.14& 80502325006&\textit{NuSTAR}&$01/04/2020~(58940.53)$&- &- &$\sim 21.6$&$0.50^{+0.05}_{-0.09}$\\
			
    &NU1.15& 80502325008&\textit{NuSTAR}&$20/04/2020~(58959.05)$ &- &- &$\sim 22.4$&$2.43^{+0.23}_{-0.97}$\\
			
    &NU1.16& 90702303001&\textit{NuSTAR}&$23/01/2021~(59237.91)$ &- &- &$\sim 25.0$&$2.77^{+0.16}_{-0.92}$\\
		
    GX 339$-$4 &NU1.17&90702303003&\textit{NuSTAR}&$05/02/2021~(59250.99)$ &- &- &$\sim 21.9$&$2.58^{+0.03}_{-0.01}$\\
		
    &AS1.18 &T03\_275T01\_9000004180&\textit{AstroSat}&$13/02/2021~(59258.04)$ & $\sim 10.6$ & $\sim 27.9$ & - & $2.06^{+0.05}_{-0.20}$ \\
		
    &NU1.19&90702303005&\textit{NuSTAR}&$20/02/2021~(59265.42)$ &- &- &$\sim 20.9$&$2.44^{+0.04}_{-0.01}$\\
			
    &AS1.20& T03\_279T01\_9000004218 & \textit{AstroSat}& $02/03/2021~(59275.04)$ & $\sim 99.8$  & $\sim 43.6$&- & $1.99^{+0.09}_{-0.54}$  \\
			
    &AS1.21& T03\_280T01\_9000004222& \textit{AstroSat}& $05/03/2021~(59278.11)$ & $\sim 9.2$ & $\sim 2.1$ & - & $1.90_{-0.14}^{+0.07}$\\
			
    &AS1.22& T03\_281T01\_9000004224 &  \textit{AstroSat}& $05/03/2021~(59278.38)$ & $\sim 0.9$ & $\sim 1.8$ & - & $2.22^{+0.52}_{-1.29}$\\
			
    &AS1.23& T03\_282T01\_9000004226 & \textit{AstroSat}& $05/03/2021~(59278.45)$ &$\sim 1.2$& $\sim 1.4$ & - & $2.01^{+0.38}_{-0.14}$\\
			
    &NU1.24&90702303007&\textit{NuSTAR}&$07/03/2021~(59280.99)$ &- &- &$\sim 23.4$&$2.35^{+0.03}_{-0.01}$\\
			
    &NU1.25&90702303009&\textit{NuSTAR}&$26/03/2021~(59299.50)$ &- &- &$\sim 15.8$&$1.26^{ +0.35}_{-0.01}$\\
		
    &AS1.26&T03\_291T01\_9000004278  &\textit{AstroSat}&$30/03/2021~(59303.06)$ &$\sim 22.3$ & $\sim 36.5$ &-&  $0.33^{+0.04}_{-0.10}$&$5,6,7$ \\
		
    &AS1.27 &T03\_291T01\_9000004278 &\textit{AstroSat}&$31/03/2021~(59304.00)$ &$\sim 23.0$&$\sim 42.4$ &-& $0.20^{+0.09}_{-0.02}$&$5$ \\
		
    &AS1.28&T03\_291T01\_9000004278& \textit{AstroSat}&$ 01/04/2021~(59305.03)$&$\sim 25.7$ & $\sim 40.9$&-& $0.29^{+0.01}_{-0.03}$&$5$\\
		
    &NU1.29&90702303011& \textit{NuSTAR}& $01/04/2021~(59305.81)$ &- & -&$\sim $14.8&$0.29^{+0.09}_{-0.07}$ \\
		
    &AS1.30&T03\_291T01\_9000004278& \textit{AstroSa}t&$02/04/2021~(59306.04)$ &$\sim 23.8$&$\sim 47.5$ &- &$0.27^{+0.07}_{-0.10}$&$5$\\
		
    &AS1.31&T03\_291T01\_9000004278&\textit{AstroSat} &$03/04/2021~(59307.50)$ &$\sim 19.9$ &$\sim 33.1$ &- &$0.25^{+0.01}_{-0.11}$&$5$\\
		
    &AS1.32&T03\_291T01\_9000004278&\textit{AstroSat} &$04/04/2021~(59308.00)$&$\sim 22.5$ &$\sim 42.2$&-&$0.22^{+0.10}_{-0.02}$&$5$\\
		
    &AS1.33&T03\_291T01\_9000004278&\textit{AstroSat}&$05/04/2021~(59309.02)$ &$\sim 27.6$ &$\sim 51.1$ &- &$0.35^{+0.09}_{-0.04}$&$5$\\
		
    &AS1.34  &T03\_291T01\_9000004278&\textit{AstroSat}&$06/04/2021~(59310.33)$& $\sim 14.8$&$\sim 21.8$ &- &$0.15^{+0.05}_{-0.20}$&$5$ \\
		
    &NU1.35  &90702303013&\textit{NuSTAR}&$23/04/2021~(59327.82)$& -&- &$\sim 20.1$ &$0.16^{+0.06}_{-0.09}$\\
		
    &NU1.36  &80601302002&\textit{NuSTAR}&$04/05/2021~(59338.96)$& -&- &$\sim 13.6$ &$0.11^{+0.04}_{-0.02}$\\
		
    &NU1.37  &80601302004&\textit{NuSTAR}&$05/05/2021~(59339.42)$& -&- &$\sim 4.1$ &$0.11 \pm 0.06$\\
		
    &NU1.38  &80601302006&\textit{NuSTAR}&$05/05/2021~(59339.62)$& -&- & $\sim 2.4$&$0.09^{+0.01}_{-0.02}$\\
		
    & NU1.39 &80702316002&\textit{NuSTAR}&$10/10/2021 ~(59497.09)$&-&-&$\sim 21.3$&$2.36^{+0.30}_{-0.24}$ &\\
		
    \rowcolor{lightgray}
		
    & NU1.40&80702316004&\textit{NuSTAR}&$20/10/2021 ~(59507.08)$&-&-&$\sim 21.3$ & $2.47^{+0.08}_{-0.07}$\\
		
    \rowcolor{lightgray}
		
    & NU1.41&80702316005&\textit{NuSTAR}&$02/11/2021 ~(59520.35)$&-&-&$\sim 17.8$ & $2.47^{+0.02}_{-0.07}$\\
		
    & AS1.42& T05\_051T01\_9000005334&\textit{AstroSat}& $07/09/2022~(59829.51)$ & $\sim 17.3$&$\sim 38.9$&-&$2.29 \pm 0.03$\\
		
    & AS1.43& T05\_053T01\_9000005338 & \textit{AstroSat}& $09/09/2022~(59831.54)$ &$\sim 34.0$&$\sim 86.9$&-&$2.95^{+0.03}_{-0.06}$ \\
		
    & NU1.44&80801341002&\textit{NuSTAR}&$11/09/2022~( 59833.22)$&$-$&$-$&$\sim 21.0$&$2.71^{+0.05}_{-0.12}$\\
		
	 \rowcolor{lightgray}
	 
    & AS1.45 &A12\_099T01\_9000005716&\textit{AstroSat}&$28/06/2023~(60123.00)$&$\sim 32.2$ &$\sim 50.8$&$-$&$0.73 \pm 0.04$\\	
    
    & NU1.46&91001302001 &\textit{NuSTAR}&$19/01/2024~( 	60328.04)$&$-$&$-$&$\sim 18.6$&$0.89^{+0.02}_{-0.04}$\\
    
    &NU1.47&91001304002&\textit{NuSTAR}&$31/01/2024~( 	60340.54)$&$-$&$-$&$\sim 17.3$&$0.16^{+0.01}_{-0.02}$\\
    
    &NU1.48&91002306002&\textit{NuSTAR}&$14/02/2024~( 	60354.70)$& $-$&$-$&$\sim 16.4$&$0.20^{+0.03}_{-0.04}$ \\
    
    &AS1.49&A05\_166T01\_9000006070&\textit{AstroSat}&$14/02/2024~(60354.74)$&$\sim 28.6$&$\sim 50.7$&$-$&$0.14^{+0.02}_{-0.04}$& \\
    
    &AS1.50$^\ddagger$ &A13\_028T01\_9000006122&\textit{AstroSat}&$10/03/2024~(60379.44)$&$-$&$\sim 84.3$&$-$&$-$\\
    
    &NU1.51$^\ddagger$ &80902342002&\textit{NuSTAR}&$11/03/2024~( 	60380.16)$&$-$&$-$&$\sim 19.5$&$-$\\
    
    &NU1.52$^\ddagger$ &80902342001&\textit{NuSTAR}&$11/03/2024~( 	 	60380.16)$&$-$&$-$&$\sim 0.36$&$-$\\
    
    &AS1.53 &T05\_187T01\_9000006132 &\textit{AstroSat}&$21/03/2024~(60390.00)$& $\sim 16.9$&$\sim 25.7$&$-$&$0.10 \pm 0.02$\\		
    
    \hline
    
    &\negthickspace AS2.01&T01\_045T01\_9000000364 &\textit{AstroSat}&$09/03/2016~(57456.40)$ &$\sim 6.4$ &$\sim 11.6$&- &$2.52^{+0.02}_{-0.04}$&$8,9$ \\
    
    &NU2.02&80202012002&\textit{NuSTAR}&$13/03/2016~(57460.07)$ &- &- & $\sim 65.9$&$2.55 \pm 0.01$
    &$10$\\ &NU2.03&80202012004&\textit{NuSTAR}&$15/03/2016~(57462.282)$ &- &- & $\sim $65.7&$2.520^{+0.002}_{-0.001}$&$10$\\
    \rowcolor{lightgray}
    
    &AS2.04&G05\_253T03\_9000000562 &\textit{AstroSat}&$26/07/2016~(57595.10)$ &$\sim 41.3 $ &$\sim 58.4 $&-&$1.44^{+0.06}_{-0.01}$  \\
    \rowcolor{lightgray}
    
    H 1743$-$322 &AS2.05& G05\_144T01\_9000000612&\textit{AstroSat}&$18/08/2016~
    (57618.62)$&$\sim 6.9$ &$\sim 19.3$&-&$1.38^{+0.08}_{-0.19}$ \\
    
    &\negthickspace AS2.06& G07\_039T01\_9000001444& \textit{AstroSat}&$08/08/2017~(57973.32)$ &$\sim 12.5$ & $\sim 14.8$&-&$2.040^{+0.040}_{-0.003}$&$8$ \\		
    
    & NU2.07&90401335002 &\textit{NuSTAR}&$19/09/2018 ~(58380.11)$ &- &- &$\sim 38.5$ &$2.950^{+0.003}_{-0.002}$&$11, 12$\\
    
    & NU2.08&80202012006&\textit{NuSTAR}&$26/09/2018 ~(58387.36)$ &- &- &$\sim 65.7$ &$3.040 \pm 0.003$&$11, 12$\\

    \hline	\hline

\end{tabular}
\end{adjustbox}
\begin{list}{}
    \item $^\dagger$Hardness Ratio is calculated as the ratio of photon flux in $6-20$ keV to $3-6$ keV from {\it LAXPC} and {\it NuSTAR} spectra. A model combination \texttt{TBabs$\times$(diskbb$+$gaussian$+$smedge$\times$Nthcomp)} is used for photon flux estimation in units of photons cm$^{-2}$ s$^{-1}$.
    \item $^\ddagger$Data are not in public domain.\\
    \noindent{\bf References:} $1$: \cite{Wang-etal2020}, $2$: \cite{Garcia-etal2019}, $3$; \cite{Debnath-etal2023}, $4$; \cite{Husain-etal2022}, $5$; \cite{Jana-etal2024}, $6$; \cite{Mondal-etal2023},  $7$; \cite{Peirano-etal2023}, $8$; \cite{Husain-etal2023}, $9$; \cite{Swadesh-etal2021}, $10$; \cite{Swadesh-etal2020}, $11$; \cite{Wang-etal2022}, $12$; \cite{Stiele-Kong2021}.	
\end{list}
\end{table*}

\begin{figure*}
    \centering
    \includegraphics[width=0.4\textwidth]{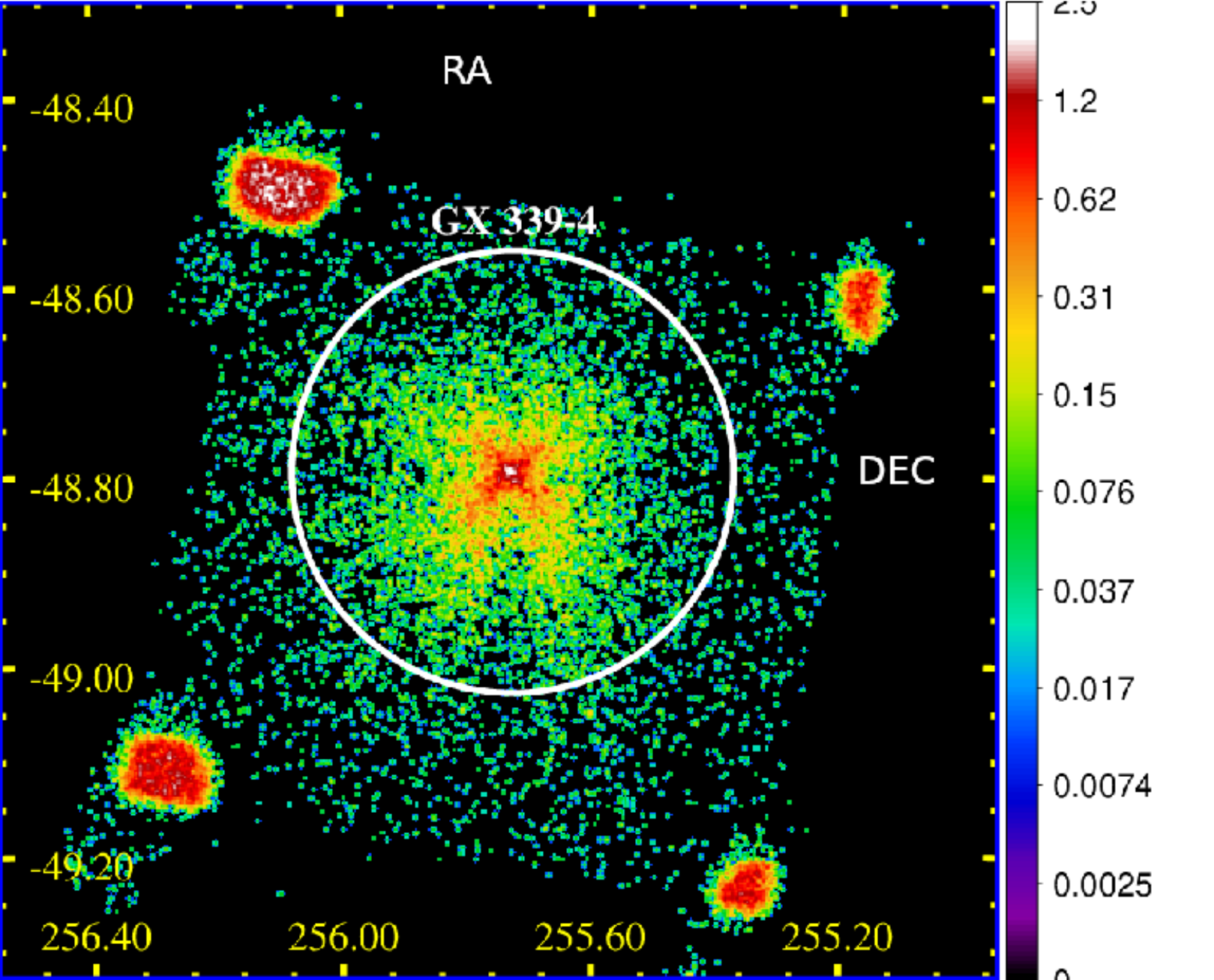}
    \hspace{0.5 cm}
    \includegraphics[width=0.4\textwidth]{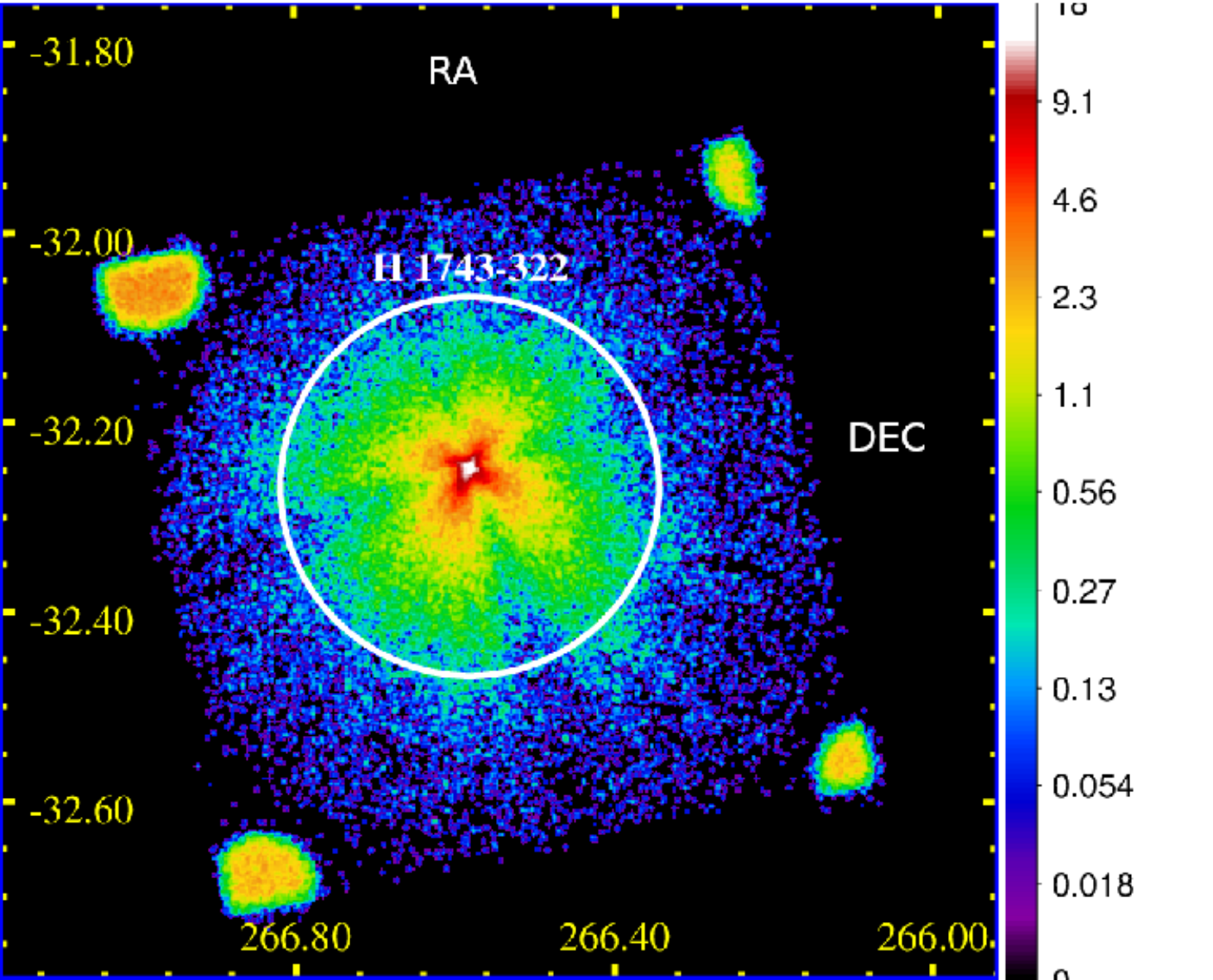}
    \caption{False color SXT image of GX 339$-$4 (left) during 4 October 2017 (MJD 58030.88) and H 1743$-$322 (right) during 8 August 2017 (MJD 57973.32). The image was generated from Photon Counting (PC) mode data. The image is smoothen out with a gaussian kernel of 3 pixel radius (1 pixel = 4.122 arcsec) using dS9. The four bright portions at the corners of the images are internal calibration sources which illuminate the corners of the CCD.
    }
    \label{fig:ds9}
\end{figure*}

\begin{figure*}
    \centering
    \includegraphics[width=\textwidth]{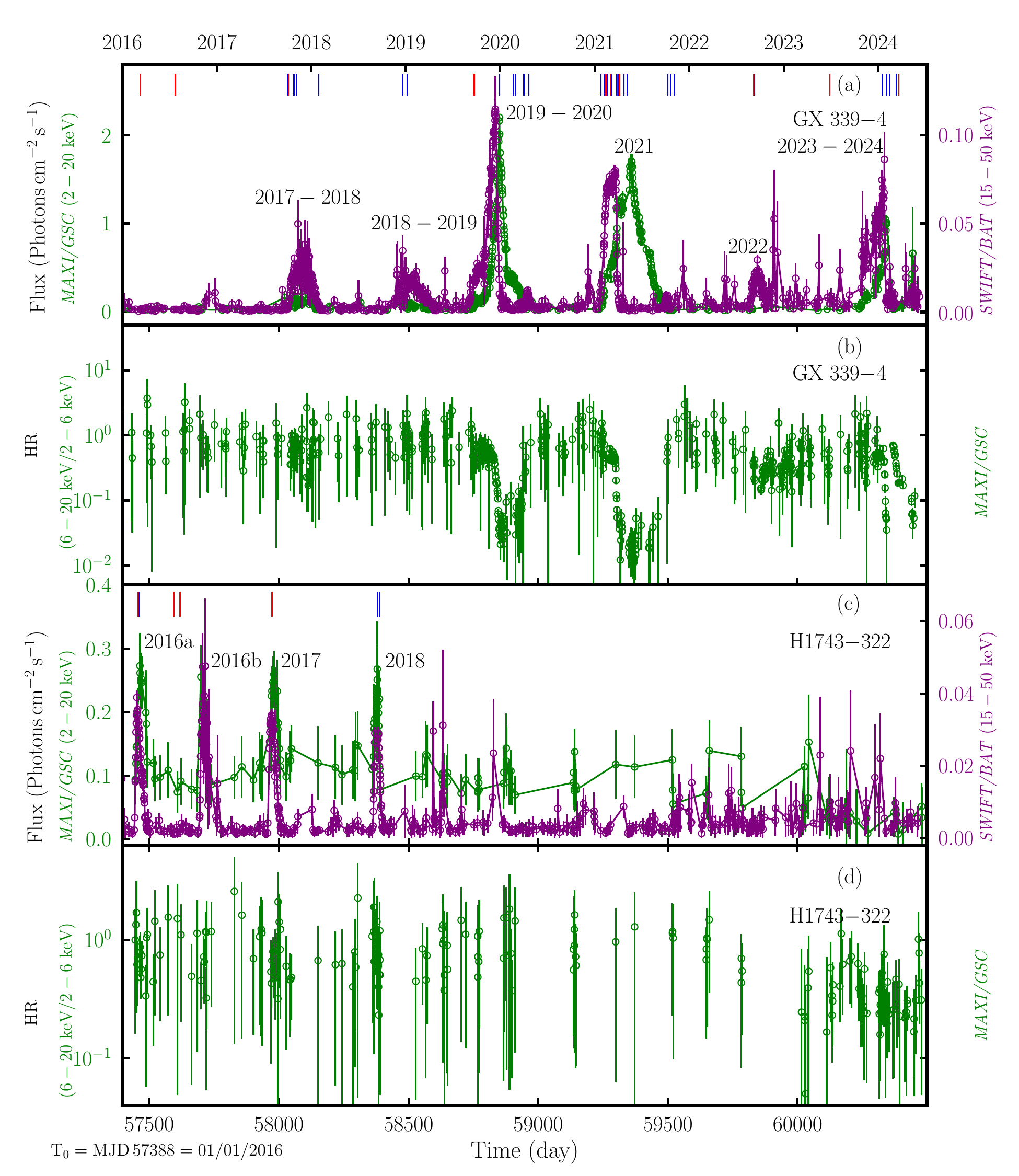}
    \caption{Variation of {\it MAXI/GSC} flux in $2-20$ keV (open circles in green) and \textit{Swift/BAT} flux in $15-50$ keV (open circles in purple) with time (day) are shown for (a) GX 339$-$4  and (c) H 1743$-$322, respectively. The corresponding HR variations of GX 339$-$4 and H 1743$-$322 are presented in panel (b) and (d). Here, HR is calculated using photon flux (in units of photons cm$^{-2}$ s$^{-1}$) values in different energy bands. Red and blue vertical bars denote \textit{AstroSat} and \textit{NuSTAR} observations. See the text for details.
    }
    \label{fig:lc-maxi-bat}  
\end{figure*}
  
\section{Observation and Data Reduction}
\label{sec:obs-red}

We analyse the publicly available observations of GX 339$-$4 and H 1743$-$322 from {\it AstroSat} and {\it NuSTAR} during $2016-2024$. \textit{AstroSat} data are obtained from Indian Space Research Organisation's (ISRO's) archival web-page\footnote{\url{https://astrobrowse.issdc.gov.in/astro\_archive/archive/Home.jsp}} hosted at Indian Space Science Data Center (ISSDC). The \textit{NuSTAR} data are publicly available at HEASARC website. All the observations under considerations are listed in Table \ref{tab:obs_log}. We make use of {\it MAXI} data obtained from {\it MAXI/GSC} on-demand web interface to depict the long term variation of lightcurve and hardness ratio (HR). For comparative study of the outburst profiles and HID of both sources, we utilize {\it MAXI/GSC} data. We also use \textit{Swift/BAT} data to study the long term evolution of the lightcurves.  
 
\subsection{AstroSat/LAXPC}

The LAXPC instrument on board {\it AstroSat} is a proportional counter operates in $3-80$ keV energy range \cite[]{Yadav-etal2016b,Antia-etal2017}. We use \textit{LAXPC} Level-1 data in Event Analysis (EA) mode available in {\it AstroSat} public archive
% \footnote{\label{note1}\url{https://webapps.issdc.gov.in/astro_archive/archive/Home.jsp.}} 
for timing and spectral analyses. We follow \cite{Sreehari-etal2019a, Sreehari-etal2020} for {\it LAXPC} data extraction procedure and analysis methods. We use the software \texttt{LaxpcSoftv3.4}\footnote{\url{http://www.tifr.res.in/~astrosat_laxpc/LaxpcSoft.html}}\citep{Antia-etal2017}, released on June 14, 2021 to process the Level-1 data to Level-2 data. We combine \textit{LAXPC10} and \textit{LAXPC20} lightcurves for the timing studies, where $10 ~\mu {\rm s}$ temporal resolution of {\it LAXPC} data is used. Since \textit{LAXPC10} has low gain and \textit{LAXPC30} is not in working condition, we use data only from \textit{LAXPC20} units for the spectral studies as it's gain remain stable throughout the entire observational period \citep{Antia-etal2021}. Using \texttt{GRPPHA}, we group the \textit{LAXPC} spectra to obtain the minimum count rate of $25$ per bin.

\subsection{AstroSat/SXT}
\label{sub:sxt-red}

The sources GX 339$-$4 and H 1743$-$322 were observed by the X-ray imaging instrument \textit{SXT} on board {\it AstroSat}, which operates in $0.3-8$ keV energy range. We obtain the Level-2 \textit{SXT} data from the Indian Space Science Data Center (ISSDC)
% \textsuperscript{\ref{note1}} 
archive. The {\it SXT} data for all these sources are available in Photon Counting (PC) mode. The data reduction processes are performed following the procedures provided by \textit{SXT} team at TIFR\footnote{\url{https://www.tifr.res.in/~astrosat\_sxt/dataanalysis.html}}. We use Julia based event merger tool (Julia\_v01) to merge the individual orbit data. From the resultant cleaned event file, we generate images of both sources using \texttt{XSELECT}. We select a circular region of radius $12^\prime$ around the source from which the source spectra and lightcurve were extracted \cite[]{Sreehari-etal2019a,Blessy-etal2021}. The {\it SXT} image of GX 339$-$4 and H 1743$-$322 are presented in the left and right panels of Fig. \ref{fig:ds9}. We check \textit{SXT} images of the sources to find the count rate and follow the guidelines provided in \textit{AstroSat} handbook\footnote{\url{https://www.issdc.gov.in/docs/as1/AstroSat-Handbook-v1.10.pdf}} for pile-up correction in our analysis. For both sources, we use background spectrum and response provided by the \textit{SXT} instrument team. The ancillary response file is created by \texttt{SXTARFMODULE}, which is provided by the \textit{SXT} team. The \textit{SXT} spectra is then grouped using \texttt{GRPPHA} to get the minimum count rate of $25$ per bin. 
 
\begin{figure*}
    \includegraphics[width=0.43\textwidth]{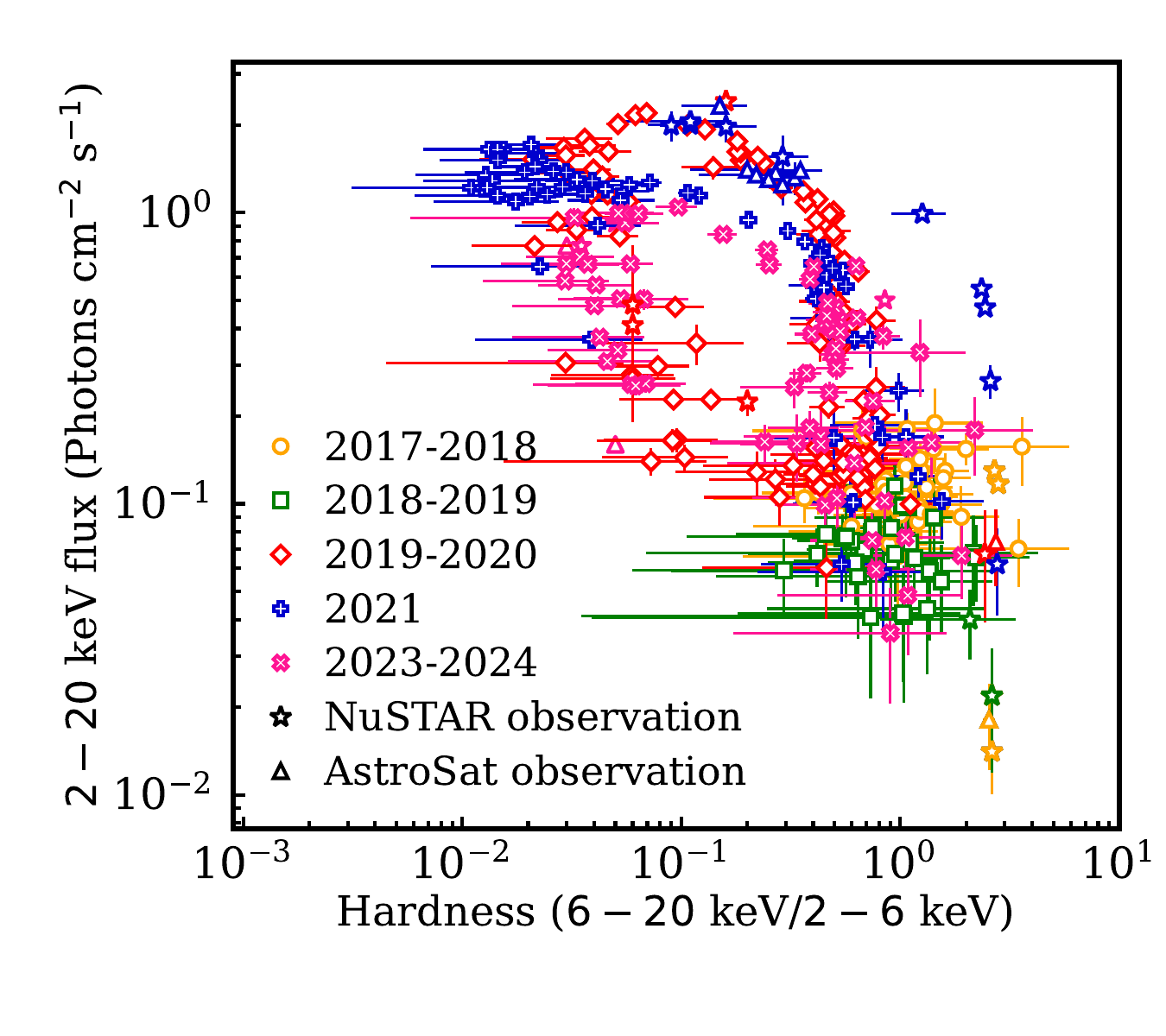}
    \hspace{0.5 cm}
    \includegraphics[width=0.43\textwidth]{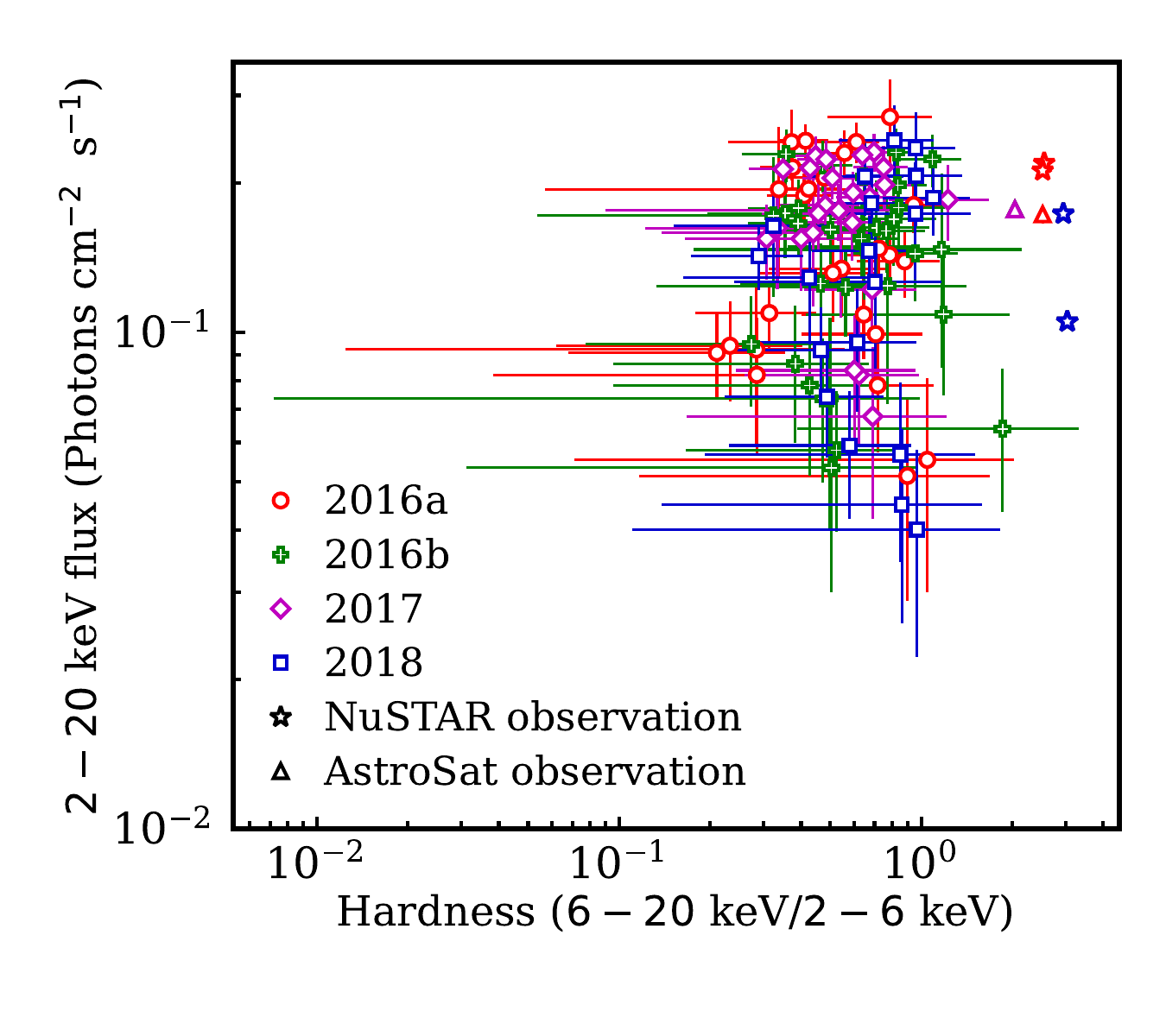}
    \caption{Hardness Intensity Diagram (HID) generated from \textit{MAXI} data are presented for GX 339$-$4 ({\it left panel}) and H 1743$-$322 ({\it right panel}). Different outbursts are shown using different symbols as marked in each panel. Asterisk and triangle denote \textit{AstroSat} and \textit{NuSTAR} observations. See the text for details.
    }
    \label{fig:HID}
\end{figure*}

\subsection{NuSTAR}

The {\it NuSTAR} is an X-ray telescope consisting of two detectors focal plane modules (FPMs), such as FPMA and FPMB, which operate in $3 - 79$ keV energy range \cite[]{Harrison-etal2013}. The {\it NuSTAR} instrument observed GX 339$-$4 and H 1743$-$322 on multiple epochs and the observation details are given in Table \ref{tab:obs_log}. To obtain the cleaned event file, we process the data using \texttt{NUPIPELINE} (version 0.4.8) available within \texttt{HEASOFT} and use the calibration database of \texttt{CALDB} version 20100101. \texttt{XSELECT} task is performed to select the source and background regions from cleaned event file. For source region, we select a circular region of radius  $40^{\prime\prime}$ centered at the peak brightness. The background is extracted from a region of same size away from the point source. We apply \texttt{NUPRODUCTS} task to generate the lightcurves, source spectra, background spectra along with their corresponding response matrices and ancillary responses. 

\subsection{MAXI/GSC and Swift/BAT}

The Gas Slit Camera (GSC) on board Monitor of All-sky X-ray Image ({\it MAXI})  \cite[]{Matsuoka-etal2009,Nakahira-etal2012} operates in $2-20$ keV energy range, and the Burst Alert Telescope (BAT) on board Neil Gehrels Swift Observatory ({\it Swift}) \cite[]{Gehrels-etal2004, Krimm-etal2013}) operates in $15-50$ keV energy range. These two instruments continuously monitored GX 339$-$4 and H 1743$-$322. In this work, we make use of both {\it MAXI/GSC} and {\it Swift/BAT} lightcurves during $2016-2024$. The one day averaged {\it MAXI/GSC} lightcurves of both sources are generated from {\it MAXI} on demand data publication system\footnote{\url{http://maxi.riken.jp/top/index.html}}. The {\it Swift/BAT} one day averaged lightcurves are downloaded from public archive\footnote{\url{https://swift.gsfc.nasa.gov/results/transients/}}.

\begin{table*}
    \centering
    \caption{Parameters of the outbursts of  GX 339$-$4 and H 1743$-$322 from the {\it MAXI/GSC} data. From left to right (1) Outburst under study; (2) Start date of outburst; (3) End date of outburst; (4) Time taken for $2-6$ keV, $6-20$ keV and $2-20$ keV lightcurve to reach the peak; (5) Peak flux in  $2-6$ keV, $6-20$ keV and $2-20$ keV lightcurve ; (6) Recurrence time; (7) Duration; (8) Outburst type. See the text for details.}
    \label{tab:lc_proper}
    \small	
    \begin{tabular}{l @{\hspace{0.2cm}} c @{\hspace{0.2cm}}c @{\hspace{0.2cm}} c @{\hspace{0.2cm}} c @{\hspace{0.2cm}} c @{\hspace{0.2cm}} c @{\hspace{0.2cm}} c @{\hspace{0.2cm}} c @{\hspace{0.2cm}} c @{\hspace{0.2cm}} c @{\hspace{0.2cm}} r}	
    \hline		\hline
    Outburst&Start date (MJD)&End date (MJD)&\multicolumn{3}{c}{${\rm T}_{\rm peak}^\dagger$} &\multicolumn{3}{c}{$\rm{F}_{\rm{peak}}^\star$}&$\rm{T}_{\rm{R}}^\boxdot$ & $\rm{T}_{\rm{D}}^\boxtimes$  & Type\\
    \cline{4-11}
    \vspace*{0.5ex}
    &&&\multicolumn{3}{c}{days} &  \multicolumn{3}{c}{$\rm{Photons \, cm^{-2} \, s^{-1}}$ }&days&days\\ 
    &&&($2-6$ keV)& ($6-20$ keV) &($2-20$ keV)&($2-6$ keV) & ($6-20$ keV)&($2-20$ keV)&\\
    \hline
    \multicolumn{12}{c}{GX 339$-$4} \\
		
    \hline
		
    2017-2018&06/09/2017 (58002)& 02/04/2018 (58210) &$42$&$42$&$42$&$0.18 \pm 0.04$& $0.06 \pm 0.01$&$0.18 \pm 0.02$&$921$&$208$ & {\it Failed}\\

    2018-2019 &23/12/2018 (58475) & 25/04/2019 (58598) &$30$ & $7$ &$7$ &$0.08 \pm 0.02$ &$0.05 \pm 0.02$ &$0.10 \pm 0.02$ &$265$& $123$ & {\it Failed} \\
		
    2019-2020&04/09/2019 (58730)& 31/05/2020 (59000) &$115$ & $95$ & $115$ & $2.24 \pm 0.03$ & $0.33 \pm 0.02$ &$2.20 \pm 0.03$ & $132$ & $270$ & {\it Successful}\\	
    
    2021& 18/12/2020 (59201)&11/10/2021 (59498)& $158$ & $100$& $157$ & $1.89 \pm 0.08$ & $0.21 \pm 0.01$ & $1.68 \pm 0.03$& $201$ & $297$ & {\it Successful} \\
		
    2022$^\ddagger$ &03/08/2022 (59794)& 03/11/2022 (59886) &$-$ & $-$ & $-$ & $-$ & $-$ & $-$ &$296$&$92$& $-$\\
		
    2023-2024 &10/10/2023 (60227)&$10/04/2024 (60410)$& $120$ &$110$ &$120$ & $1.07 \pm 0.07$ & $0.24 \pm 0.02$ & $1.04 \pm 0.05$ &$341$&$183$&${\it Successful}$\\
    \hline
    \multicolumn{12}{c}{H 1743$-$322} \\	
    \hline
    2016a & 12/02/2016 (57430)& 02/05/2016 (57510)& $24$ & $24$ & $24$& $0.11 \pm 0.06$&$ 0.08 \pm 0.02$ &$0.20 \pm 0.04 $&$191$ & $80$  & {\it Successful}\\
		
    2016b & 19/10/2016( 57680) &31/01/2017 (57784)&$24$ & $26$ & $24$& $0.17 \pm 0.02$ & $0.10 \pm 0.02$& $0.24 \pm 0.03$ & $170$ & $104$ & {\it Failed}\\
		
    2017 & 16/07/2017 (57950)	& 24/09/2017 (58020) & $23$ & $22$ & $22$ & $0.22 \pm 0.03$& $0.12 \pm 0.02$ & $0.24 \pm 0.04$ & $166$ &$70$ & {\it Failed}\\
		
    2018 &19/09/2018 (58380)&  	19/10/2018 (58410) & $8$ & $8$ & $8$ & $0.12 \pm 0.03$ & $0.10 \pm 0.03$ & $0.24 \pm 0.04$ & $360$ & $30$  & {\it Failed}\\
		
    \hline	\hline
    \end{tabular}
    \begin{list}{}{}
    \item $\dagger$ Time taken to reach peak flux in different energy bands.
    \item $\star$ The peak flux in different energy band.
    \item $\boxdot$ The recurrence time (the time interval between the onset of this outburst and the end of the previous outburst).
    \item $\boxtimes$ The duration of outburst. 
    \item $^\ddagger$ Lack of monitoring with {\it MAXI}
    \end{list}
	
\end{table*}

\section{Outburst profile and HID}
\label{sec:HID}

In this section, we examine the long term variability of lightcurves for GX 339$-$4 and H 1743$-$322 using \textit{MAXI/GSC} and \textit{SWIFT/BAT} data during {$2016-2024$}. Towards this, in Fig. \ref{fig:lc-maxi-bat}a and Fig. \ref{fig:lc-maxi-bat}c, we show the lightcurves of GX 339$-$4 and H 1743$-$322, which are obtained from \textit{MAXI/GSC} in $2-20$ keV and \textit{Swift/BAT} in $15-50$ keV energy range. Using \textit{MAXI/GSC} data, we calculate the hardness ratio (HR) defined as the ratio of photon counts in $6-20$ keV to $2-6$ keV, and plot it in Fig. \ref{fig:lc-maxi-bat}b and Fig. \ref{fig:lc-maxi-bat}d for GX 339$-$4 and H 1743$-$322, respectively.

\subsection{GX 339$-$4}

During the {\it AstroSat} era starting from January $2016$, GX 339$-$4 experienced six outbursts in $2017-2018$, $2018-2019$, $2019-2020$, $2021$, $2022$ and $2023-2024$ as shown in Fig. \ref{fig:lc-maxi-bat}a. In Fig. \ref{fig:lc-maxi-bat}b, we show the HR variation during all the outbursts including quiescent phases. We observe significant HR variability during $2019-2020$, $2021$ and $2023-2024$ outbursts, whereas nearly steady HR variability ($\sim 1.3$) is seen during $2017-2018$, $2018-2019$ and $2022$ outbursts. In particular, we find that HR changes significantly from $1.2$ to $0.01$ during $2019-2020$, $2021$ and $2023-2024$ outbursts indicating clear evidence of the spectral state transitions. This motivates us to carry out in-depth spectral analysis of this source presented in section \ref{sec:spec_ana}. 

In the left panel of Fig. \ref{fig:HID}, we present the HIDs of different outbursts generated from {\textit MAXI/GSC} data for GX 339$-$4. We find that during $2017-2018$ and $2018-2019$ outbursts, the source remained in the hard state without much variation of hardness. However, during $2019-2020$, $2021$ and $2023-2024$ outbursts, the source underwent spectral state transition from hard state (initial phase) to soft state (peak of the outburst). The reverse trend is observed during the decay phase of both outbursts. Overall, we observe canonical `q'-shape in HID which is common in BH-XRBs \citep{Nandi-etal2012,Radhika-etal2018,Sreehari-etal2019b,Blessy-etal2021,Katoch_etal2021,Geethu-etal2022}.

We generate {\textit{MAXI}} lightcurves in $2-6$ keV, $6-20$ keV and $2-20$ keV energy ranges for each outburst and compare the obtained results among different outbursts. The timing characteristics, such as peak flux, rising time, recurrence time and outburst duration determined from individual lightcurves are tabulated in Table \ref{tab:lc_proper}. The $2017-2018$ and $2018-2019$ outbursts appeared less luminous ($F_{\rm peak} \sim 0.2$ Photons cm$^{-2}$ s$^{-1}$  $\sim 50$ mCrab\footnote{$1$ Crab $=3.8$ Photons cm$^{-2}$ s$^{-1}$ in the $2-20$ keV band of {\textit{MAXI/GSC}}}) and have short duration ($T_{\rm D} \sim 120$ days) compared to $2019-2020$ and $2021$ outbursts ($F_{\rm peak} \sim 2$ Photons cm$^{-2}$ s$^{-1}$ $\sim 530$ mCrab and $T_{\rm D}=300$ days). Needless to mention that outbursts in $2017-2018$ and $2018-2019$ show shorter rising time ($T_{\rm peak} \sim 7-42$ days) compared to $2019-2020$, $2021$ and $2023-2024$ outbursts ($T_{\rm peak} \sim 115-157$ days). We also find that $2-20$ keV lightcurve lags $15-50$ keV lightcurve typically by $ \sim 13-58$ days during $2018-2019$, $2019-2020$, $2021$ and $2023-2024$ outbursts, whereas no such time delay is observed during $2017-2018$ outburst. 

With the above considerations, we categorise $2017-2018$ and $2018-2019$ outbursts as {\it failed} one, and consider $2019-2020$, $2021$ and $2023-2024$ outbursts as {\it successful} one. These findings are in agreement with the results reported earlier \cite[]{Garcia-etal2019,Paice-etal2019,Wang-etal2020,Liu-etal2021,Liu-etal2022}. However, we point out that $2022$ outburst is not monitored continuously by {\it MAXI} and hence, it is not straightforward to understand whether the outburst was {\it successful} or {\it failed}. 
 
\subsection{H 1743$-$322}

H 1743$-$322 displayed four outbursts, namely $2016a$, $2016b$, $2017$ and $2018$, during the period of $2016-2024$ as shown in Fig. \ref{fig:lc-maxi-bat}c. We present the HR variation in Fig. \ref{fig:lc-maxi-bat}d, and find that HR does not show noticeable variations during the entire observation period. Further, we plot the HID in the right panel of Fig. \ref{fig:HID} and observe insignificant HID evolution even in the $2016a$ successful outburst \cite[]{Swadesh-etal2020}. We also carry out the spectral analysis of H 1743$-$322 to examine the spectral properties during the state transition as discussed in section \ref{sec:spec_ana}. 

Similar to GX 339$-$4, we perform the comparative analysis of outburst characteristics ($i.e.$, peak flux, rising time, recurrence time and outburst duration) of H 1743$-$322 and obtained results are tabulated in Table \ref{tab:lc_proper}. All four outbursts yield similar peak flux $F_{\rm peak} \sim 0.2$ Photons cm$^{-2}$ s$^{-1}$ $\sim 50$ mCrab with shorter rising time as well as smaller duration ($T_{\rm D} \sim 30-100$ days). Note that we do not observe any time lag between lightcurves from $2-20$ keV and $15-50$ keV energy ranges.

\section{Spectral Analysis and Results}

\label{sec:spec_ana}

We generate wide band energy spectra ($0.7-60$ keV) combining {\it SXT} ($0.7-6$ keV) and {\it LAXPC20} ($4-60$ keV) data of all {\it AstroSat} observations for GX 339$-$4 and H 1743$-$322 sources. However, we consider spectral data upto $20$ keV during the quiescent states of these sources. For {\it NuSTAR} observations, we use spectral data from  FPMA in $3-60$ keV energy band.

The spectral analysis is done using \texttt{XSPEC V 12.11.1} \cite[]{Arnaud1996}) of \texttt{HEASOFT V 6.28}. We include $2\%$ systematics \cite[]{Sreehari-etal2019a, Antia-etal2021} for the spectral fitting of {\it AstroSat} observations. We incorporate gain corrections to all {\it SXT} and {\it LAXPC} fittings by using \texttt{gain fit} commands in \texttt{XSPEC} with fixed slope of $1$ to correct the instrumental residue at $1.8$ keV, and $2.2$ keV. For few {\it SXT} observations, we simultaneously use both \texttt{gain fit} and \texttt{edge} models at $1.8$ keV and $2.2$ keV. While fitting the {\it LAXPC} spectra, we use \texttt{gain fit} and \texttt{gaussian} at $\sim 30$ keV to account for the Xenon edge arising from the instrument \cite[]{Antia-etal2017,Antia-etal2021,Antia-etal2022}. Also, \texttt{TBabs} \cite[]{Wilms-etal2000} is used to account for the interstellar absorption of soft X-rays. We estimate the hydrogen column density $N_{\rm H}$ following \cite{Wilms-etal2000}. The relative cross-normalization between \textit{SXT} and \textit{LAXPC} is taken care by multiplying the model combination with a \textit{constant}. The value of cross-normalization constant is set to unity for {\it SXT}, however, it is freely varied for {\it LAXPC} data, which is found as $\sim 1$ for most of the spectral fits. 

The spectral modelling of GX 339$-$4 is carried out using basic thermal Comptonization model \texttt{Nthcomp} \cite[]{Zdziarski-etal1996,Zycki-etal1999}. In \texttt{Nthcomp} model, photon index {$\Gamma_{\rm {nth}}$}, electron temperature ($kT_{\rm e}$) and seed photon temperature ($kT_{\rm bb}$) are treated as free parameters, and seed photons are modelled as multi-color disc blackbody (\texttt{diskbb}). For example, we fit the epoch AS$1.01$ (see Table \ref{tab:obs_log}) spectrum (using \texttt{Nthcomp}) which belongs to the quiescent state resulted in a comparatively good fit with $\chi^{2}_{\rm red}= 1.24~(54 ~{\rm dof})$. On the contrary, the spectral fitting of epoch AS$1.04$ (LHS) requires an additional \texttt{smedge} component at $6.9$ keV leads to the acceptable fit with $\chi^{2}_{\rm red}= 0.95~(105~{\rm dof})$. Epoch AS$1.26$ belongs to intermediate state (IMS) which requires an additional thermal disc component \texttt{diskbb} along with \texttt{Nthcomp} and \texttt{smedge} for spectral fitting and good fit is obtained with $\chi^{2}_{\rm red}= 1.13~(595~{\rm dof})$. Similarly, we fit the {\it NuSTAR} spectrum (HSS; epoch NU$1.35$) using \texttt{diskbb}, \texttt{Nthcomp} and \texttt{gaussian} that resulted a good fit with {$\chi^{2}_{\rm red}=1.04~(542~{\rm dof})$}. Note that we use $N_{\rm H} = 0.5 \times 10^{22} ~ {\rm atoms} \, {\rm cm}^{-2} $ \cite[]{Mendez-etal1997} while fitting the spectra. However, for few observations, we treat $N_{\rm H}$ as free parameter and find that $N_{\rm H} \sim 0.40-0.67 \times 10^{22} ~ {\rm atoms} \, {\rm cm}^{-2}$. 

\begin{figure*}
    \centering
    \includegraphics[width=\columnwidth]{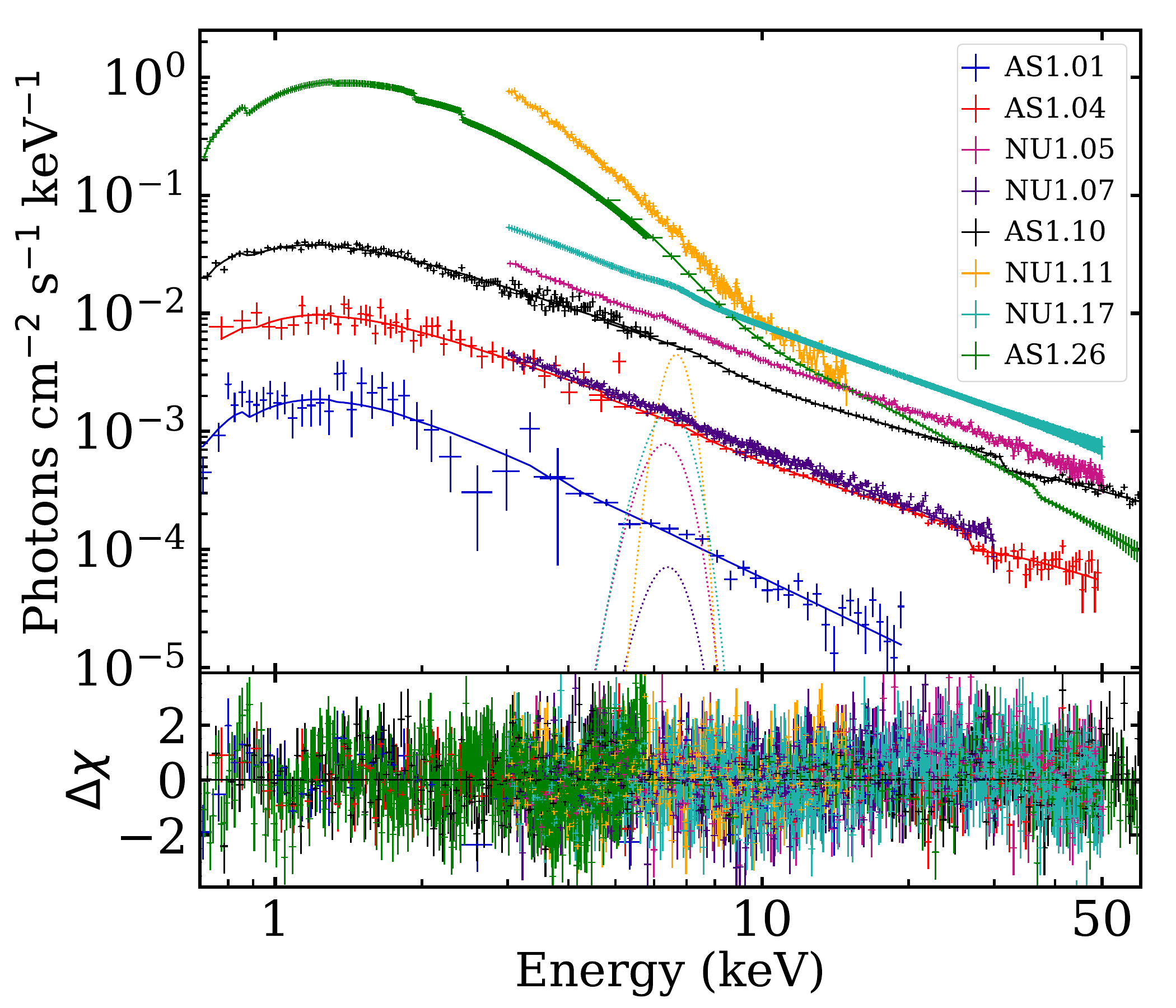}
    \includegraphics[width=\columnwidth]{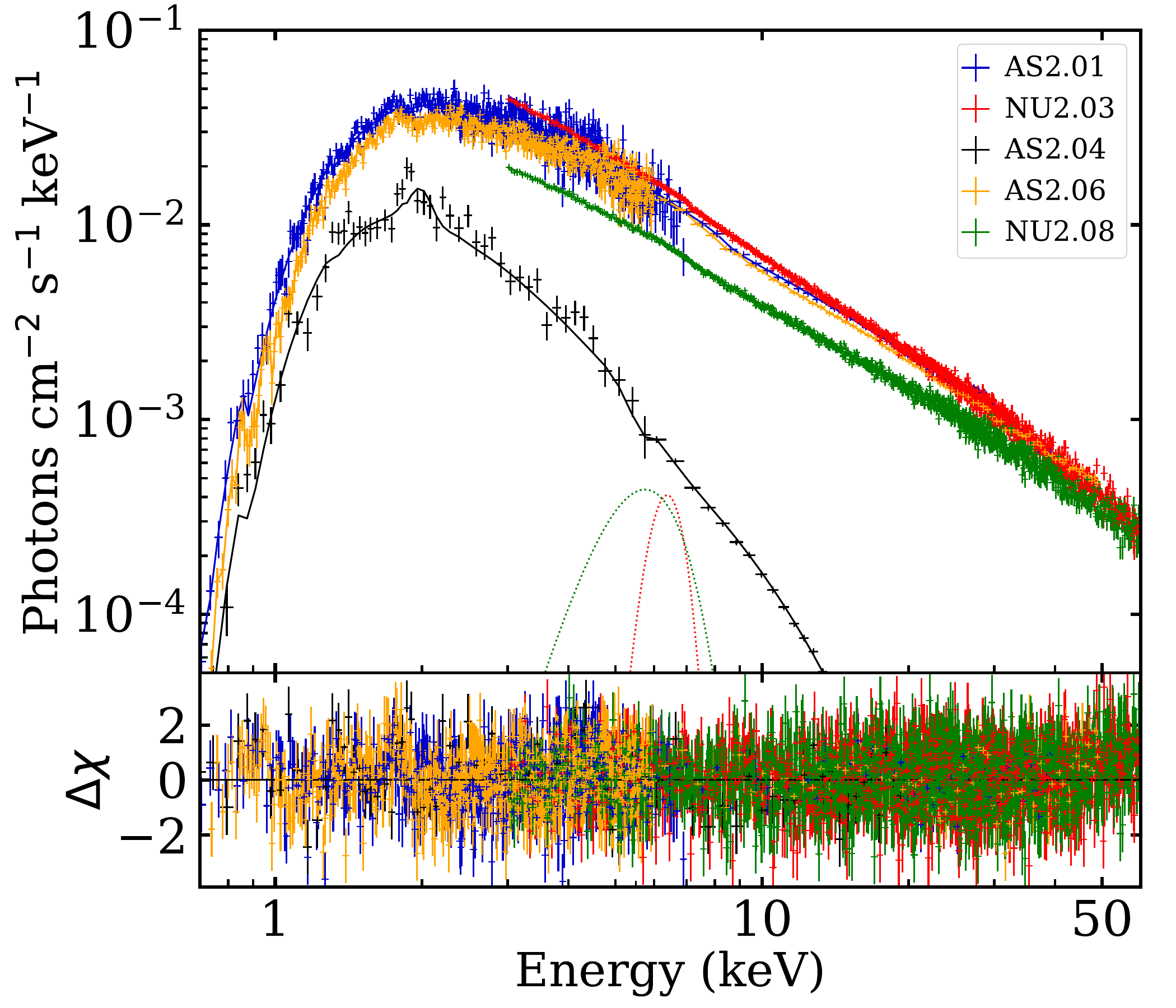}
    \caption{{\it Left panel:} Spectra of GX 339$-$4 during quiescence $-$ AS1.01 (blue), hard $-$ AS1.04 (red); NU1.05 (magenta); NU1.07 (navy blue); AS1.10 (black); NU1.17 (sky blue), intermediate $-$ AS1.26 (green), and soft $-$ NU1.11 (orange) states are shown. {\it Right panel:} Spectra of H 1743$-$322 during quiescence $-$ AS2.04 (black), and hard $-$ AS2.01 (blue); NU2.03 (red); AS2.06 (orange); NU2.08 (green) states are shown. Corresponding guassian emission line components are also shown. In both figures, bottom panels show residual variation of spectral fitting. See the text for details.}
    \label{fig:spec}
\end{figure*}

In our spectral analyses, we employ a uniform model combination as \texttt{TBabs$\times$(diskbb+gaussian+smedge$\times$Nthcomp}) to compare the spectral model parameters in different spectral states. In reality, different spectral states are associated with the different accretion scenarios, and hence, alteration of the above model combination is often required based on a given spectral state. For example, \texttt{diskbb} component is required to describe the thermal emission during IMS and HSS of GX 339$-$4, whereas LHS is satisfactorily described without \texttt{diskbb} component. Notably, Fe line feature at $6.4$ keV is prominently observed in {\it NuSTAR} spectra which is modelled with \texttt{gaussian} component. However, Fe line feature is hardly visible in {\it AstroSat} spectra possibly due to the poor resolution of the instrument, and {\it AstroSat} spectra are well fitted with \texttt{smedge} model component. Indeed, we find few exceptions, where additional \texttt{smedge} component is required for modelling the {\it NuSTAR} observations of GX 339$-$4. Note that the seed photon temperature ($kT_{\rm bb}$) and inner disc temperature ($kT_{\rm in}$) are tagged while modelling the combined spectra using \texttt{diskbb} and \texttt{Nthcomp} components. Following this, we carry out the spectral modelling of all the remaining observations in various epochs (see Table \ref{tab:obs_log}) and present the best fitted model parameters in Table \ref{tab:par_log}. For the purpose of representation, we depict the model fitted spectra of various spectral states of GX 339$-$4 in the left panel of Fig. \ref{fig:spec}. 

We carry out the spectral analyses of H 1743$-$322 using the identical model combinations as used for GX 339$-$4 source. In epoch AS$2.01$, H 1743$-$322 was in LHS and the best fit spectral modelling is obtained using \texttt{smedge} component at $7.93$ keV along with \texttt{Nthcomp} component, where $\chi^{2}_{\rm red}= 1.07~(500~{\rm dof})$. We fit {\it NuSTAR} spectrum (LHS, epoch NU$2.02$) using \texttt{Nthcomp}, \texttt{gaussian} at $6.25$ keV and \texttt{smedge} around $8$ keV to obtain an acceptable fit with {$\chi^{2}_{\rm red}= 1.02~(1413~{ \rm dof})$}. However, during QS of {\it AstroSat} observation (epoch AS$2.05$), we fit the spectrum using \texttt{Nthcomp} and \texttt{smedge} around $7$ keV that yields a good fit with $\chi^{2}_{\rm red}= 1.35~ (45~{\rm dof})$. Similarly, we carry out the spectral modelling of all the observations and present the extracted model fitted parameters in Table \ref{tab:par_log}. Note that for most of the observations, we treat $N_{\rm H}$ as free parameter yielding $N_{\rm H} \sim 1.8-2.7 \times 10^{22} \, {\rm atoms} \, {\rm cm}^{-2}$ for acceptable fit. We show the model fitted spectra of various states of H 1743$-$322 in the right panel of Fig. \ref{fig:spec}.

Using convolution model \texttt{cflux} in \texttt{XSPEC}, we estimate \texttt{diskbb} flux and \texttt{Nthcomp} flux in the energy range $0.7-50$ keV. Similarly, we also compute the bolometric flux ($F_{\rm bol}$) in $0.3-100$ keV range and obtain the bolometric luminosity as $L_{\rm bol} = F_{\rm bol} \times 4\pi d^{2}$, where $d$ is the source distance. We express $L_{\rm bol}$ in unit of Eddington luminosity ($L_{\rm Edd}$)\footnote{Eddington luminosity $\rm L_{\rm Edd}=1.26 \times 10^{38} {(M_{\rm BH}}/{\rm M_\odot})\, \rm erg\, s^{-1}$ for BH of mass $M_{\rm BH}$ \citep{Frank-etal2002}.}. We use system parameters as $M_{\rm BH}=9 {\rm M}_{\odot}$, $D=8.5$ kpc \citep{Parker-etal2016} for GX 339$-$4 and $M_{\rm BH}=11.21 {\rm M}_{\odot}$ \citep{Molla-etal2017}, $D=8.5$ kpc \citep{Steiner-etal2012} for H 1743$-$322 to estimate the flux values and bolometric luminosity, which are summarized in Table \ref{tab:par_log}.

In order to understand the nature of the Comptonizing medium in the vicinity of the source, we  estimate the optical depth ($\tau$) of the medium considering diffusion regime ($ \tau \gg 1$). Following \cite{Lightman-Zdziarski1987,Titarchuk-Lyubarskij1995,Zdziarski-etal1996}, we calculate the spectral index $\alpha ~(= \Gamma-1$) which is given by,  
\begin{equation}
	\label{eq:tau}
	\alpha=\bigg[\frac{9}{4}+\frac{1}{(kT_{e}/m_{e}c^{2})\tau(1+\tau/3)}\bigg]^{1/2}-\frac{3}{2},
\end{equation}
where $kT_{\rm e}$ is the electron temperature, $m_e$ is the electron mass, and $c$ refers speed of light. We also estimate the Compton y-parameter (y-par) \cite[see][]{Agrawal-etal2018,2021MNRAS.505.3785C}. For each observation under consideration, we tabulate both $\tau$ and y-par in Table \ref{tab:par_log}. 

\begin{table*}
    \centering
    \caption{Spectral parameters obtained from best fit modelling of GX 339$-$4 and H 1743$-$322 using \textit{AstroSat} and \textit{NuSTAR} observations during $2016-2024$. A model combination \texttt{TBabs$\times$(diskbb$+$gaussian$+$smedge$\times$Nthcomp)} is used for the spectral fitting of GX 339$-$4 and H 1743$-$322.
    The errors are quoted with $90\%$ confidence. { From left to right are (1) Epoch, (2) normalization constant for \texttt{diskbb} (3) electron temperature, (4) seed photon temperature, (5) power-law index, (6) Gaussian energy, (7) gaussian width, (8) gaussian normalisation, (9) smedge energy, (10) smearing width, (11) $\chi^{2}_{\rm red}$ ($\rm \chi^{2}/dof$, dof =degree of freedom) (12) unabsorbed disc flux in $0.7-50$ keV, (13) unabsorbed \texttt{Nthcomp} flux in $0.7-50$ keV, (14) bolometric luminosity in $0.3-100$ keV, (15) optical depth, (16) Compton y-parameter and (17) spectral states, respectively}. See the text for details.}
    
    \label{tab:par_log}
    \normalsize	
    \begin{adjustbox}{width=\textwidth,center=\textwidth}
    \Huge
    \begin{tabular}{l@{\hspace{0.2cm}}c@{\hspace{0.2cm}}c@{\hspace{0.2cm}}c@{\hspace{0.2cm}}c@{\hspace{0.2cm}}c@{\hspace{0.2cm}}c@{\hspace{0.2cm}}c@{\hspace{0.2cm}}c@{\hspace{0.2cm}}c@{\hspace{0.2cm}}c@{\hspace{0.2cm}}c@{\hspace{0.3cm}}c@{\hspace{0.4cm}}c@{\hspace{0.3cm}}c@{\hspace{0.2cm}}c@{\hspace{0.2cm}}r}
    \hline	\hline
			
    Epoch&\multicolumn{9}{c}{Model fitted parameters} & \multicolumn{5}{c}{Estimated parameters} & & Spectral State \\
    \cline{2-10}
    \cline{12-16}
    &$\rm N_{diskbb}$& $\rm kT_{e}$ & $\rm kT_{bb}^{*}$ &$\rm \Gamma_{nth}$  &E$_{\rm gaussian}$&W$_{\rm gaussian}$&Norm$_{\rm gaussian}$&E$_{\rm smedge}$&W$_{\rm smedge}$&$\rm \chi^{2}_{\rm red}$ (dof) & $\rm  F_{diskbb}$ &  $\rm F_{nth}$&$\rm L_{\rm bol}$ & $\tau$ & y-par&\\
			
    &(km$^2)$& (keV)  &  (keV) & &(keV)&(keV)& $\times 10^{-3}$Photons cm$^{-2}$s$^{-1}$&(keV) & (keV) & &($0.7-50$ keV) &($0.7-50$ keV) &($0.3-100$ keV) \\
    &	& & &  & & & & & & &  \textrm{$\times 10^{-9} {\rm erg ~cm}^{-2}~ {\rm s}^{-1}$} &$\times 10^{-9} {\rm erg ~cm}^{-2}~ {\rm s}^{-1}$ & ({$\% \,  L_{\rm Edd}$}) \\
    \hline  \hline\\
			
    \multicolumn{17}{c}{GX 339$-$4} \\\\
    \hline
			
    \rowcolor{lightgray}
			AS1.01& $-$ &$10 \dagger$ &$0.30$ $\dagger$&$2.10^{+0.10}_{-0.09}$ &$-$ & $-$ &$-$ &$-$ &$-$ &$1.24~(54)$ & $-$&$0.04$ & $0.03$ & $-$ & $-$ &QS\\
			\rowcolor{lightgray}
			
			AS1.02 & $-$ &$10 \dagger$&  $0.15 \dagger$ &$2.48^{+0.22}_{-0.12}$  &$-$ & $-$ &$-$ &$-$ &$-$ &  $1.32~(107)$& $-$ &$0.05$&$0.05$ & $-$ & $-$ &QS  \\
			
			NU1.03 & $-$ &$27.54^{+9.68}_{-5.73}$&$0.35 \pm 0.95$ & $1.66_{-0.03}^{+0.02}$&$6.42^{-0.68}_{+0.10}$&$0.19 \pm 0.07$&$0.09 \pm 0.02$&$-$&$-$&$1.06~ (542)$& $-$ & $0.27$  &  $2.78$& $4$ &	$2.68 \pm 0.02$&LHS\\
			
			AS1.04 & $-$ &$50^{\dagger} $&$0.14^{**}$ & $1.60\pm 0.04$&$-$&$-$&$-$&$6.90_{-0.84}^{+1.03}$&$4.02_{-0.78}^{+0.99}$&$0.95~(105)$& $-$ &$0.45$  &  $4.32$ & $-$ & $-$ &LHS\\
			
			NU1.05 & $-$ &$22.30^{+7.37}_{-5.56}$&$0.213 \pm {0.003}$ & $1.570 \pm {0.003}$&$6.33 \pm 0.10 $&$0.32 \pm 0.26$ & $0.25 \pm 0.08$&$7.08 \pm 0.09$&$5\dagger$&$1.10~(1292)$& $-$ &$2.83 $  &  $2.58$ & $5$	& $3.54 \pm 0.10$&LHS\\
			
			NU1.06 & $-$ &$36.02^{+16.05}_{-7.08}$&$0.214 \pm 0.003$ & $1.582 \pm 0.007$&$6.39_{-0.15}^{+0.12}$&$0.31 \pm 0.08$&$0.5 \pm 0.1$&$-$&$-$&$1.00~(1165)$& $-$ &$2.43$  &  $2.68$ &$4$ &	$3.37 \pm 0.14$&LHS\\
			NU1.07 & $-$ &$30.34^{+7.59}_{-5.02}$&$0.40 \pm 0.006$ & $1.64_{-0.02}^{+0.01}$&$6.52 \pm 0.07$&$0.38 \pm 0.19$&$0.09 \pm 0.03$&$-$&$-$&$1.10~(668)$& $-$ &$0.43$  &  $4.39$& $3$ &	$2.73 \pm 0.69$&LHS\\
			
			NU1.08& $-$ &$4.33 \pm 0.83$&$0.43 \pm 0.03$ & $1.74^{+0.08}_{-0.09}$&$6.66 \pm 0.30$&$0.36 \pm 0.22$&$0.39 \pm 0.18$&$-$&$-$&$1.34~(218)$& $-$ &$0.51$  &  $4.16$ & $9$ &	$3.33 \pm 0.64$&LHS\\
			
			NU1.09& $-$ &$19.89^{+4.61}_{-3.18}$&$0.38 \pm 0.01$ & $1.57^{+0.08}_{-0.09}$&$6.56 \pm 0.13$&$0.34_{-0.12}^{+0.15}$&$0.48 \pm 0.14$&$-$&$-$&$1.33 (219)$& $-$ &$1.56$  &  $1.81$ & $5$	& $3.64 \pm 0.88$&LHS\\
			
			AS1.10 & $-$ &$33.18 \pm 13.63$&$0.43 \pm 0.03$ & $1.543 \pm 0.006$&$-$&$-$&$-$&$8.09_{-0.10}^{+0.18}$&$2.00 \pm 0.41$&$1.09~(480)$& $-$ &$2.07$  &  $2.08$ & $4$ &	$3.41 \pm 1.40$&LHS\\
			
			NU1.11  & $1634^{+67}_{-72}$ &$10\dagger$&$0.88_{-0.02}^{+0.03}$ & $2.69 \pm 0.02$&$6.67 \pm 0.12$&$0.4^\dagger$&$4.49 \pm 1.81$&$-$&$-$&$1.11~(293)$&$17.40$  &$6.07$  &  $30.06$ & $-$ &$-$&HSS\\
			
			NU1.12 & $2103 \pm 5.21$&$50 \dagger$&$0.66 \pm 0.02$& $2.11 \pm 0.07$&$6.40 \pm 0.08$ &$1.02 \pm 0.04$&$1.87 \pm 0.06$&$-$&$-$  &$1.08~ (540)$ &$6.37$&$0.22$  & $6.29$ & $-$ & $-$ &IMS\\
			
			NU1.13  & $2334^{+123}_{-101}$&$50^\dagger$&$0.64 \pm 0.01$& $2.27 \pm 0.01$ &$6.40_{-0.25}^{+0.18}$&$0.88^{+0.15}_{-0.12}$&$1.61 \pm 1.00$&$-$&$-$&$1.20~(417)$   &$5.76$&$0.20$  & $5.78$ &  $-$& $-$&IMS\\
			
			NU1.14   &  $1083_{-68}^{+76}$ &$50^\dagger$&$0.63 \pm 0.03$ & $2.13 \pm 0.02$&$6.02 \pm 0.04$&$1.00^\dagger$&$0.4 \pm 0.1$&$-$&$-$&$1.13~(543)$& $2.47$ &$0.52$  &  $2.90$ & $-$ & $-$&IMS\\
			
			NU1.15 & $-$&$60.75_{-17.30}^{+38.32}$&$0.355 \pm 0.004$& $1.700 \pm 0.002$&$6.39 \pm 0.26$&$0.49 \pm 0.13$&$0.23 \pm 0.06$&$-$&$-$&$1.20~(918)$& $-$ &$1.27$  &  $1.31$ & $2$ & $1.83 \pm 0.53$&LHS\\
			
			NU1.16& $-$ &$34.37^{+15.14}_{-12.73}$&$0.22 \pm 0.01$ & $1.58\pm 0.01$&$6.45^{+0.06}_{-0.07}$&$0.18 \pm 0.08$&$0.20 \pm 0.04$&$6.79 \pm 0.31$&$0.20\dagger$&$0.98~(915)$& $-$ &$1.32$  &  $1.45$ & $3$ & $2.81 \pm 0.25$&LHS\\
			
			NU1.17  & $-$  &$32.45^{+2.88}_{-2.53}$ &$0.120 \pm 0.003$&$1.613 \pm 0.002$  &$6.40 \pm 0.09$&$0.13 \pm 0.07$&$0.47 \pm 0.10$&$7.36 \pm 0.08$&$2\dagger$& $1.11~(1290)$ & $-$& $5.41$ & $5.70$ & $3$ & $2.86 \pm 0.22$&LHS\\
			
			AS1.18 & $-$ & $18.11_{-2.37}^{+3.20}$ &$0.22 \pm 0.01$ & $1.675 \pm 0.002$ &$-$&$-$&$-$&$7.41_{-0.57}^{+0.53}$&$4.41 \pm 0.85$& $1.23~(544)$ & $-$ &$4.91$ & $6.86$ & $5$ & $2.91 \pm 0.38$& LHS\\
			
			NU1.19& $-$  &$27.11^{+4.54}_{-2.85}$ &$0.203 \pm 0.004$&$1.657 \pm 0.001$  &$6.48 \pm 0.09$&$0.26 \pm 0.09$&$1.02 \pm 0.30$&$7.37 \pm 0.09$&$5\dagger$& $1.06~(1165)$ & $-$ & $9.57$ & $9.34$ & $4$ & $2.93 \pm 0.10$ &LHS\\
			
			AS1.20& $-$&$50\dagger$& $0.25 \pm 0.07$ & $1.65 \pm 0.01$&$-$&$-$&$-$&$6.99 \pm 0.10$&$5\dagger$&$1.23~(588)$&$-$ &$4.00$ &$9.13$ & $-$ & $-$& LHS\\
			
			AS1.21& $-$ & $50\dagger$ &$0.25 \pm 0.15$ &$1.666 \pm 0.001$&$-$&$-$&$-$&$7.14 \pm 0.11$&$5\dagger$ & $1.30~(410)$ & $-$ &$5.91$ & $6.88$ & $-$ &$-$ &LHS\\
			
			AS1.22& $-$& $71.99 \pm 14.21$&$0.25 \pm 0.01$&$1.671 \pm 0.003$&$-$&$-$&$-$&$6.97 \pm 0.33$&$5^\dagger$  &$1.33~(339)$ &$-$ & $6.09 $& $7.07$ &$2$ & $1.83 \pm 0.37$& LHS\\
			
			AS1.23& $-$ & $106.82 \pm 19.06$ & $0.16 \pm 0.02$ & $1.68 \pm 0.01$&$-$&$-$&$-$&$6.94 \pm 0.29$&$4.74 \pm 3.33$&$1.15~(374)$ &$-$ &$6.78$ & $6.95$ & $1$ & $1.46 \pm 1.11$&LHS\\
			
			NU1.24& $-$  &$39.27 \pm 19.09$ &$0.29^{+0.03}_{-0.05}$&$1.68 ^{+0.17}_{-0.20}$&$6.42_{-0.07}^{+0.05}$&$0.27 \pm 0.07$&$1.0 \pm 0.2$&$7.32 \pm 0.07$&$4.15 \pm 1.98$  & $1.19~(914)$ &$-$& $10.32$ & $11.56$ & $3$ & $2.28 \pm 0.28$&LHS\\
			
			NU1.25& $1262^{+36}_{-25}$  &$50^\dagger$ &$0.61 \pm 0.03$&$2.13 \pm 0.02$  &$6.09 \pm 0.12$&$0.60\dagger$&$6.75 \pm 0.23$&$7.71 \pm 0.04$&$5^\dagger$& $1.10~(916)$ &$2.62$& $11.10$ & $12.12$ & $-$ & $-$&IMS\\
			
			AS1.26 & $1208^{+40}_{-55}$  &$50^\dagger$ &$0.879^{+0.002}_{-0.003}$&$2.40 \pm 0.04$  &$-$&$-$&$-$&$7.67_{-0.11}^{+0.26}$&$5\dagger$& $1.13~(595)$ &$12.41$& $4.65$ & $14.84$ &$-$ & $-$&IMS\\
			
			AS1.27&$1427 \pm 33$   & $50\dagger$ &$0.84 \pm 0.02$& $2.43 \pm 0.14$&$-$&$-$&$-$&$7.61 \pm 0.12$ &$5\dagger$& $1.13~(583)$&$11.95$ &$3.43$  &  $12.93$ & $-$ &$-$ &IMS\\
			
			AS1.28&$1450^{+38}_{-47}$  &$50\dagger$ &$0.84 \pm 0.03$ &$2.37 \pm 0.05$&$-$&$-$&$-$&$7.81 \pm 0.36$&$5\dagger$ &$1.12~(598)$ & $11.90$ & $2.97$  &  $12.77$ & $-$ & $-$& IMS\\
			
			NU1.29&$2132^{+52}_{-50}$  &$50\dagger$ &$0.81^{+0.04}_{-0.17}$ &$2.19 ^{+0.02}_{-0.03}$&$6.40 \pm 0.11$ &$1.02 \pm 0.04$&$11.2 \pm 0.40$&$-$&$-$ &$1.08~(542)$ & $15.61$ & $0.24$  &  $16.09$ & $-$ & $-$ &IMS\\
			
			AS1.30&$1166^{+36}_{-43}$  & $50^\dagger$ &$0.83 \pm 0.01 $ & $2.27 \pm 0.04$&$-$&$-$&$-$&$7.45 \pm 0.32$ &$5\dagger$ &$1.29~(596)$ &$9.49$&$1.68$  & $10.26$ & $-$ &$-$&IMS \\
			
			AS1.31  &  $1562 \pm 12$&  $11.43^{+4.19}_{-4.12}$ &$0.56 \pm 0.03$& $2.17 \pm 0.12$ &$-$&$-$&$-$&$6.95 \pm 0.32$&$5\dagger$& $1.18~(572)$&$13.69$&$3.07$& $14.17$ & $4$ &$1.39 \pm 0.24$ &IMS\\
			
			AS1.32&$1754^{+28}_{-15}$    & $50^\dagger$ & $0.58^{+0.12}_{-0.33}$ & $1.97^{+0.04}_{-0.02}$&$-$&$-$&$-$&$6.71 \pm 0.72$&$5^\dagger$&$1.11~(582)$ &$14.73$&$1.65$  & $13.47$ & $1$ & $0.87 \pm 0.08$&IMS\\
			
			AS1.33 &$1712^{+23}_{-32}$  & $37.16 \pm 9.62$ &$0.8634 \pm 0.0004$&$1.78^{+0.22}_{-0.41}$&$-$&$-$&$-$&$7.89\pm 0.16$&$0.100 \pm 0.003$ &$1.14~(595)$ &$16.21$&$0.06$  & $14.89$ & $1$ & $1.06 \pm 0.76$ & IMS \\
			
			AS1.34  &$2602^{+30}_{-34}$  & $16.78^{+0.31}_{-1.03}$ &$0.852 \pm 0.003$&$1.82^{+0.13}_{-0.03}$&$-$&$-$&$-$&$7.40_{-0.32}^{+0.24}$&$5\dagger$ &$1.18~(593)$ &$26.06$&$0.08$  & $22.60$ & $4$ & $2.21 \pm 0.10$& HSS \\
			
			NU1.35 &$2538^{+41}_{-39}$  & $32.36 \pm 12.78$ &$0.83 \pm 0.01$&$1.99^{+0.03}_{-0.04}$ &$6.40 \pm 0.11$&$0.78 \pm 0.06$ &$4.79 \pm 0.60$&$-$&$-$&$1.04~(542)$ &$20.09$&$0.82$  & $19.59$ & $2$ & $1.30 \pm 0.54$&HSS\\
			
			NU1.36 &$2939^{+24}_{-15}$  & $5.18^{+0.18}_{-0.17}$ &$0.82 \pm 0.03$&$1.46^{+0.09}_{-0.10}$&$6.18_{-0.35}^{+0.23}$&$0.78_{-0.16}^{+0.21}$&$4.70 \pm 0.23$&$-$&$-$&$1.14~(541)$ &$22.95$&$0.02$  & $20.88$ & $12$ & $6.05 \pm 0.22$&HSS \\
			
			NU1.37 &$2953_{-24}^{+190}$  & $5.79^{+0.24}_{-0.84}$ &$0.82^{+0.02}_{-0.05}$&$1.75^{+0.37}_{-0.21}$&$6.16_{-0.79}^{+0.33}$&$0.6^\dagger$&$4.38 \pm 0.06$&$-$&$-$&$1.10~(366)$ &$23.25$&$0.03$  & $21.26$ & $8$ & $3.13 \pm 0.18$&HSS \\
			
			NU1.38 &$2747 \pm 16$  & $1.63^{+0.07}_{-0.03}$ &$0.823^{+0.004}_{-0.002}$&$3.26 \pm 0.05$ &$-$&$-$&$-$&$-$&$-$ &$1.06~(173)$ &$20.35$&$1.41$  & $20.19$ & $8$ & $0.72 \pm 0.04$&HSS\\
			
			NU1.39& $-$& $50^{\dagger}$&$0.250 \pm 0.003$&$1.698 \pm 0.001$&$6.24 \pm 0.21$&$0.64 \pm 0.24$&$0.22 \pm 0.08$&$-$&$-$&$1.08~(532)$& $-$ &$0.88$&$0.91$& $-$ & $-$&LHS\\
			
			\rowcolor{lightgray}
			NU1.40& $-$  & $10.69^{+3.36}_{-1.87}$ &$0.12\dagger$&$1.71 \pm 0.02$&$-$&$-$&$-$&$-$&$-$  &$1.08~(495)$ & $-$&$0.08$  & $0.06$ & $6$ &$3.04 \pm 0.06$&QS\\
			
			\rowcolor{lightgray}
			NU1.41 & $-$  & $4.79^{+0.60}_{-0.41}$ &$0.20 \dagger$&$1.76 \pm 0.01$ &$-$&$-$&$-$&$-$&$-$ &$1.07~(344)$ &$-$&$0.06$  & $0.04$ & $8$ & $3.17 \pm 0.05$&QS\\
			
			AS1.42&$-$&$15.48_{-1.15}^{+1.62}$& $0.57_{-0.02}^{+0.01}$&$1.58 \pm 0.01$&$-$&$-$&$-$&$8.02 \pm 0.12$&$5\dagger$&$1.28~(529)$ &$-$&$1.12$ &$1.58$&$6$&$3.75 \pm 0.28$&LHS\\
			
			AS1.43&$-$&$32.90^{+12.55}_{-5.75}$&$0.773 \pm 0.001$&$1.602 \pm 0.011$&$-$&$-$&$-$&$7.75 \pm 0.15$&$5\dagger$ &$1.11~(527)$&$-$&$0.55$&$1.85$&$3$&$2.92 \pm 0.12$&LHS\\
			
			NU1.44&$-$& $55.72_{-6.49}^{+10.22}$ &$0.33 \pm 0.01$&$1.582 \pm 0.001$&$6.34^{+0.11}_{-0.14} $&$0.23_{+0.14}^{-0.15}$&$0.34 \pm 0.11$&$-$&$-$&$1.26~(752)$&$-$&$2.29$&$2.51$&$2$&$2.58 \pm 0.30$&LHS\\
			
			\rowcolor{lightgray} AS1.45 &$-$&$10^{\dagger}$&$0.14^{\dagger}$&$2.79_{-0.23}^{+0.26}$&$-$&$-$&$-$&$-$&$-$&$1.02~(87)$&$-$&$0.01$&$0.01$&$-$&$-$&QS\\

		      NU1.46&$-$&$24.98^{+2.22}_{-1.44}$&$0.22 \pm 0.08$&$1.722 \pm 0.003$& $6.42^{+0.08}_{-0.07}$ &$0.34 \pm 0.08$& $1.99 \pm 0.34$ &$7.62 \pm 0.04$ &$5^\dagger$&$1.07 ~(1165)$&$-$&$9.02$&$9.61$&$4$&$2.41 \pm 0.22$&LHS \\
		      NU1.47&$1920_{-90}^{+97}$ &$50^\dagger$&$0.80 \pm 0.01$ &$2.48 \pm 0.03$&$6.17 \pm 0.03$&$0.6^\dagger$&$6.20 \pm 0.60$ &$7.98 \pm 0.05$&$5^\dagger$& $1.11~(791)$ &$13.35$ &$4.31$&$15.69$&$-$&$-$&IMS\\

                NU1.48&$1449_{-26}^{+22}$&$50^\dagger$&$0.83 \pm 0.03$ &$2.43 \pm 0.02$ &$6.20 \pm 0.05$ &$0.6^\dagger$ &$7.43_{-0.30}^{+0.45}$ &$7.51 \pm 0.11$ &$5\dagger$ &$1.22~(791)$ &$11.51$ &$5.45$ &$14.91$ &$-$&$-$&IMS\\

                AS1.49&$1772^{+107}_{-102}$&$50^\dagger$&$0.77 \pm 0.02$&$2.49^{+0.10}_{-0.09}$&$-$&$-$&$-$&$7.46 \pm 0.45$ &$5^\dagger$&$0.82~(527)$&$10.13$&$6.92$&$14.42$&$-$&$-$&IMS\\
		
                AS1.53&$2134_{-78}^{+80}$&$50^{\dagger}$&$0.61 \pm 0.61$&$2.66_{-0.08}^{+0.07}$&$-$&$-$&$-$&$7.31 \pm 0.16$&$5^\dagger$&$1.11~(495)$&$4.44$&$0.72$&$4.78$ &$-$&$-$ &IMS\\
  
			\hline\\

                \multicolumn{17}{c}{H 1743$-$322} \\\\
			\hline
			AS2.01 & $-$ &$17.37^{+5.12}_{-3.05}$  &$0.73^{+0.01}_{-0.02}$& $1.66 \pm 0.02$ &$-$&$-$&$-$&$7.93 \pm 0.76$&$0.70 \pm 0.01$& $1.07~(500)$ & $-$ &$3.37$  &$2.66$ & $5$ &$3.03 \pm 0.16$ & LHS\\
			
			NU2.02& $-$ &$19.17^{+0.90}_{-0.83}$  &$0.246^{+0.002}_{-0.001}$  & $1.700 \pm 0.002$ &$6.25 \pm 0.05$&$0.47 \pm 0.07$&$0.80 \pm 0.10$&$8.13 \pm 0.18$&$2.48 \pm 1.18$& $1.02 ~(1413)$ & $-$&$4.57$  &$3.48$ &$4$ &$2.72 \pm 0.12$& LHS \\
			
			NU2.03&  $-$&$19.06^{+1.04}_{-1.11}$  &$0.26 \pm 0.01$  & $1.71 \pm 0.01$ &$6.38 \pm 0.02$&$0.50 \pm 0.11$&$0.50 \pm 0.10$& $7.57_{-0.40}^{+0.50}$&$4.31 \pm 0.28$&$1.04 ~(1413)$ & $-$&$4.40$  &$3.33$ &$4$ &$2.66 \pm 0.10$ & LHS\\
			
			\rowcolor{lightgray}
			
			AS2.04  & $-$  &$16.13 \pm 3.77$& $0.130 \pm 0.005$ &$2.22^{+0.12}_{-0.07}$&$-$&$-$&$-$&$6.89 \pm 0.35$ &$5^{\dagger}$ &$1.36~(179)$  & $-$& $0.23$ &$0.16$& $3$ & $1.17 \pm 0.25$ &QS\\
			
			\rowcolor{lightgray}
			AS2.05 & $-$ &$3.61_{-0.19}^{+0.32}$&$0.13 \pm 0.03$&$1.56 \pm 0.03$&$-$&$-$&$-$&$7.64 \pm 0.47$&$5\dagger$&$1.35~(45)$& $-$ &$0.13$  &$0.10$ & $13$ & $4.90 \pm 0.26$ &QS \\
			
			AS2.06 & $-$ & $50^{\dagger}$ &$0.96 \pm 0.05$  &$1.68 \pm 0.01$&$-$&$-$&$-$&$7.29_{-0.44}^{+0.48}$&$5\dagger$&$1.01~(561)$ &$-$&  $3.24$  & $2.62$ & $-$ &$-$& LHS\\
			
			NU2.07& $-$&$18.22^{+0.86}_{-0.77}$  &$0.640 \pm {0.018}$  & $1.60 \pm 0.01$&$6.31_{-0.20}^{+0.84}$&$0.52_{-0.19}^{+0.24}$&$0.56^{+0.03}_{-0.04}$ &$7.61 \pm 0.31$&$1.46_{-0.42}^{+0.58}$& $1.08~(1413)$ &$-$&$4.11$  &$3.21$ & $5$ &$3.50 \pm 0.15$ & LHS\\
			
			NU2.08& $-$&$22.15^{+1.64}_{-1.63}$  &$0.492 \pm 0.002$  & $1.57 \pm 0.01$&$6.25^{+0.23}_{-0.31}$&$0.39^{+0.01}_{-0.06}$&$0.22_{-0.03}^{+0.08}$&$7.03 \pm 0.31$&$2.97 \pm 0.08$ & $1.09~(1416)$ & $-$&$2.37$  &$2.08$ &$5$ & $3.51 \pm 0.10$& LHS  \\
			\hline	
		\end{tabular}		
	\end{adjustbox}
	\small
	\begin{list}{}{}
	\item $\dagger$ Frozen parameter. { * kT$_{\rm in}$ and kT$_{\rm bb}$ are tagged}. ** Error values are insignificant.
	\end{list}
	
\end{table*}

\subsection{GX 339$-$4}

During {\it AstroSat} and {\it NuSTAR} observations, GX 339$-$4 underwent spectral transitions from quiescence to outburst phases and vice versa. We notice that during the outburst phases, the source transits between hard to soft states via intermediate states (IMS). The spectral modelling of different states results in $\tau \sim 1-12$. These findings infer the presence of optically thick corona surrounding the source. Moreover, we obtain y-par in the range $0.72-6.05$. In the subsequent sections, we present the spectral properties of the source in different spectral states.
  
\subsubsection{Quiescent State (Epoch AS1.01-1.02,  NU1.40-1.41, AS1.45)}

During the five epochs of QS, the photon index ($\Gamma_{\rm nth}$) varies in the range $1.71 - 2.48$ and the corresponding HR value lies in between $1.13-2.47$. This evidently indicates that the spectral nature of the source tends towards the harder state. From {\it NuSTAR} observations during epoch NU1.40 and NU1.41, the electron temperature $kT_{\rm e}$ is found to be $10.69$ keV and $4.79$ keV, respectively, whereas $kT_{\rm e}$ is kept frozen to $10$ keV during {\it AstroSat} observations (see Table \ref{tab:par_log}). Subsequently, we obtain \texttt{Nthcomp} flux as $(0.04-0.08) \times 10^{-9}\,\rm  erg \, cm^{-2} \, s^{-1}$, and the source luminosity ($L_{\rm bol}$) is found to be low as $(0.03-0.06) \% {\rm L}_{\rm Edd}$ during quiescent phase. 

\subsubsection{Hard State (Epoch NU1.03, AS1.04, NU1.05-1.09, AS1.10, NU1.15-1.17, AS1.18, NU1.19, AS1.20-1.23, NU1.24, NU1.39, AS1.42-1.43, NU1.44, NU1.46)}

GX 339$-$4 was observed in LHS during twenty three epochs of {\it AstroSat} and {\it NuSTAR}. Spectral analyses reveal that $ \Gamma_{\rm nth} \sim 1.54-1.74$ and hardness $\sim 0.89-2.95$. Moreover, we find that the seed photon temperature $kT_{\rm bb}$ varies as $\sim 0.12-0.77$ keV without any strong signature of disc emission, and  $kT_{\rm e} \sim 4.33 - 106.82$ keV. We observe Fe line feature comprising of the Gaussian width and normalisation that vary as $0.13-0.64$ keV and  $\rm (0.09-1.02) \times 10^{-3}~ photons \, cm^{-2} \, s^{-1}$, respectively. We find that the {\it Nthcomp} flux is varied in the range $\rm (0.27-10.32) \times 10^{-9}~ erg \, cm^{-2} \, s^{-1}$, and obtain $L_{\rm bol} \sim (0.91-11.56) \% \, {\rm L}_{\rm Edd}$ which is significantly higher than QS state. The best fit spectral parameters are tabulated in Table \ref{tab:par_log}.

\subsubsection{Intermediate State (Epoch NU1.12-1.14, NU1.25, AS1.26-1.28, NU1.29, AS1.30-1.33, NU1.47-1.48, AS1.49, AS1.53)}

{\it AstroSat} and {\it NuSTAR} observed GX 339$-$4 sixteen times in IMS which render $kT_{\rm in} \sim 0.56-0.88$ keV signifying hotter disc, $\Gamma_{\rm nth} \sim 1.76 - 2.66$ and HR $\sim 0.13-1.26$. In addition, electron temperature exhibits variation as $kT_{\rm e} \sim 11.43-50.0$ keV. We find that the disc temperature reaches its maximum value ($ kT_{\rm in} \sim 0.88$ keV) during the epoch AS1.26. As in LHS, Fe line signature is also observed in IMS, however with stronger features having width as $0.60-1.02$ keV and normalization as $(0.40-11.20) \times 10^{-3} ~{\rm photons \, cm^{-2} \, s^{-1}}$. Further, we compute fluxes corresponding to \texttt{diskbb} and \texttt{Nthcomp}, which are obtained as $(2.47-16.21) \,\times \rm  10^{-9} \, \rm erg \, cm^{-2} \, s^{-1}$ and $(0.06-11.10) \, \times \rm  10^{-9} \, erg \, cm^{-2} \, s^{-1}$, respectively. During IMS, the source luminosity is seen to vary in the range $L_{\rm bol}\sim (2.90-16.09) \, \% {\rm L}_{\rm Edd}$. The details of extracted spectral fitted parameters are presented in Table \ref{tab:par_log}. 
   
\subsubsection{Soft State (Epoch NU1.11, AS1.34, NU1.35-1.38)}

During six epochs of {\it AstroSat} and {\it NuSTAR} observations, GX 339$-$4 was observed in HSS. We find $kT_{\rm in} \sim 0.82-0.88$ keV and $\Gamma_{\rm nth} \sim 1.46-3.26$. The corona temperature is found to vary between $kT_{\rm e} \sim 1.63-32.36$ keV. We observe that the hardness ratio (see Table \ref{tab:lc_proper}) varies as $0.09-0.35$ confirming the source spectral nature as soft state. Notably, in HSS, we find Fe line signature as in IMS. The source luminosity is estimated as $L_{\rm bol} \sim (19.59-30.06)\%{\rm L}_{\rm Edd}$, and the fluxes corresponding to $\texttt{diskbb}$ and \texttt{Nthcomp} are $(17.40-26.06) \, \times \rm  10^{-9} \, erg \, cm^{-2} \, s^{-1}$ and $(0.02-6.07) \, \times \rm  10^{-9} \, erg \, cm^{-2} \, s^{-1}$, respectively. The model fitted and estimated parameters are mentioned in Table \ref{tab:par_log}. 

\subsection{H 1743$-$322}

H 1743$-$322 mostly remained in quiescence and occasionally underwent outburst (see Fig. \ref{fig:lc-maxi-bat}). Interestingly, during the outburst phase, the source was seen to be in the hard state only. The detail spectral analysis infer $\tau \sim 3-13$ suggesting the presence of an optically thick corona in the vicinity of the source. In addition, we observe the y-par to vary between $1.17-4.90$.

\subsubsection{Quiescent State (Epoch AS2.04 and AS2.05)}

{\it AstroSat} observed the source two times during QS and the source spectra are characterized by a steep hard tail with $\Gamma_{\rm nth}$ varied in the range $1.56 -2.22$. We estimate the \texttt{Nthcomp} flux in between  $(0.13-0.23) \, \times \rm  10^{-9} \, erg \, cm^{-2} \, s^{-1}$. The source seemed to be in a low luminous state with the bolometric luminosity $L_{\rm bol} \sim (0.10-0.16) \% \,{\rm L}_{\rm Edd}$. We tabulate the fitted and estimated spectral parameters during these epochs in Table \ref{tab:par_log}.

\subsubsection{Hard State (Epoch AS2.01, NU2.02, NU2.03, AS2.06, NU2.07, NU2.08)}

{\it AstroSat} and {\it NuSTAR} observed H 1743$-$322 during six epochs in LHS. The spectral analyses of these observations yield $\Gamma_{\rm nth} \sim 1.57-1.71$ along with HR $\sim 2.04 - 3.04$ signifying the hard spectral state. Note that we find weak signature of seed photon temperature $kT_{\rm bb} \sim 0.25 -0.96$ keV without any thermal disc component. We also observe Fe line feature at $\sim 6.4$ keV. We find that the source luminosity is enhanced compared to the QS as $L_{\rm bol} \sim (2.08-3.48)\% \, {\rm L}_{\rm Edd}$. All the fitted and estimated parameters are summarized in Table \ref{tab:par_log}.

\section{Timing Analysis and Results}
\label{sec:timing}

In order to examine the timing variability of GX 339$-$4 and H 1743$-$322, we generate PDS using $5$ ms lightcurves in different energy bands of {\it LAXPC} ({\it LAXPC10} and {\it LAXPC20}) and {\it NuSTAR} ({\it FPMA} and {\it FPMB}) observations. We use \texttt{powspec} tool to generate PDS in $0.001-30$ Hz frequency range. For GX 339$-$4 source, we divide {\it LAXPC} lightcurve into intervals of $8192$ time bins and construct the PDS for each of these intervals. We generate Leahy-normalised PDS using these segments and eventually average them to obtain the final PDS. Further, the average PDS is binned geometrically by a factor of $-1.02$ in the frequency space. Similarly, for {\it NuSTAR} data, we divide the lightcurve into $16384$ time bins and obtain the resultant PDS after re-binning geometrically by a factor of $-1.03$. In case of H 1743$-$322, we divide both {\it LAXPC} and {\it NuSTAR} lightcurves into $8192$ time bins and rebin the resultant PDS with a geometrical rebinning factor of $-1.03$. 

We carry out the model fitting of the PDS using a \texttt{constant}, zero-centered \texttt{Lorentzian} and \texttt{Lorentzian} to identify the QPO features \cite[]{Belloni-Hasinger1990}. In doing so, we avoid correcting the Poisson noise contribution. The functional form of \texttt{Lorentzian} used in \texttt{XSPEC} is given by,
\begin{equation}
L(\nu)= \frac{K}{\pi} \frac{\Delta}{(\nu -\nu_{\rm c})^2+\Delta^2} ,
\end{equation}
where $\nu$ is the frequency and $\nu_{\rm c}$ is the central frequency representing the characteristic QPO frequency. Here, $\Delta$ is the half width at half maximum (HWHM) of the line and $K$ is the normalization. For each observation describing QPO, we estimate $\nu_{\rm c}$, $\Delta$, and $K$. Subsequently, we compute quality factor ($Q=\nu_{\rm c}/2\Delta$) and significance (ratio of \texttt{Lorentzian} normalization to its negative error) \cite[see][and references therein]{Alam-etal2014,Sreehari-etal2019a} that yield confirmed QPO detection provided $Q \geq 3$ and significance $\sigma \geq 3$. Further, the temporal variability in the lightcurve is parametrized by means of fractional root mean square (rms) variability. When PDS is Leahy normalised \cite[]{Leahy-etal1983}, the fractional rms amplitude of each QPO is calculated as the square root of the definite integral of
the Lorentzian representing the QPO in frequency–power space. Subsequently, the total rms of the PDS is calculated by integrating the \texttt{Lorentzian} over the entire frequency. 

\begin{table*}
    \centering
    \caption{Details of the best fitted PDS parameters using {\it NuSTAR} and {\it LAXPC} observations of GX 339$-$4 and H 1743$-$322 during $2016-2024$. PDS of both sources are fitted adopting a model combination $CO + \sum_{i=1}^{6} L_{i}$, where CO denotes \texttt{constant} and $L_{i}~ (i = 1, 2, 3, 4, 5  {\rm \, and} ~6)$ refers multiple \texttt{Lorentzian} used to obtain the best fit.  $\sigma_{1}$, $\sigma_{2}$ and $\sigma_{3}$ denote the significance of QPOs.  $\rm QPO_{\rm rms}\%$ and  $\rm Total_{\rm rms}\%$ represent the rms percentage of the QPO features and the entire PDS. The centroid frequency (LC) (in bold font), FWHM (LW) and normalization (LN) of the detected QPOs are mentioned. All errors are computed with $68 \%$  confidence. See the texts for details.}
    \label{table:PDSparameters}	
    \begin{adjustbox}{width=0.975\textwidth,center=\textwidth}
    \large	
    \begin{tabular}{l  @{\hspace{0.2cm}} c @{\hspace{0.2cm}} c @{\hspace{0.2cm}} c @{\hspace{0.2cm}} c @{\hspace{0.2cm}} c @{\hspace{0.2cm}} c @{\hspace{0.2cm}} c @{\hspace{0.2cm}} c @{\hspace{0.2cm}} c @{\hspace{0.2cm}} c @{\hspace{0.2cm}} c @{\hspace{0.2cm}} c  @{\hspace{0.2cm}} c @{\hspace{0.2cm}} c @{\hspace{0.2cm}} c @{\hspace{0.2cm}} c @{\hspace{0.2cm}} c @{\hspace{0.2cm}} c @{\hspace{0.2cm}} c}
		
    \hline\hline \\
    & \multicolumn{6}{c}{Model Parameters} & & \multicolumn{8}{c}{Estimated Parameters} & \\
		
    \cline{2-9}
    \cline{11-18}\\
		
		Epoch &CO& & $\rm L_1$ & $\rm L_2$ & $\rm L_3$ & $\rm L_4$ & $\rm L_5$ &$\rm L_6$ &$\rm \chi^{2}_{\rm red}$ (dof)  & $\sigma_{1}$ & $\rm QPO_{\rm rms1}\%$ & $\sigma_{2}$ & $\rm QPO_{\rm rms2}\%$ & $\sigma_{3}$ & $\rm QPO_{\rm rms3}\%$ & $\rm Total_{\rm rms}\%$ &QPO type \\ \\
		
		\hline \\
		\multicolumn{18}{c}{\large GX 339$-$4} \\
		\hline
		\hline
		\rowcolor{lightgray}
		&&LC&$0.0$&$-$&$-$&$-$&$-$&$-$ & & & & & & & && \\
            \rowcolor{lightgray}
		AS1.01&$2.05 \pm 0.01$&LW& $1.852 \pm 1.010$&$-$&$-$&$-$&$-$&$-$&$0.91~ (122)$&$-$&$-$&$-$&$-$&$-$&$-$&$2.82 \pm 0.02$ &$-$ \\
		\rowcolor{lightgray}
		&&LN &$0.09 \pm 0.06$&$-$&$-$&$-$ &$-$&$-$ & & & & & & & & & \\
		\hline
		
            \rowcolor{lightgray}
		&&LC&$0.0$&$0.42_{-0.08}^{+0.09}$&$-$&$-$&$-$&$-$ &&&  & &&  & && \\
		\rowcolor{lightgray}
		AS1.02&$2.062 \pm 0.003$&LW&$0.78 \pm 0.10$&$0.62 \pm 0.12$&$-$&$-$&$-$&$-$&$1.31 ~(184)$ &$-$&$-$&$-$&$-$&$-$&$-$&$9.93 \pm 0.01$&$-$\\
		\rowcolor{lightgray}
		&&LN &$0.68 \pm 0.14$&$0.43 \pm 0.12$&$-$&$-$&$-$&$-$ & & &  & &&  & &&\\
		\hline
		
		&&LC&$0.0$&$0.0$&$-$&$-$&$-$&$-$ & \\
		NU1.03&$1.976 \pm 0.003$& LW&$1.04^{+0.16}_{-0.19}$&$0.09^{+0.01}_{-0.02}$ &$-$&$-$&$-$&$-$& $1.10~(156)$&$-$&$-$&$-$&$-$&$-$&$-$&$42.61 \pm 0.001$&$-$ \\
		&&LN &$0.62 \pm 0.05$&$0.66 \pm 0.05$ &$-$& $-$& $-$& $-$&$-$& \\
		\hline
		
            &&LC&$0.0$&$0.0$& $-$&$-$ &$-$ &$-$&& & &\\
		AS1.04&$2.059 \pm 0.004$&LW& $0.16 \pm 0.02$&$1.89^{+0.25}_{-0.36}$ &$-$& $-$ &$-$&$-$&$1.07~ (183)$&$-$&$-$&$-$&$-$&$-$&$-$&$14.82 \pm 0.01$&$-$ \\
		&&LN &$1.98 \pm 0.12$&$1.57 \pm 0.09$ &$-$&$-$&$-$ &$-$& & \\
		\hline
		
            & &LC &$0.0$&$0.0$&$0.0$&$-$&$-$&$-$& \\
		NU1.05&$1.771 \pm  0.004$&LW &$3.42 \pm 0.22$&$0.067^{+0.004}_{-0.005}$&$0.84_{-0.08}^{+0.09}$ &$-$&$-$&$-$&$1.12~(137)$&$-$&$-$&$-$&$-$&$-$&$-$&$47.55^{+0.81}_{-0.07}$&$-$ \\
		&&LN &$3.12 \pm 0.15$&$5.79 \pm 0.27$&$1.66 \pm 0.19 $&$-$&$-$&$-$& & \\
		\hline
		
            &&LC&$0.0$&$0.0$&$-$&$-$&$-$&$-$& & \\
		NU1.06&$1.879 \pm 0.004$&LW& $4.24_{-0.38}^{0.43}$&$0.54 \pm 0.06$ &$-$& $-$&$-$&$-$&$0.91~(99)$ &$-$&$-$&$-$&$-$&$-$&$-$&$24.92^{+0.08}_{-0.09}$& $-$ \\
		&& LN &$1.81 \pm 0.10$&$1.79 \pm 0.09$ &$-$&$-$&$-$ &$-$& & \\
		\hline
		
            &&LC&$0.0$&$0.0$&$-$ &$-$ & $-$&$-$& & &\\
		NU1.07&$1.978 \pm 0.003$&LW& $3.32_{-0.67}^{+0.88}$&$0.63 \pm 0.10$&$-$& $-$& $-$&$-$&$1.04~(120)$&$-$&$-$&$-$&$-$&$-$&$-$&$29.17^{+0.12}_{-0.10}$&$-$ \\
		&&LN &$0.37 \pm 0.07$&$0.54 \pm 0.05$ &$-$&$-$&$-$ &$-$& & &\\
		\hline	
		
            && LC&$0.0$&$-$&$-$ &$-$ & $-$&$-$& & \\
		NU1.08&$1.95 \pm 0.02$&LW& $0.32_{-0.10}^{+0.09}$&$-$ &$-$& $-$&$-$&$-$&$1.35~ (161)$&$-$&$-$&$-$&$-$&$-$&$-$&$26.21 \pm 0.01$ &$-$\\
		&&LN &$1.54 \pm 0.26$&$-$&$-$&$-$&$-$&$-$&& \\
		\hline					
		
            &&LC&$0.0$&$0.0$&$0.0$&$-$&$-$&$-$& & \\
		NU1.09&$1.90 \pm 0.01$ &LW&$2.78^{+0.78}_{-0.55}$&$0.21 \pm 0.10$ &$0.015^{+0.006}_{-0.004}$&$-$&$-$&$-$&$1.10~(224)$&$-$&$-$&$-$&$-$&$-$&$-$&$38.28^{+0.11}_{-0.13}$&$-$ \\
		&&LN &$2.13 \pm 0.19$&$1.25 \pm 0.22$ &$2.35 \pm 0.68$&$-$&$-$ &$-$& & \\
		\hline	
		
            && LC&$0.0$&$0.0$&$-$&$-$ &$-$&$-$& & \\
		AS1.10&$2.012 \pm 0.003$&LW& $2.45^{+0.14}_{-0.15}$&$0.23 \pm 0.02$ &$-$&$-$&$-$ &$-$&$0.99~(372)$&$-$&$-$&$-$&$-$&$-$&$-$&$15.220 \pm 0.002$&$-$ \\
		&&LN&$4.24 \pm 0.12$&$3.39 \pm 0.15$&$-$&$-$&$-$ &$-$& & \\
		\hline
		
            &&LC&$0.0$&$-$&$-$ &$-$ & $-$&$-$& & \\
		NU1.11&$1.48 \pm 0.01$&LW& $0.57^{+0.28}_{-0.90}$&$-$ &$-$& $-$&$-$&$-$&$1.16 ~(164)$&$-$&$-$&$-$&$-$&$-$&$-$&$1.64 \pm 0.01$&$-$\\
		&&LN &$0.12 \pm 0.07$&$-$&$-$&$-$& $-$&$-$&& \\
		\hline	
		
            &&LC&$0.0$&$-$&$-$ &$-$ & $-$&$-$& & \\
		NU1.12&$1.907 \pm 0.004$& LW& $0.03^{+0.02}_{-1.07}$&$-$&$-$&$-$&$-$&$-$&$1.07 ~(181)$&$-$&$-$&$-$&$-$&$-$&$-$&$1.50 \pm 0.71$&$-$\\
		&&LN &$0.012 \pm 0.008$&$-$&$-$&$-$&$-$&$-$& & \\
		\hline
		
            &&LC&$0.0$&$-$&$-$&$-$&$-$&$-$& & \\
		NU1.13&$1.920 \pm 0.002$&LW&$0.53 \pm {0.22}$&$-$&$-$&$-$&$-$&$-$&$1.05 ~(228)$&$-$&$-$&$-$&$-$&$-$&$-$&$2.89 \pm 0.51 $&$-$\\
		&&LN &$0.04 \pm 0.02$&$-$&$-$&$-$&$-$ & $-$&& \\
		\hline
		
            &&LC&$0.0$&$-$&$-$&$-$&$-$&$-$&$-$& \\
		NU1.14&$1.939 \pm 0.004$&LW&$9.48 \pm 0.01$&$-$ &$-$&$-$&$-$&$-$& $1.10 ~(392)$&$-$&$-$&$-$&$-$&$-$&$-$&$7.50 \pm 0.03$&$-$\\
		&&LN &$0.17 \pm 0.08$&$-$&$-$&$-$&$-$& $-$&$-$& \\
		\hline
		
            &&LC&$0.0$&$0.0$&$-$ &$-$ & $-$&$-$& & \\
		NU1.15&$1.938 \pm 0.004$&LW& $1.79_{-0.15}^{+0.18}$&$0.27 \pm 0.03$&$-$& $-$& $-$&$-$&$1.07~(226)$&$-$&$-$&$-$&$-$&$-$&$-$&$19.04 \pm 0.21$&$-$\\
		&&LN &$0.76 \pm 0.06$&$0.37 \pm 0.03$&$-$&$-$&$-$ &$-$& & \\
		\hline
		
            &&LC&$0.0$&$0.0$&$0.0$&$-$&$-$&$-$& & \\
		NU1.16&$1.928 \pm 0.003$ &LW& $4.94^{+0.57}_{-0.67}$&$0.039_{-0.003}^{+0.004}$ &$0.81 \pm 0.10$&$-$&$-$ &$-$&$1.18 ~ (224)$&$-$&$-$&$-$&$-$&$-$&$-$&$35.63 \pm 0.39$&$-$\\
		&&LN &$1.03 \pm 0.11$&$1.99_{-0.07}^{+0.08}$&$0.99 \pm 0.07$&$-$&$-$ &$-$& & \\
		\hline
		
            &&LC&$0.0$&$0.0$&$0.0$&$-$&$-$ &$-$&  \\
		NU1.17&$1.970 \pm 0.003$&LW& $0.13 \pm 0.01$&$3.75_{-0.81}^{+1.20}$ &$0.0004 \pm 0.0001$&$-$&$-$&$-$&$1.65 ~ (272)$&$-$&$-$&$-$&$-$&$-$&$-$&$23.52 \pm 0.78$ &$-$\\
		&&LN &$0.31 \pm 0.02$&$0.27 \pm 0.05$&$0.008 \pm 0.003$&$-$&$-$&$-$& \\
		\hline
		
            &&LC& $0.0$ & $0.0$&$0.0$&$\textbf{0.097} \pm 0.002$ & $0.138 \pm 0.004$ &$-$&\\
		AS1.18&$1.998 \pm 0.004$&LW&$4.97 \pm 0.11$ &$0.23 \pm 0.01$ & $0.03 \pm 0.01$ & $0.017 \pm 0.004$ &$0.03 \pm 0.01$&$-$&$1.04~(499)$ & $4.93$ & $5.53 \pm 0.32$ &$-$&$-$&$-$&$-$&$26.67 \pm 1.10$ & Type-C\\
		&&LN &$17.28 \pm 0.22$& $20.08 \pm 0.65$ & $2.98 \pm 0.08$ & $1.87 \pm 0.36$& $1.16 \pm 0.28$&$-$& \\
		\hline
		
            &&LC&$0.0$&$0.0$&$0.11 \pm 0.01$ &$\textbf{0.133} \pm 0.002$&$-$&$-$& & \\
		NU1.19&$1.650 \pm 0.004$& LW& $5.09_{-0.23}^{+0.24}$&$0.06 \pm 0.01$ &$0.26 \pm 0.01$& $0.012^{+0.005}_{+0.004}$&$-$&$-$&$1.37 ~(223)$&$3.75$&$3.22 \pm 0.37$&$-$&$-$&$-$&$-$&$19.38 \pm 1.08$&Type-C \\
		&&LN &$0.75 \pm 0.05$&$2.19 \pm 0.30$&$0.21 \pm 0.10$&$1.02_{-0.22}^{+0.30}$&$-$&$-$& & \\
		\hline
		
            &&LC&$0.0$ & $0.0$&$0.16 \pm 0.01$&$\textbf{0.18} \pm 0.01$&$-$&$-$& && \\
		AS1.20& $2.00 \pm 0.01$ & LW&$6.10 \pm 0.15$ &$0.14 \pm 0.01$ & $0.35 \pm 0.02$ & $0.03 \pm 0.01$ &$-$&$-$& $1.29~(451)$ & $5.65$ &$4.66 \pm 0.23$ &$-$&$-$& $-$&$-$&$24.10 \pm 0.55$&Type-C\\
		&&LN &$16.61 \pm 0.26$ & $9.21 \pm 0.71$ &$13.09 \pm 3.31$ & $1.52 \pm 0.27$ &$-$&$-$& \\
		\hline
	
            &&LC&$0.0$ & $0.0$ & $0.25 \pm 0.02$ & $-$& $-$ &$-$& \\
	    AS1.21&$2.002 \pm 0.006$&LW & $6.56 \pm 0.39$ & $0.34 \pm 0.05$ & $0.16 \pm 0.05$ & $-$ & $-$&$-$&$1.19~(372)$ & $-$ & $-$& $-$& $-$&$-$&$-$&$22.58 \pm 0.23$&$-$\\
	    &&LN & $13.23 \pm 0.24$ & $6.36 \pm 0.52$ &$12.01 \pm 0.95$& $-$& $-$&$-$& & \\
	    \hline
	
            &&LC& $0.0$& $0.0$ &$0.19 \pm 0.02$ &$-$& $-$&$-$& &\\
	    AS1.22&$2.02 \pm 0.02$ & LW & $5.44^{+0.61}_{-0.53}$ & $0.11^{+0.03}_{-0.04} $& $0.25 \pm 0.04$ & $-$& $-$&$-$& $0.97~(185)$&$-$&$-$&$-$&$-$&$-$&$-$&$22.28 \pm 1.22$&$-$ \\
	    &&LN & $13.79 \pm 0.74$ & $8.34 \pm 1.64$& $11.07 \pm 0.89$ & $-$& $-$&$-$&  \\
	    \hline
     
	    &&LC& $0.0$& $0.0$ & $-$& $-$&$-$&$-$& & & & & \\
	    AS1.23&$1.99 \pm 0.02$ &LW& $7.18_{-0.79}^{+0.90}$ & $0.38 \pm 0.04$&$-$&$-$&$-$&$-$& $1.17~(307)$ & $-$&$-$&$-$&$-$&$-$&$-$&$22.61 \pm 0.60$&$-$\\
	    &&LN &$12.99 \pm 0.87$ &$21.24 \pm 1.59$&$-$& $-$ & $-$&$-$& && &  & &  \\
	    \hline
	
            &&LC&$0.0$&$0.0$&$\textbf{0.238} \pm 0.003$&$0.36 \pm 0.01$ &$-$&$-$&& & \\
		NU1.24&$1.632 \pm 0.002$& LW&$0.30 \pm 0.01$&$7.01 \pm 0.38$ &$0.07 \pm 0.01$& $0.25 \pm 0.02$& $-$&$-$&$1.90~(269)$&$3.30$ &$5.09 \pm 0.30$&$-$&$-$&$-$&$-$&$17.071^{+0.001}_{-0.002}$&Type-C\\
		&&LN &$3.76 \pm 0.09$&$2.01 \pm 0.06$&$0.64 \pm 0.05$&$0.79 \pm 0.05$&$-$&$-$&& \\
		\hline
		
            &&LC&$0.0$&$0.0$&$0.40 \pm 0.03$&$\textbf{2.64} \pm 0.02$ & $-$&$-$&&& \\
		NU1.25&$1.581\pm 0.003$& LW&$6.85_{-0.48}^{+0.54}$&$0.51 \pm 0.05$ &$0.41_{-0.05}^{+0.04}$&$0.29 \pm 0.04$& $-$&$-$&$1.28 ~(223)$&$4.80$&$2.88 \pm 0.35$&$-$&$-$&$-$&$-$&$8.73 \pm 1.97 $& Type-C\\
		&&LN &$1.36 \pm 0.10$&$0.45 \pm 0.04$&$0.34 \pm 0.04$&$0.29 \pm 0.03$&$-$&$-$& & \\

		\hline
		
		&&LC&$0.0$&$0.0$&$2.86_{-0.11}^{+0.08}$ &$\textbf{5.37} \pm 0.02$&$9.83_{-0.37}^{+0.33}$&$-$& & \\
		AS1.26&$2.002 \pm 0.004$& LW&$3.75 \pm 1.13$&$0.015 \pm 0.001$ &$0.78_{-0.14}^{+0.18}$&$1.21 \pm 0.06$&$4.00^{+1.13}_{-0.87}$ &$-$& $1.18 ~(174)$&$3.66$&$2.08 \pm 0.22$&$4.44$&$6.77 \pm 0.14$&$3.95$&$3.23 \pm 0.20$&$12.21 \pm 0.20$&Type-B\\
		&&LN &$1.13 ^{+0.37}_{-0.33}$&$5.37_{-0.35}^{+0.33}$&$0.33_{-0.09}^{+0.13}$&$3.48^{+0.69}_{-0.13}$&$0.79 \pm 0.20$&$-$& \\
		\hline
	
	\end{tabular}
\end{adjustbox}
\end{table*}

\begin{table*}
	\begin{adjustbox}{width=0.975\textwidth,center=\textwidth}	
	\large
\begin{tabular}{l  @{\hspace{0.2cm}} c @{\hspace{0.2cm}} c @{\hspace{0.2cm}} c @{\hspace{0.2cm}} c @{\hspace{0.2cm}} c @{\hspace{0.2cm}} c @{\hspace{0.2cm}} c @{\hspace{0.2cm}} c @{\hspace{0.2cm}} c @{\hspace{0.2cm}} c @{\hspace{0.2cm}} c @{\hspace{0.2cm}} c  @{\hspace{0.2cm}} c @{\hspace{0.2cm}} c @{\hspace{0.2cm}} c @{\hspace{0.2cm}} c @{\hspace{0.2cm}} c @{\hspace{0.2cm}} c @{\hspace{0.2cm}} c}
	\hline	
	&&LC&$0.0$&$\textbf{4.85}^{+0.01}_{-0.02}$ &$9.78^{+0.24}_{-0.28}$&$-$&$-$ &$-$& \\
	AS1.27&$2.002 \pm 0.003$& LW& $0.61 \pm 0.14$&$0.54 \pm 0.03$ &$1.40 \pm 0.43$& $-$&$-$&$-$&$1.05 ~(179)$&$21.96$&$4.48 \pm 0.05$&$2.65$&$1.66 \pm 0.02$&$-$&$-$&$5.03 \pm 0.55$&Type-B\\
	&&LN &$0.16\pm 0.03$&$1.41 \pm 0.06$&$0.20 \pm 0.07$&$-$&$-$&$-$& & \\	
		\hline	
		
            &&LC&$0.0$&$-$&$-$&$-$&$-$&$-$ & & \\
		AS1.28&$2.011 \pm 0.004$ &LW& $1.83^{+0.56}_{-0.36}$&$-$&$-$&$-$&$-$& $-$&$0.98 ~(219)$&$-$&$-$&$-$&$-$&$-$&$-$ &$1.77 \pm 0.05$&$-$\\
		&&LN &$0.21_{-0.05}^{+0.06}$&$-$&$-$&$-$&$-$&$-$ && \\
		\hline
		
            &&LC&$0.0$&$-$&$-$&$-$&$-$&$-$ && \\
		NU1.29&$1.655 \pm 0.003$ & LW& $0.80 ^{+0.11}_{-0.10}$&$-$&$-$&$-$&$-$&$-$ &$0.99 ~(129)$&$-$&$-$&$-$&$-$&$-$&$-$&$3.45 \pm 0.26$&$-$\\
		&&LN &$0.31 \pm 0.03$&$-$&$-$&$-$&$-$&$-$ & & \\
		\hline
		
            &&LC&$0.0$&$0.0$&$-$&$-$&$-$&$-$ & & \\
		AS1.30&$1.949 \pm 0.003$ &LW& $1.83 ^{+0.56}_{-0.37}$&$0.15 \pm 0.10$ &$-$&$-$&$-$&$-$ & $1.23~(285)$&$-$&$-$&$-$&$-$&$-$&$-$&$4.53 \pm 0.12$&$-$\\
		&&LN &$1.11 \pm 0.05$&$0.08 \pm 0.01$&$-$&$-$&$-$&$-$ && \\
		\hline
		
            &&LC&$0.0$&$0.0$&$-$&$-$&$-$&$-$ && \\
		AS1.31&$2.007 \pm 0.003$&LW& $8.86 ^{+3.53}_{-2.36}$&$0.27^{+0.18}_{-0.10}$ &$-$&$-$&$-$&$-$ & $1.22~(372)$&$-$&$-$&$-$&$-$&$-$&$-$&$4.09 \pm 0.34$&$-$\\
		&&LN &$0.89 \pm 0.20$&$0.18 \pm 0.03$&$-$&$-$&$-$&$-$ & & \\
		\hline
		
            &&LC&$0.0$&$-$&$-$&$-$&$-$&$-$ && \\
		AS1.32&$2.006 \pm 0.004$&LW& $7.36 ^{+1.20}_{-1.06}$&$-$&$-$&$-$&$-$&$-$ & $1.08~(372)$&$-$&$-$&$-$&$-$&$-$&$-$&$4.43 \pm 0.01$&$-$\\
		&&LN &$1.21_{-0.14}^{+0.16}$&$-$&$-$&$-$&$-$&$-$ && \\
		\hline
		
            &&LC&$0.0$&$-$&$-$&$-$&$-$&$-$ & & \\
		AS1.33&$2.011 \pm 0.001$& LW& $0.25 \pm 0.14$&$-$&$-$&$-$&$-$&$-$ & $0.97~(374)$&$-$&$-$&$-$&$-$&$-$&$-$&$2.26 \pm 0.01$&$-$\\
		&&LN &$0.03 \pm 0.02$&$-$&$-$&$-$&$-$&$-$ && \\
		\hline	
		
            &&LC&$0.0$&$-$&$-$&$-$&$-$&$-$ && \\
		AS1.34&$2.011 \pm 0.003$& LW& $0.34 ^{+0.26}_{-0.18}$&$-$&$-$&$-$&$-$&$-$ &$1.04 ~(374)$&$-$&$-$&$-$&$-$&$-$&$-$&$0.75 \pm 0.01$&$-$\\
		&&LN &$0.04 \pm 0.01$&$-$&$-$&$-$&$-$&$-$ & & \\
		\hline	
		
            &&LC&$0.0$&$-$&$-$&$-$&$-$&$-$ && \\
		NU1.35&$1.623 \pm 0.003$& LW& $0.59_{-0.07}^{+0.08}$&$-$&$-$&$-$&$-$&$-$ & $1.24 ~(326)$&$-$&$-$&$-$&$-$&$-$&$-$&$2.63 \pm 0.01$&$-$\\
		&&LN &$0.21 \pm 0.02$&$-$&$-$&$-$&$-$&$-$ && \\
		\hline	
		\break
		
            &&LC&$0.0$&$-$&$-$&$-$&$-$&$-$ & & \\
		NU1.36&$1.635 \pm 0.004$& LW& $0.60 \pm 0.10$&$-$&$-$&$-$&$-$&$-$ &$1.10 ~(129)$&$-$&$-$&$-$&$-$&$-$&$-$&$2.56 \pm 0.01$&$-$\\
		&&LN &$0.19 \pm 0.02$&$-$&$-$&$-$&$-$&$-$ && \\
		\hline	
		
            &&LC&$0.0$&$-$&$-$&$-$&$-$&$-$ && \\
		NU1.37&$1.63 \pm 0.01$& LW& $0.48 \pm 0.13$&$-$&$-$& $-$&$-$&$-$ &$1.22 ~(194)$&$-$&$-$&$-$&$-$&$-$&$-$&$2.43 \pm 0.01$&$-$\\
		&&LN &$0.18 \pm 0.04$&$-$&$-$&$-$&$-$&$-$ && \\
		\hline	
		
            &&LC&$0.0$&$-$&$-$&$-$&$-$&$-$&$-$ & \\
		NU1.38&$1.62 \pm 0.01$& LW& $0.33_{-0.17}^{+0.28}$&$-$ &$-$& $-$&$-$&$-$ &$1.13~(326)$&$-$&$-$&$-$&$-$&$-$&$-$&$2.03 \pm 0.01$&$-$\\
		&&LN &$0.12 \pm 0.04$&$-$&$-$&$-$&$-$&$-$ && \\
		\hline
		
		&&LC&$0.0$&$0.0$&$-$&$-$&$-$&$-$ && &&&&&&& \\
		NU1.39&$1.954 \pm 0.002$&LW&$2.545_{-0.26}^{+0.30}$&$0.42 \pm 0.03$&$-$&$-$&$-$&$-$ &$1.12~(193)$ &$-$&$-$&$-$&$-$&$-$&$-$&$59.36^{+0.07}_{-0.05}$ &$-$\\
		&&LN&$0.43 \pm 0.04$ &$0.33 \pm 0.02$&$-$&$-$&$-$ & \\
		\hline
		
		\rowcolor{lightgray}
		
		&&LC&$0.0$&$0.0$&$-$&$-$&$-$&$-$ && &&&&&&&\\
		\rowcolor{lightgray}
		NU1.40&$1.983 \pm 0.004$& LW& $0.13_{-0.03}^{+0.02}$&$1.06^{+0.24}_{-0.38}$ &$-$& $-$&$-$&$-$ &$1.00~(71)$&$-$&$-$&$-$&$-$&$-$&$-$&$26.25 \pm 0.01$&$-$\\
		\rowcolor{lightgray}
		&&LN &$0.23 \pm 0.03$&$0.22 \pm 0.04$&$-$&$-$&$-$&$-$ &&&&&&&&& \\
		\hline
		
		\rowcolor{lightgray}
		&&LC&$0.0$&$-$&$-$&$-$&$-$&$-$ && &&&&&&&\\
		\rowcolor{lightgray}
		NU1.41&$1.997 \pm 0.004$&LW& $0.05_{-0.01}^{+0.02}$&$-$&$-$&$-$&$-$&$-$ & $1.12 ~(194)$&$-$&$-$&$-$&$-$&$-$&$-$&$12.43 \pm 0.01$&$-$\\
		\rowcolor{lightgray}
		&&LN &$0.09 \pm 0.02$&$-$&$-$&$-$&$-$&$-$ &&&&&&&&& \\
		\hline
		
		&&LC&$0.0$&$0.0$&$0.0$&$-$&$-$ & \\
		AS1.42 &$2.013 \pm 0.004$&LW&$9.97^{+3.85}_{-2.20}$&$1.08_{-0.04}^{+0.05}$&$0.03 \pm 0.01$&$-$&$-$&$-$ &$0.94~(140)$&$-$&$-$&$-$&$-$&$-$&$-$&$33.01 \pm 0.33$ &$-$\\
		&&LN&$1.14 \pm 0.20$ &$8.89 \pm 0.27$&$4.97 \pm 0.85$&$-$&$-$ & \\
		\hline
		
		&&LC&$0.0$&$0.0$&$0.0$& $-$&$-$&$-$ & \\
		AS1.43&$2.01 \pm 0.01$ &LW&$5.54_{-0.44}^{+0.50}$ &$0.038 \pm 0.004$ &$0.85 \pm 0.10$ &$-$&$-$&$-$ &$0.87 ~(109)$ &$-$&$-$&$-$&$-$&$-$&$-$&$37.14 \pm 0.22$&$-$\\
		&&LN& $3.46 \pm 0.18$ &$7.31 \pm 0.65$ &$3.49 \pm 0.20$&$-$&$-$&$-$ &\\
		\hline
		
		&&LC&$0.0$&$0.0$&$0.0$& $-$& $-$&$-$ & \\
		NU1.44&$1.878 \pm 0.002$&LW&$4.76_{-0.29}^{+0.31}$&$0.053 \pm 0.002$&$1.05_{-0.04}^{+0.05}$&$-$&$-$&$-$ &$1.16 ~(191)$&$-$&$-$&$-$&$-$&$-$&$-$&$55.57 \pm 0.10$ &$-$\\
		&&LN&$1.36 \pm 0.07$ &$2.86 \pm 0.08$ &$1.65 \pm 0.06$ &$-$&$-$&$-$ &\\
		\hline
			% &&&\\
		% AS1.45$^\boxtimes$ &&&&&&& $-$& \\
		% &&&\\
		% \hline
  
		&&LC&$0.0$&$0.0$&$0.067 \pm 0.005$& $\bf 0.464 \pm 0.002$& $0.63 \pm 0.01$&$-$ && \\
	    NU1.46 &$1.698 \pm 0.01$ &LW&$0.82 \pm 0.05$ &$0.103 \pm 0.002$ &$0.046 \pm 0.002$&$0.060 \pm 0.004$&$0.126 \pm 0.050$&$-$ &$1.38~(133)$&$8.41$ & $5.18 \pm 0.23$&$-$&$-$&$-$&$-$&$28.91 \pm 1.22$&Type-C\\
		&&LN&$3.76 \pm 0.25$&$0.46 \pm 0.02$ &$0.22 \pm 0.02$ &$0.47 \pm 0.03$ &$0.11 \pm 0.03$$-$ & \\
		\hline
		
            &&LC&$0.0$&$-$& $-$& $-$&$-$&$-$ & \\
	    NU1.47&$1.640 \pm 0.004$ &LW&$0.76^{+0.12}_{-0.14}$&$-$&$-$&$-$&$-$&$-$ &$0.94~(176)$ &$-$&$-$&$-$&$-$&$-$&$-$&$3.32 \pm 0.02$&$-$\\
		&&LN&$0.22 \pm 0.02$ &$-$&$-$&$-$&$-$$-$ &\\
		\hline
		
            &&LC&$0.0$&$0.0$& $\bf 4.49 \pm 0.09$& $4.84 \pm 0.09$&$-$$-$ & \\
	    NU1.48&$1.611 \pm 0.005$&LW&$0.60 \pm 0.08$ &$0.04 \pm 0.01$ &$0.34 \pm 0.10$ &$0.64^{+0.03}_{-0.02} $ &$-$&$-$ &$1.16~(136)$&$2.10$&$3.68 \pm 0.12$&$-$&$-$&$-$&$-$&$13.85 \pm 0.89$&Type-B\\
		&&LN&$0.24 \pm 0.02$ &$0.06 \pm 0.01$ &$0.23 \pm 0.11$&$0.33 \pm 0.13$&$-$&$-$ &\\
		\hline
		
            &&LC&$0.0$&$0.0$& $\bf 4.43 \pm 0.31$& $4.72 \pm 0.01$&$5.02^{+0.03}_{-0.05}$&$9.32 \pm 0.10$ \\
	    AS1.49 &$1.995 \pm 0.003$&LW&$0.91_{-0.30}^{0.21}$&$0.04 \pm 0.01$&$0.23 \pm 0.05$&$0.25 \pm 0.02$&$0.65 \pm 0.05$&$1.73 \pm 0.05$&$1.37~(146)$ &$6.05$&$9.97 \pm 0.40$&$8.71$&$7.26 \pm 0.34$&$-$&$-$&$47.10 \pm 1.31$&Type-B\\
		&&LN&$0.18 \pm 0.03$& $0.30 \pm 0.03$&$1.15_{-0.19}^{+0.22}$&$1.03 \pm 0.12$&$1.50 \pm 0.10$&$0.61 \pm 0.07$ \\
		  \hline
		
            &&LC&$0.0$&$-$&$-$& $-$& $-$& $-$ &\\
            AS1.53&$2.013 \pm 0.002$ &LW &$0.81 \pm 0.12$ &$-$&$-$&$-$&$-$&$-$ &$1.06 ~(160)$&$-$&$-$&$-$&$-$&$-$&$-$&$1.22 \pm 0.02$&$-$\\
            &&LN&$0.03 \pm 0.02$&$-$&$-$& $-$& $-$&$-$ &&\\			
	    \hline	
		
            \hline \\
			\multicolumn{18}{c}{\large H 1743$-$322} \\
		\hline
		\hline
  
		&&LC&$0.0$&$0.0$&$0.14 \pm 0.01$&$\textbf{0.600} \pm 0.002$ &$1.20^{+0.04}_{-0.02}$ &$-$ &&& \\
		AS2.01&$2.013 \pm 0.004$&LW& $2.92 \pm 0.08$&$19.36 \pm 0.20$&$0.062 \pm 0.004$ &$0.14 \pm 0.02$& $0.16^{+0.04}_{-0.03}$&$-$ &$1.36~ (193)$&$17.87$&$12.21 \pm 0.03$&$7.50$ &$4.68 \pm 0.05$&$-$&$-$&$23.63 \pm 0.35$& Type-C\\
		&&LN &$4.50 \pm 0.26$&$15.03 \pm 0.35$&$2.52 \pm 0.20$&$8.72 \pm 0.43$&$15.01 \pm 0.41$&$-$ && \\
		\hline				
		
            &&LC&$0.0$&$0.18 \pm 0.02$&$\textbf{0.947} \pm 0.002$ &$1.06 \pm 0.01$&$2.13 \pm 0.06$&$-$ && \\
		NU2.02&$1.801 \pm 0.002$&LW& $6.38_{-0.55}^{+0.62}$&$0.57 \pm 0.02$ &$0.13 \pm 0.01$&$0.21 \pm 0.01$&$1.34 \pm 0.02$&$-$ &$0.94 ~(107)$&$8.31$&$7.31 \pm 0.03$&$-$&$-$&$-$&$-$&$46.23 \pm 0.45$& Type-C\\
		&&LN &$0.85 \pm 0.12$&$0.75 \pm 0.03$&$0.62 \pm 0.03$&$0.53 \pm 0.03$&$0.34 \pm 0.04$&$-$ && \\
		\hline				
		
            &&LC&$0.0$&$0.18 \pm 0.17$&$\textbf{1.01} \pm 0.01$&$1.19 \pm 0.01$&$2.23 \pm 0.08$ &$-$ && & & \\
		NU2.03&$1.811 \pm 0.002$&LW& $11.38_{-3.30}^{+5.26}$&$0.62 \pm 0.03$ &$0.21 \pm 0.01$&$0.22 \pm 0.01$&$2.64 \pm 0.21$&$-$ &$1.21~(182)$&$8.57$&$10.12 \pm 0.25$&$-$&$-$&$-$&$-$& $23.53 \pm 0.34$& Type-C\\
		&&LN &$0.27 \pm 0.11$&$0.74 \pm 0.02$&$0.67 \pm 0.02$&$0.48 \pm 0.02$&$0.69 \pm 0.06$&$-$ &&&& \\
		\hline
		
            \rowcolor{lightgray}		
		&&LC&$0.0$&$-$&$-$&$-$&$-$&$-$ &&&&&&&&& \\
		\rowcolor{lightgray}
		AS2.04&$2.051 \pm 0.003$&LW&$1.09 \pm 0.61$&$-$ &$-$&$-$&$-$&$-$ &$1.00~(378)$&$-$&$-$&$-$&$-$&$-$&$-$& $2.160 \pm 0.001$&$-$\\
		\rowcolor{lightgray}
		&&LN &$0.07 \pm 0.04$&$-$&$-$&$-$&$-$&$-$ &&&&&&&&& \\
		\hline
		
            \rowcolor{lightgray}
		&&LC&$0.0$&$-$&$-$&$-$&$-$ &$-$ & && && && &&  \\
		\rowcolor{lightgray}
		AS2.05&$2.055 \pm 0.003$&LW&$0.103 \pm 0.08$&$-$&$-$&$-$&$-$&$-$ &$1.04~(309)$&$-$&$-$&$-$&$-$& $-$&$-$&$1.41 \pm 0.23$&$-$\\
		\rowcolor{lightgray}
		&&LN &$0.030 \pm 0.019$&$-$&$-$&$-$&$-$&$-$ &&&& && &&& \\
		\hline
  
		&&LC&$0.0$&$0.0$&$\textbf{0.434} \pm 0.002$&$0.87 \pm 0.01$&$0.94 \pm 0.03$&$-$ & & \\
		AS2.06&$2.020 \pm 0.004$&LW&$0.92 \pm 0.03$&$12.78_{-0.68}^{+0.74}$&$0.051 \pm 0.004$&$0.09 \pm 0.02$&$0.86 \pm 0.02$&$-$ &$ 1.61~(208)$&$19.74$&$11.37 \pm 0.43$&$4.84$&$3.82 \pm 0.03$&$-$&$-$& $25.19 \pm 3.25$& Type-C\\
		&&LN &$7.73 \pm 0.24$&$13.30 \pm 0.49$&$6.93 \pm 0.37$&$0.79 \pm 0.16$&$5.25 \pm 0.35$&$-$ && \\
		\hline
		
            &&LC&$0.0$&$0.11 \pm 0.01$&$\textbf{0.318} \pm 0.001$&$0.66 \pm 0.02$&$-$&$-$ &&& \\
		NU2.07&$1.833 \pm 0.003$&LW&$1.59 \pm 0.07$&$0.15 \pm 0.01$&$0.056 \pm 0.003$&$0.11 \pm 0.03$&$-$&$-$ &$1.21 ~(185)$&$8.57$&$10.12^{+0.02}_{-0.03}$&$2.71$&$2.74 \pm 0.04$&$-$&$-$&$21.27 \pm 0.04$& Type-C\\
		&&LN &$1.66 \pm 0.06$&$0.37 \pm 0.03$&$0.89 \pm 0.03$&$0.07 \pm 0.02$&$-$&$-$ & & \\
		\hline
		
            &&LC&$0.0$&$0.0$&$\textbf{0.218} \pm 0.002$&$-$&$-$&$-$ &&& \\
		NU2.08&$1.884 \pm 0.002$&LW&$0.82 \pm 0.04$&$7.07 \pm 1.20$&$0.056 \pm 0.003$&$-$&$-$&$-$ &$1.67 ~(108)$&$13.63$&$7.31^{+0.02}_{-0.03}$&$-$&$-$&$-$&$-$&$18.81^{+0.0003}_{-0.002}$& Type-C\\
		&&LN &$0.35 \pm 0.05$&$1.37 \pm 0.04$&$0.31 \pm 0.02$&$-$&$-$&$-$ && \\
		\hline
		\hline
	\end{tabular}
\end{adjustbox}
\end{table*}

\begin{figure*}
    \centering
    \includegraphics[angle=0,width=0.95\columnwidth]{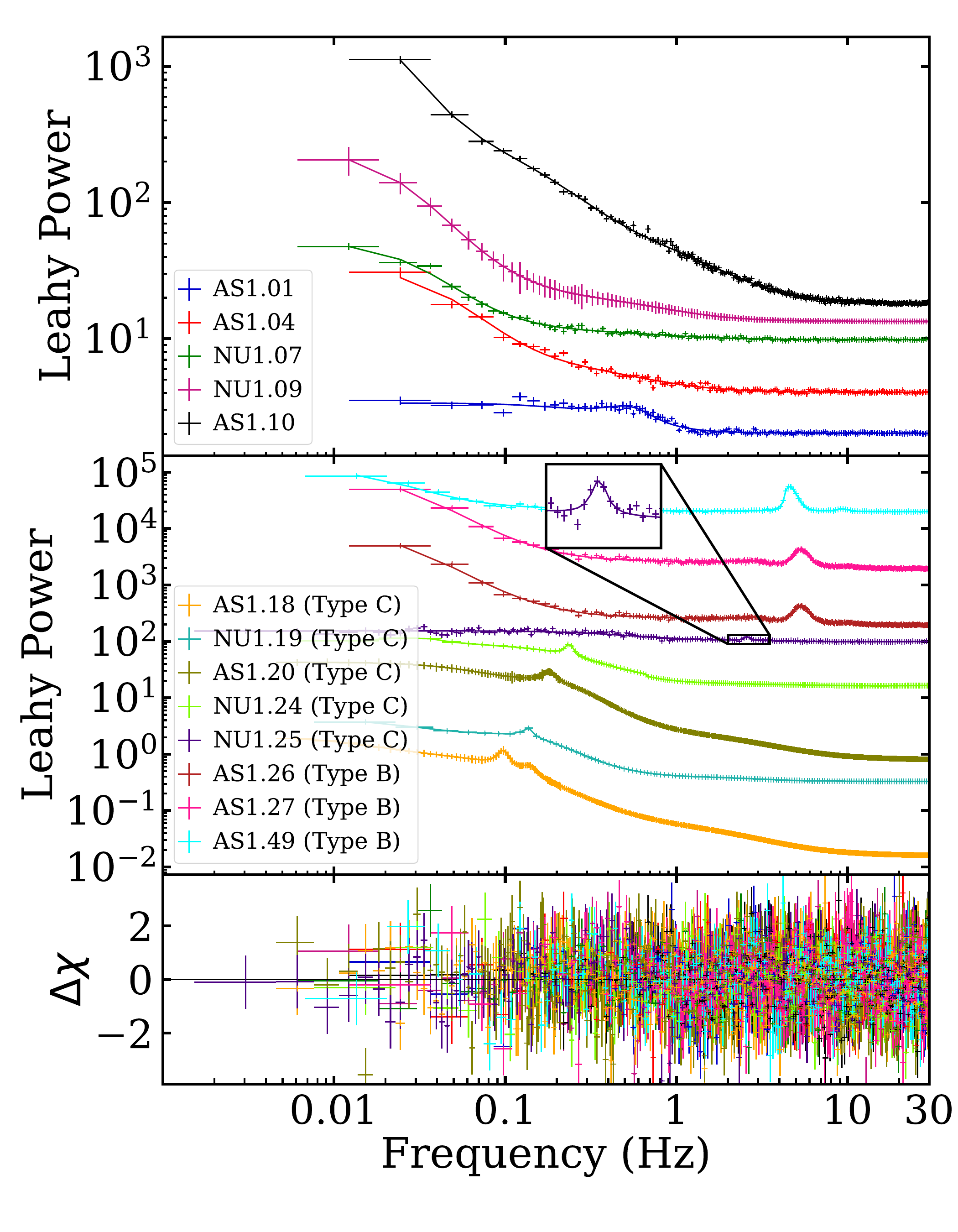}
    \hskip 0.5 cm
    \includegraphics[angle=0,width=0.95\columnwidth]{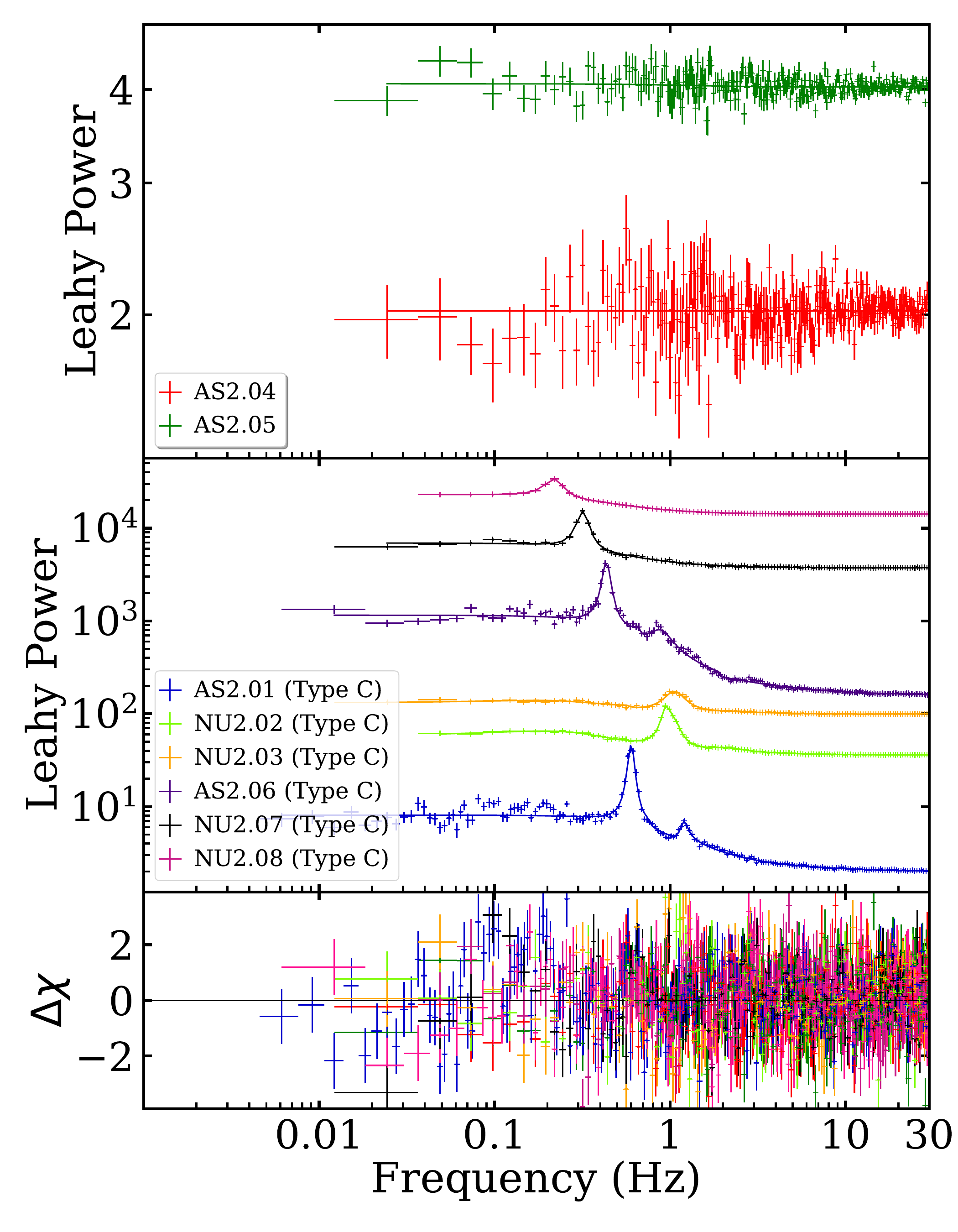}
    \caption{Power density spectra of GX 339$-$4 ({\it left panel}) and H 1743$-$322 ({\it right panel}) during different observations in the frequency range of $0.001-30$ Hz. {\it Left panel}: PDS of GX 339$-$4 during quiescence $-$ AS1.01 (blue), hard state $-$ AS1.04 (red), NU1.07 (green), NU1.09 (magenta), AS1.10 (black), AS1.18 (orange), NU1.19 (sky blue), AS1.20 (olive), NU1.24 (lime), and intermediate state $-$ NU1.25 (violet), AS1.26 (brown), AS1.27 (pink), AS1.49 (cyan) are shown. {\it Right panel}: PDS of H 1743$-$322 during quiescence $-$ AS2.04 (red), AS2.05 (green) and hard state $-$ AS2.01 (blue), NU2.02 (lime), NU2.03 (orange), AS2.06 (violet), NU2.07 (black) and NU2.08 (magenta). In both figures, bottom panels show residual variation of PDS fitting. The power corresponding to different epochs are scaled for the purpose of clarity. See the text for details.
    }
    \label{fig:pds}
\end{figure*}
 
For GX 339$-$4, we fit the PDS obtained from {\it NuSTAR} observation during epoch NU1.19 using three \texttt{Lorentzian} and a \texttt{constant} that yields a positive residual around $0.13$ Hz with $\chi^{2}_{\rm red}= 1.54~(223)$. Hence, we include an additional \texttt{Lorentzian} to account the QPO feature that renders $\chi^{2}_{\rm red}= 1.37 ~(223)$. Similarly, we fit the PDS of {\it AstroSat} observation during epoch AS1.26 with two \texttt{Lorentzian} and a \texttt{constant} that results residuals in the form of QPO near $2.86$ Hz, $5.37$ Hz and $9.83$ Hz. Accordingly, we include three additional \texttt{Lorentzian} to model the QPO features and the best fit is obtained with $\chi^{2}_{\rm red}=1.18~(174)$. We follow the same procedure to model PDS of other epochs as well. For H 1743$-$322, the PDS of epoch AS2.01 is modelled using one \texttt{constant}, two zero centroid \texttt{Lorentzian} and a \texttt{Lorentzian} around $0.14$ Hz. In order to model QPO features around $0.60$ Hz and $1.20$ Hz, we add two additional \texttt{Lorentzian} around those frequencies that provides good fit with $\chi^{2}_{\rm red}=1.36~(193)$. In addition, we model PDS generated from \textit{NuSTAR} lightcurve of epoch NU2.02 with one \texttt{Lorentzian} and a \texttt{constant}. We add two \texttt{Lorentzian} to account the QPO feature around $0.95$ Hz that gives $\chi^{2}_{\rm red}= 4.64~(113)$. In order to improve the fitting, we further include two additional \texttt{Lorentzian} around $0.18$ Hz and $2.13$ Hz and obtain  $\chi^{2}_{\rm red}= 0.94~(107)$. We follow similar fitting routine for the remaining PDS of other epochs of both sources in $0.001-30$ Hz frequency range and present the model fitted parameters of power spectra in Table \ref{table:PDSparameters}.

The detection of a QPO and its harmonic in the PDS of both sources prompted us to study the energy-dependent analysis of the power spectra. While doing so, we consider lightcurves in different energy bands and carry out the PDS analysis. For GX 339$-$4, we consider energy bands as $3-10$ keV, $10-20$ keV and $20-60$ keV, whereas for H 1743$-$322, we choose $3-10$ keV, $10-20$ keV, $20-30$ keV, $30-40$ keV and $40-60$ keV energy ranges. We extract model parameters from the best fit which are tabulated in Table \ref{tab:qpo_enedep}. We also calculate the rms spectrum following the procedure described above.

\begin{table*}
    \centering
    \caption{Energy-dependent QPO parameters of GX 339 - 4 and H 1743$-$322  from {\it NuSTAR} and {\it LAXPC} observations in the period $2016- 2024$. Power spectra have been modeled in Leahy space. The parameters corresponding to the fundamental component are indicated with f, sub-harmonic with sub and second harmonic as $I^{st}$.}
    \label{tab:qpo_enedep}
    \small
    \begin{adjustbox}{width=1\textwidth}
    \begin{tabular}{l @{\hspace{0.3cm}} c @{\hspace{0.3cm}} c @{\hspace{0.3cm}}c @{\hspace{0.3cm}} c  @{\hspace{0.3cm}} c @{\hspace{0.3cm}} c @{\hspace{0.3cm}} c @{\hspace{0.3cm}} c @{\hspace{0.3cm}} c @{\hspace{0.3cm}} c @{\hspace{0.3cm}} c @{\hspace{0.3cm}} c @{\hspace{0.3cm}} c @{\hspace{0.3cm}} c @{\hspace{0.3cm}} c @{\hspace{0.3cm}} c @{\hspace{0.3cm}} c @{\hspace{0.3cm}} c @{\hspace{0.3cm}} c @{\hspace{0.3cm}} r}	
	\hline
	\hline
	Epoch&E-band&QPO (f)&FWHM (f) & Norm (f) & Q (f) & sig (f) & rms (f) &QPO (sub)&FWHM (sub) & Norm (sub) & Q (sub) & sig (sub) & rms (sub) & QPO ($I^{st}$)& FWHM ($I^{st}$) & Norm ($I^{st}$) & Q ($I^{st}$) & sig ($I^{st}$) & rms ($I^{st}$)\\
	  \hline
        &keV&Hz&Hz & & & &(per cent)&Hz&Hz & & & &(per cent)&&Hz& &     & &(per cent)\\

        \hline\\
	\multicolumn{19}{c}{\LARGE GX 339$-$4} \\ \\
	\hline
	
        &$3-10$& $0.097 \pm 0.004$ & $0.00105 \pm 0.00008$ & $0.08 \pm 0.03$ &$8.32$&$2.66$&$25.39 \pm 0.02$&$-$&$-$&$-$&$-$&$-$&$-$&$-$&$-$&$-$&$-$&$-$&$-$\\
	AS1.18	&$10-20$&$0.09 \pm 0.01$ & $0.0013 \pm 0.0006$&$0.02 \pm 0.01$&$7.39$ &$2.00$&$23.61 \pm 0.20$&$-$&$-$&$-$&$-$&$-$&$-$&$-$&$-$&$-$&$-$&$-$&$-$\\
	&$20-60$&$-$& $-$& $-$ &$-$&$-$&$-$&$-$&$-$&$-$&$-$&$-$&$-$&$-$&$-$&$-$&$-$&$-$&$-$\\
	\hline
			
        &$3-10$& $0.133 \pm 0.002$ & $0.007^{+0.003}_{-0.001}$ & $0.15 \pm 0.04$ &$16.34$&$4.08$&$18.73 \pm 0.21$&$-$&$-$&$-$&$-$&$-$&$-$&$-$&$-$&$-$&$-$&$-$&$-$\\
	NU1.19&$10-20$&$0.14 \pm 0.01$ &$0.08 \pm 0.01 $&$0.22 \pm 0.03$&$1.80$&$7.58$&$18.05 \pm 0.22$ &$-$&$-$&$-$&$-$&$-$&$-$&$-$&$-$&$-$&$-$&$-$&$-$&\\
	&$20-60$&$-$&$-$&$-$ &$-$&$-$&$-$&$-$&$-$&$-$&$-$&$-$&$-$&$-$&$-$&$-$&$-$&$-$&$-$\\
	\hline
		
        &$3-10$& $0.18 \pm 0.02$ & $0.011^{+0.003}_{-0.001}$ & $0.50 \pm 0.34$ &$16.22$&$1.46$&$3.24 \pm 0.21$&$-$&$-$&$-$&$-$&$-$&$-$&$-$&$-$&$-$&$-$&$-$&$-$\\
        AS1.20&$10-20$&$-$&$-$&$-$&$-$&$-$&$-$&$-$&$-$ &$-$&$-$&$-$&$-$&$-$&$-$&$-$&$-$&$-$&$-$&\\
	&$20-60$&$-$&$-$&$-$ &$-$&$-$&$-$&$-$&$-$&$-$&$-$&$-$&$-$&$-$&$-$&$-$&$-$&$-$&$-$\\
	\hline
 
	&$3-10$& $0.231 \pm 0.003$ & $0.02 \pm 0.01$ & $0.11 \pm 0.03$ &$11.92$&$3.51$&$17.64 \pm 0.17$&$-$&$-$&$-$&$-$&$-$&$-$&$-$&$-$&$-$&$-$&$-$&$-$&\\
        NU1.24&$10-20$&$0.242\pm 0.001$ & $0.070_{-0.020}^{+0.001}$&$0.11_{-0.02}^{+0.05}$&$3.45$&$5.35$&$16.88 \pm 0.76 $&$-$&$-$&$-$&$-$&$-$&$-$&$-$&$-$&$-$&$-$&$-$&$-$&\\
	&$20-60$&$-$&$-$&$-$ &$-$&$-$&$-$&$-$&$-$&$-$&$-$&$-$&$-$&$-$&$-$&$-$&$-$&$-$&$-$&\\		
	\hline
 			
        &$3-10$& $2.62_{-0.03}^{+0.04}$ & $0.33_{-0.11}^{+0.17}$ & $0.40_{-0.04}^{+0.03}$ &$7.79$&$7.01$&$14.79 \pm 0.56$&$-$&$-$&$-$&$-$&$-$&$-$&$-$&$-$&$-$&$-$&$-$&$-$&\\
	NU1.25&$10-20$&$-$&$-$ &$-$&$-$&$-$&$-$&$-$&$-$&$-$&$-$&$-$&$-$&$-$&$-$&$-$&$-$&$-$&$-$\\
	&$20-60$&$-$&$-$&$-$ &$-$&$-$&$-$&$-$&$-$&$-$&$-$&$-$&$-$&$-$&$-$&$-$&$-$&$-$&$-$\\		
	\hline
			
        &$3-10$& $5.249 \pm 0.004$ & $1.13 \pm 0.25$ & $3.96 \pm 0.38$ &$4.64$&$10.42$&$7.20 \pm 0.25 $&$2.66 \pm 0.16$&$1.32 \pm 0.36$&$0.81 \pm 0.17$&$2.01$&$4.76$&$3.25 \pm 0.45$&$10.86_{-0.76}^{+0.38}$&$1.17 \pm 0.82$&$0.28 \pm 0.15$&$9.28$&$1.87$ &$1.92 \pm 0.35$\\
	AS1.26&$10-20$&$5.56 \pm 0.16$ 	&$0.88_{-0.57}^{+0.32}$&$0.43_{-0.13}^{+0.15}$&$6.32$&$3.31$&$7.15 \pm 0.77$&$3.04 \pm 0.19$&$0.33 \pm 0.22$&$0.11 \pm 0.06$&$9.21$&$1.83$&$3.61 \pm 0.62$&$-$&$-$&$-$&$-$&$-$&$-$\\
	&$20-60$&$-$&$-$&$-$ &$-$&$-$&$-$&$-$&$-$&$-$&$-$&$-$&$-$&$-$&$-$&$-$&$-$&$-$&$-$&\\		
	\hline
			
        &$3-10$& $4.89 \pm 0.02$ & $0.50_{-0.02}^{+0.03}$ & $4.24 \pm 0.16$ &$9.78$&$26.50$&$7.56 \pm 0.06$&-&-&-&-&-&-&$9.76 \pm 0.20$&$1.38 \pm 0.38$&$0.55 \pm 0.12$&$7.07$&$4.58$&$2.72 \pm 0.65$\\
	AS1.27&$10-20$&$4.81 \pm 0.04$ &$0.52_{-0.09}^{+0.10}$&$0.73 \pm 0.09$&$9.25$&$8.11$&$9.78 \pm 0.16$&$-$&$-$&$-$&$-$&$-$&$-$&$-$&$-$&$-$&$-$&$-$&$-$&\\
	&$20-60$&$-$&$-$&$-$ &$-$&$-$&$-$&$-$&$-$&$-$&$-$&$-$&$-$&$-$&$-$&$-$&$-$&$-$&$-$&\\		
	\hline
			
        &$3-10$& $0.463 \pm 0.002$ &$0.07 \pm 0.02$& $0.38 \pm 0.10$ & $6.61$ &$3.8$&$4.56 \pm 0.20$&$-$&$-$&$-$&$-$&$-$&$-$&-&$-$&$-$&$-$&$-$&$-$\\
	NU1.46&$10-20$& $0.465 \pm 0.004$ &$0.06 \pm 0.01$& $0.07 \pm 0.01$ & $7.75$ &$7.0$&$1.89 \pm 0.02$&$-$&$-$&$-$&$-$&$-$&$-$&$-$&$-$&$-$&$-$&$-$&$-$\\
	&$20-60$& $-$ &$-$& $-$ & $-$ &$-$&$-$&$-$&$-$&$-$&$-$&$-$&$-$&$-$&$-$&$-$&$-$&$-$&$-$\\
	\hline
			
        &$3-10$& $4.52 \pm 0.03$ &$0.38 \pm 0.11$& $0.21 \pm 0.09$ & $11.89$ &$2.33$&$3.49 \pm 0.12$&$-$&$-$&$-$&$-$&$-$&$-$&-&$-$&$-$&$-$&$-$&$-$\\
        NU1.48&$10-20$& $4.59 \pm 0.12$ &$0.68 \pm 0.12$& $0.13_{-0.04}^{+0.03}$ & $6.74$ &$3.25$&$2.56 \pm 0.12$&$-$&$-$&$-$&$-$&$-$&$-$&$-$&$-$&$-$&$-$&$-$&$-$\\
	&$20-60$& $-$ &$-$& $-$ & $-$ &$-$&$-$&$-$&$-$&$-$&$-$&$-$&$-$&$-$&$-$&$-$&$-$&$-$&$-$\\
	\hline
			
        &$3-10$& $4.50 \pm 0.01$ &$0.27 \pm 0.03$& $1.09 \pm 0.19$ & $16.66$ &$5.73$&$7.33 \pm 0.32$&$-$&$-$&$-$&$-$&$-$&$-$&$9.26 \pm 0.07$&$1.47_{-0.33}^{+0.41}$&$0.41 \pm 0.07$&$6.27$&$5.85$&$4.49 \pm 0.42$&\\
	AS1.49&$10-20$& $4.67 \pm 0.02$ &$0.66 \pm 0.05$&      $0.55_{-0.07}^{+0.05}$ & $7.00$ &$7.86$&$5.16 \pm 0.23$&$-$&$-$&$-$&$-$&$-$&$-$&$9.51_{-0.41}^{+0.69}$&$1.36 \pm 0.72$&$0.12 \pm 0.06$&$6.99$&$1.89$&$2.41 \pm 0.12$\\
	&$20-60$& $-$ &$-$& $-$ & $-$ &$-$&$-$&$-$&$-$&$-$&$-$&$-$&$-$&$-$&$-$&$-$&$-$&$-$&$-$\\
	\hline\\
		\multicolumn{19}{c}{\LARGE H 1743$-$322} \\\\
	\hline
	
        &$3-10$& $0.599 \pm 0.003$ & $0.058 \pm 0.004$ & $2.79 \pm 0.16$ &$10.32$&$17.43$ & $13.31 \pm 0.08$ &$-$&$-$&$-$&$-$&$-$&$-$ &$1.22 \pm 0.01$&$0.14_{-0.03}^{+0.06}$&$0.51 \pm 0.06$&$8.71$&$8.50$&$5.69 \pm 0.47$ \\
	&$10-20$&$0.602 \pm 0.003$ &$0.068 \pm 0.006$&$1.21 \pm 0.08$&$8.85$&$15.13$&$13.51 \pm 0.09$ &$-$&$-$&$-$&$-$&$-$&$-$ &$1.22 \pm 0.02$ &$0.16_{-0.07}^{+0.11}$&$0.19_{-0.06}^{+0.09}$&$7.63$&$3.17$& $5.35 \pm 0.56$\\
	AS2.01&$20-30$&$0.602 \pm  0.007$ &$0.08 \pm 0.02$& $0.19_{-0.03}^{+0.04}$ &$7.53$&$6.33$&$16.79 \pm 0.96$&$-$&$-$&$-$&$-$&$-$&$-$&$-$&$-$&$-$&$-$&$-$&$-$&\\		
	&$30-40$&$0.60 \pm 0.02$&$0.08 \pm 0.03$&$0.07 \pm 0.02$&$7.39$&$3.29$&$7.88 \pm 0.20$&$-$&$-$&$-$&$-$&$-$&$-$&$-$&$-$&$-$&$-$&$-$&$-$& \\
	&$40-60$&$-$&$-$&$-$ &$-$&$-$&$-$&$-$&$-$&$-$&$-$&$-$&$-$&$-$&$-$&$-$&$-$&$-$&$-$&\\		
	\hline
	
        &$3-10$& $0.981 \pm 0.006$ & $0.20_{-0.01}^{+0.02}$ & $0.86 \pm 0.02$ &$4.78$&$43.38$&$16.68 \pm 0.10$&$-$&$-$&$-$&$-$&$-$&$-$&$-$&$-$&$-$&$-$&$-$&$-$&\\
	&$10-20$&$0.97 \pm 0.01$ &$0.22 \pm 0.04$&$0.48 \pm 0.03$&$4.49$&$23.96$&$35.72 \pm 0.28$&$-$&$-$&$-$&$-$&$-$&$-$&$-$&$-$&$-$&$-$&$-$&$-$&\\
	NU2.02&$20-30$&$ 0.95 \pm 0.03$ &$0.19^{+0.11}_{-0.07}$& $0.05 \pm 0.02$ &$5.00$&$2.78$&$14.50 \pm 0.44$&$-$&$-$&$-$&$-$&$-$&$-$&$-$&$-$&$-$&$-$&$-$&$-$&\\
	&$30-40$&$-$&$-$&$-$ &$-$&$-$&$-$&$-$&$-$&$-$&$-$&$-$&$-$&$-$&$-$&$-$&$-$&$-$&$-$&\\		
	&$40-60$&$-$&$-$& $-$&$-$&$-$&$-$&$-$&$-$&$-$&$-$&$-$&$-$&$-$&$-$&$-$&$-$&$-$&$-$&\\		
	\hline
 
	&$3-10$& $1.07 \pm 0.01$ & $0.22_{-0.02}^{+0.03}$ &$0.51 \pm 0.02$ &$4.60$&$25.57$&$16.68 \pm 0.10$ &$-$&$-$&$-$&$-$&$-$&$-$&$-$&$-$&$-$&$-$&$-$&$-$&\\
	&$10-20$&$1.09 \pm 0.01$ &$0.42 \pm 0.03$&$0.34 \pm 0.02$&$2.53$&$18.87$&$14.02 \pm 0.21$&$-$&$-$&$-$&$-$&$-$&$-$&$-$&$-$&$-$&$-$&$-$&$-$&\\
	NU2.03&$20-30$&$1.08 \pm 0.05$ &$0.31 \pm 0.11$& $0.05 \pm 0.01$ &$3.48$&$5.00$&$12.55 \pm 0.45$&$-$&$-$&$-$&$-$&$-$&$-$&$-$&$-$&$-$&$-$&$-$&$-$&\\	
	&$30-40$&$-$&$-$& $-$&$-$&$-$&$-$&$-$&$-$&$-$&$-$&$-$&$-$&$-$&$-$&$-$&$-$&$-$&$-$&\\	
	&$40-60$&$-$&$-$&$-$&$-$&$-$&$-$&$-$&$-$&$-$&$-$&$-$&$-$&$-$&$-$&$-$&$-$&$-$&$-$&\\		
	\hline

        &$3-10$& $0.433 \pm 0.002$ & $0.05 \pm 0.01$ & $2.61 \pm  0.17$ &$8.66$&$15.35$&$13.47 \pm 0.10 $ &$-$&$-$&$-$&$-$&$-$&$-$ &$0.90_{-0.01}^{+0.02}$&$0.19_{-0.04}^{+0.06}$&$0.81_{-0.08}^{+0.10}$&$4.74$&$10.11$& $7.50 \pm 0.16$ \\
	&$10-20$&$0.434 \pm 0.002$ &$0.05 \pm 0.01$&$1.07 \pm 0.08$&$8.68$&$13.38$&$13.02 \pm 0.11 $ &$-$&$-$&$-$&$-$&$-$&$-$ &$0.88_{-0.02}^{+0.03}$&$0.29_{-0.11}^{+0.13}$&$0.31_{-0.16}^{+0.11}$&$3.03$&$1.93$& $7.01 \pm 0.62$\\
	AS2.06&$20-30$&$0.44 \pm 0.01$ &$0.05_{-0.01}^{+0.02}$& $0.15 \pm 0.03$ &$8.80$&$5.00$&$10.14 \pm 0.28$&$-$&$-$&$-$&$-$&$-$&$-$&$-$&$-$&$-$&$-$&$-$&$-$&\\		
	&$30-40$&$0.43 \pm 0.01$&$0.03 \pm 0.01$&$0.03 \pm 0.01$&$12.59$&$3.11$&$5.48 \pm 0.20$&$-$&$-$&$-$&$-$&$-$&$-$&$-$&$-$&$-$&$-$&$-$&$-$&\\
	&$40-60$&$-$&$-$&$-$ &$-$&$-$&$-$&$-$&$-$&$-$&$-$&$-$&$-$&$-$&$-$&$-$&$-$&$-$&$-$&\\		
	\hline
 
	&$3-10$& $0.317 \pm 0.002$ & $0.06 \pm 0.01$ & $1.11 \pm 0.04$ &$5.79$&$27.87$&$35.15 \pm 0.13$&$-$&$-$&$-$&$-$&$-$&$-$&$-$&$-$&$-$&$-$&$-$&$-$&\\
	&$10-20$&$0.32 \pm 0.01$ &$0.05_{-0.01}^{+0.02}$&$0.28  \pm 0.03$&$6.29$&$9.95$&$23.37 \pm 0.31$&$-$&$-$&$-$&$-$&$-$&$-$&$-$&$-$&$-$&$-$&$-$&$-$&\\
	NU2.07&$20-30$&$0.32 \pm 0.01$ &$0.04 \pm 0.02$& $0.03 \pm 0.01$ &$7.64$&$3.66$&$17.00 \pm 0.42$&$-$&$-$&$-$&$-$&$-$&$-$&$-$&$-$&$-$&$-$&$-$&$-$&\\
	&$30-40$&$-$&$-$& $-$&$-$&$-$&$-$&$-$&$-$&$-$&$-$&$-$&$-$&$-$&$-$&$-$&$-$&$-$&$-$&\\		
	&$40-60$&$-$&$-$& $-$&$-$&$-$&$-$&$-$&$-$&$-$&$-$&$-$&$-$&$-$&$-$&$-$&$-$&$-$&$-$&\\		
	\hline
 
	&$3-10$& $0.218 \pm 0.002$ & $0.06 \pm 0.01$ & $0.22_{-0.01}^{+0.03}$ &$3.05$&$16.21$&$19.62 \pm 0.26$&$-$&$-$&$-$&$-$&$-$&$-$&$-$&$-$&$-$&$-$&$-$&$-$&\\
	&$10-20$&$0.22 \pm  0.01$ &$0.04 \pm 0.01$&$0.06  \pm 0.01$&$3.14$&$5.32$&$17.44 \pm 0.47$&$-$&$-$&$-$&$-$&$-$&$-$&$-$&$-$&$-$&$-$&$-$&$-$&\\
	NU2.08&$20-30$&$0.23 \pm 0.01$ &$0.03 \pm 0.01$& $0.013 \pm 0.004$ &$7.34$&$3.19$&$13.73 \pm 0.25$&$-$&$-$&$-$&$-$&$-$&$-$&$-$&$-$&$-$&$-$&$-$&$-$&\\		
	&$30-40$&$-$&$-$& $-$&$-$&$-$&$-$&$-$&$-$&$-$&$-$&$-$&$-$&$-$&$-$&$-$&$-$&$-$&$-$&\\
	&$40-60$&$-$&$-$& $-$&$-$&$-$&$-$&$-$&$-$&$-$&$-$&$-$&$-$&$-$&$-$&$-$&$-$&$-$&$-$&\\		
	\hline
\end{tabular}
\end{adjustbox}
\end{table*}

\subsection{GX 339$-$4}

It is observed that for GX 339$-$4, QPO frequency increases with the decrease of total PDS power during $2021$ outburst. Moreover, the source renders total variability as $2.82-26.25\%$ rms, $14.82-59.36\%$ rms, $1.22-47.10\%$ rms and $0.75-2.63\%$ rms in quiescent, hard, intermediate and soft states respectively. During seven epochs (AS1.18, NU1.19, AS1.20, NU1.24, NU1.25, AS1.26 and AS1.27) in the $2021$ outburst, GX 339$-$4 exhibits QPO features, where fundamental QPO frequency ($\nu_{\rm QPO}$) varies in the range $0.097-5.37$ Hz with significance ($\sigma$) as $3.3-21.96$ and ${\rm QPO}_{\rm rms}\%$ as $2.08-5.53$. In addition, we find that $\nu_{\rm QPO}$ increases monotonically as time progresses which is shown in the middle panel of Fig. {\ref{fig:pds}} (left panel). A rapid evolution is observed between epoch AS1.26 to AS1.27 (separated by only a day), where $\nu_{\rm QPO}$ evolves from $5.37 \pm 0.02$ Hz to $4.85^{+0.01}_{-0.02}$ Hz. During epoch AS1.26, GX 339-4 shows sub-harmonic and second harmonic around $2.86$ Hz and $9.83$ Hz, respectively. Interestingly, in epoch AS1.27, the sub-harmonic disappears and second harmonic is seen at $\sim 9.78$ Hz. The results of similar kind are also reported in the literature \cite[]{Mondal-etal2023,Peirano-etal2023}. Recently, GX 339$-$4 has shown QPOs during three epochs (NU1.46, NU1.48 and AS1.49) of $2023-2024$ outburst with $\nu_{QPO}$ varies in the range $0.46-4.49$ Hz with $\sigma$ in the range $2.10-8.41$ and $\rm QPO_{\rm rms}\%$ as $3.68-9.97$. We observe second harmonic during AS1.49 around $9.32$ Hz. We classify QPOs belonging to AS1.18, NU1.19, AS1.20, NU1.24, NU1.25, NU1.46 as type-C, and those in AS1.26, AS1.27, NU1.48, AS1.49 as type-B \cite[see][for details]{Motta-etal2011}. Needless to mention that the observed QPOs are associated with either LHS or IMS.

\begin{figure}
    \centering
    \includegraphics[angle=0,width=0.95\columnwidth]{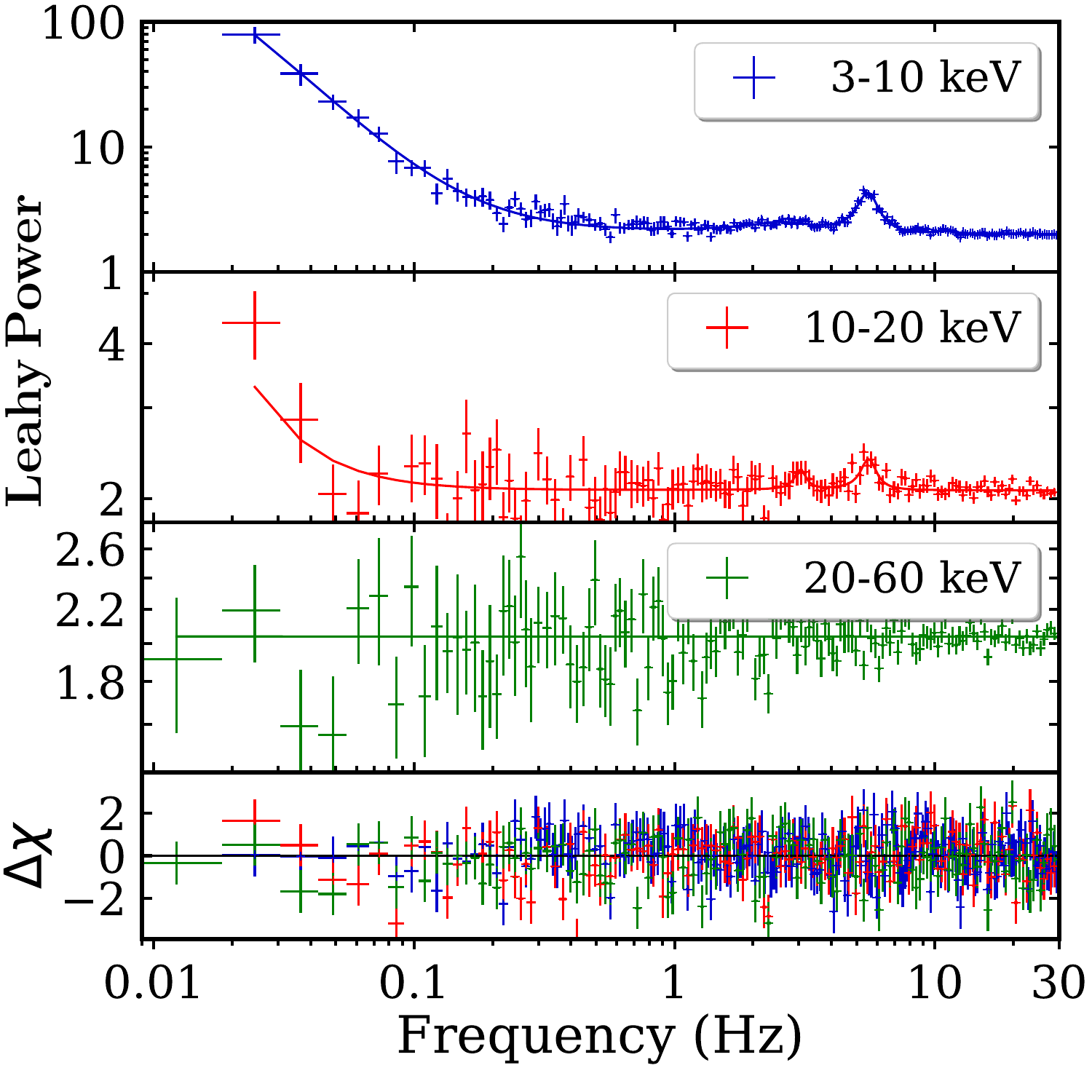}
    \caption{Energy-dependent PDS ({\it LAXPC}) of GX 339$-$4 in Leahy space during epoch AS1.26 (MJD 59303.06). PDS are shown in the energy range $3-10$ keV (blue), $10-20$ keV (red) and $20-60$ keV (green) starting from top panel, respectively. The residual variation is presented in the bottom panel. See the text for details.}
    \label{fig:GX_pds_ene}
\end{figure}

In Table \ref{tab:qpo_enedep}, we present the energy dependent model fitted parameters for each epochs where QPO is present. In all epochs, we find that QPO feature disappears in $20-60$ keV energy band. Moreover, we do not find any significant detection of QPO after $10$ keV for epoch AS1.20 and epoch NU1.25. During epoch AS1.26, $\nu_{\rm QPO}$ and its sub-harmonic are absent beyond $20$ keV, and the second harmonic is seen in $3-10$ keV energy range. On the other hand, we notice that during epoch AS1.27, the second harmonic is present only in $3-10$ keV range and the QPO signature disappears above $20$ keV, as shown in Fig. \ref{fig:GX_pds_ene}. 

\subsection{H 1743$-$322}

\begin{figure}
    \centering
    \includegraphics[angle=0,width=0.95\columnwidth]{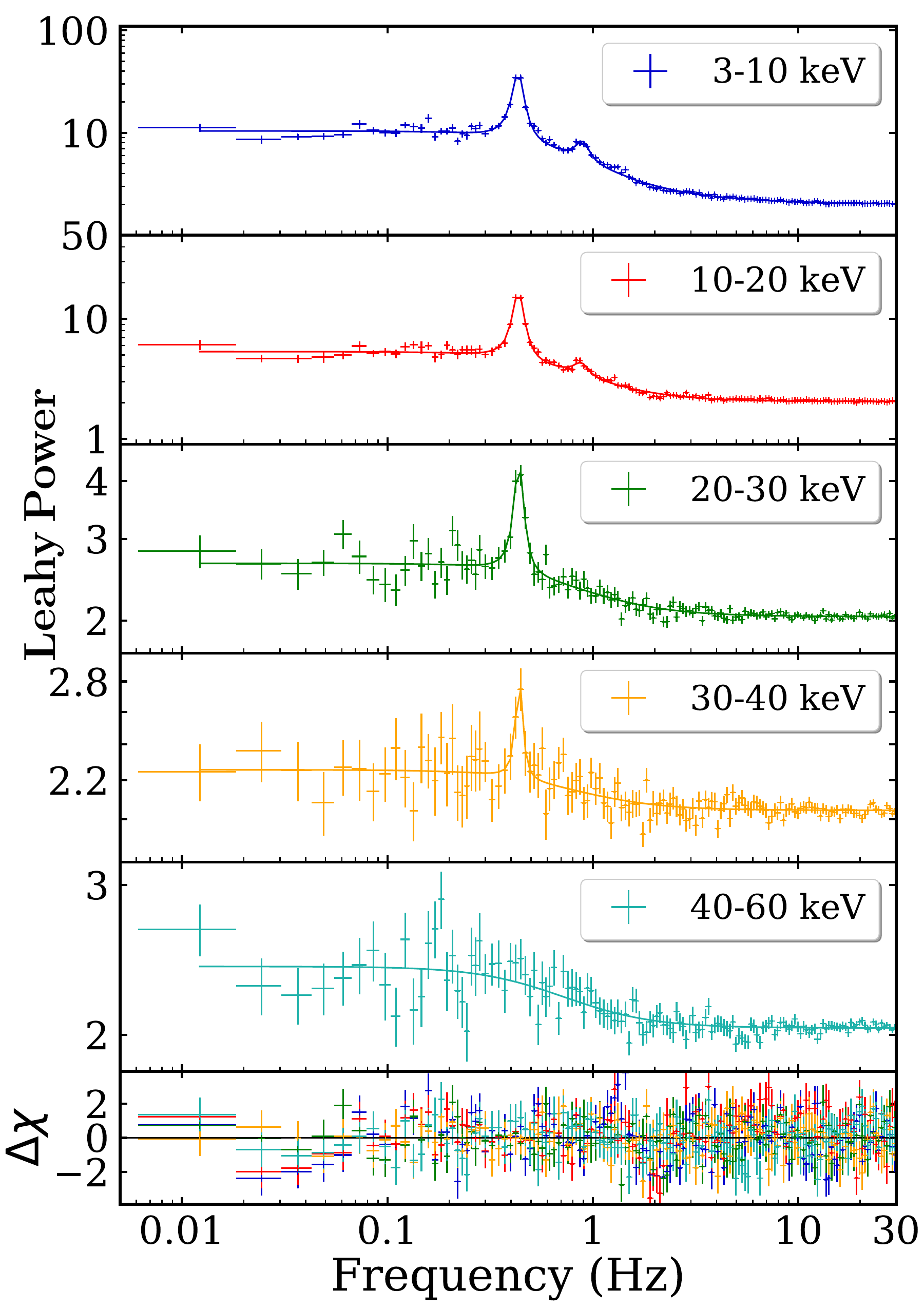}
    \caption{Energy-dependent PDS ({\it LAXPC}) of H 1743$-$322 in Leahy space during epoch AS2.06 (MJD 57973.32). PDS are shown in the energy range $3-10$ keV (blue), $10-20$ keV (red), $20-30$ keV (green), $30-40$ keV (orange) and  $40-60$ keV (sky blue) starting from top panel, respectively. The residual variation is presented in the bottom panel. See the text for details.}
    \label{fig:H_pds_ene}
\end{figure}

In Fig. \ref{fig:pds} (right panel), we present the PDS for selected epochs. In particular, we find QPO features during six epochs, namely AS2.01, NU2.02, NU2.03, AS2.06, NU2.07 and NU2.08, respectively, where the fundamental QPO frequency ($\nu_{\rm QPO}$) varies in the range $0.22-1.03$ Hz with significance ($\sigma$) as $8.31-19.74$ and ${\rm QPO}_{\rm rms}\%$ as $7.31-12.21$. What is more is that we observe the presence of fundamental QPO ($\nu_{\rm QPO}$) along with its harmonic during three epochs. For epochs AS2.01, AS 2.06 and NU2.07, $\nu_{\rm QPO}$ and its harmonic are estimated as $0.6$ Hz and $1.20$ Hz, $0.43$ Hz and $0.87$ Hz, and $0.32$ Hz and $0.66$ Hz, respectively. We observe that H 1743$-$322 shows total variability of $1.41-2.16 \%$ rms and $18.81-46.23 \%$ rms during the quiescent and hard states. Further, we carry out the energy dependent analysis of PDS and depict the obtained results in Fig. \ref{fig:H_pds_ene}. We find that QPO feature continues to present in the energy range $3-40$ keV for the epochs AS2.01 and AS2.06, whereas harmonics disappears for energy beyond $20$ keV. For the remaining epochs, QPO is seen in $3-30$ keV energy range only. All the model extracted parameters are tabulated in Table \ref{tab:qpo_enedep}.

\section{Discussions and Conclusions}
\label{sec:Dis-con}

In this work, we present the comprehensive analyses of wide-band spectro-temporal variabilities of two BH-XRBs, namely GX 339$-$4 and H 1743$-$322, using {\it AstroSat} and {\it NuSTAR} observations during $2016-2024$. Since their discovery, it was observed that both sources underwent frequent outbursts \cite[]{Homan-etal2005,Belloni-etal2005,Tomsick-etal2008,Santo-etal2009,Shaposhnikov-Titarchuk2009,Capitanio-etal2010,Debnath-etal2010,Chen-etal2010,Debnath-etal2013,Nandi-etal2012,Sreehari-etal2018,Sreehari-etal2019a}. In particular, during the observation period of our interest, H 1743$-$322 and GX 339$-$4 exhibited four and six outbursts, respectively, followed by the long term evolution as shown in Fig. \ref{fig:lc-maxi-bat}. Meanwhile, \cite{Aneesha-etal2019,Aneesha-Mandal2020,Bhowmick-etal2021} examined the outburst triggering mechanism in these sources. These authors indicated that during quiescence phase, matter accumulates inside the disc and slowly diffuses inwards. Accordingly, the disc mass builds up until at some disc radius where the local accretion rate increases above its critical limit. It is believed that, the sudden rise of viscosity due to the increase in accretion rate can lead to the triggering of outburst in BH-XRBs \cite[]{Ebisawa-etal1996,Lasota-2001,Seifina-Titarchuk2010,Seifina-etal2014,Titarchuk-Seifina2017,Titarchuk-Seifina2021}.

After triggering of the outburst, the outer part of the disc becomes hot due to the strong irradiation originated from the inner disc \cite[]{vanParadijs1996,King-Ritter1998,Shahbaz-etal1998}. Because of this, viscosity is enhanced that allows the Keplerian disc to move towards the central BH \citep{Chakrabarti-1990}. Since the supply of soft photons is increased, more and more seed photons are up-scattered leading the accreting system to transit towards the soft state via intermediate states. As time evolves, disc starts getting depleted due to the reduction of matter supply from the companion \cite[]{Hameury-etal1986,Hameury-Lasota2014} and hence, the temperature of the disc is decreased. At this point, outburst declines and possibly, source transits towards the quiescent accretion phase.

Both GX 339$-$4 and H 1743$-$322 experienced {\it successful} as well as {\it failed} outbursts during $2016-2024$ (see \S 3). The $2019-2020$, $2021$ and $2023-2024$ outbursts of GX 339$-$4 appeared as {\it successful} one that shows canonical HID patterns \cite[left panel of Fig. \ref{fig:HID}; see also][]{Sreehari-etal2018,Nandi-etal2018,Blessy-etal2021,Dong-etal2021,Geethu-etal2022,Wang-etal2022a} and also exhibited longer duration with slow rise slow decay profile. These two outbursts are the brightest and they perhaps occurred due to the high mass accretion rate ($L_{\rm bol} \gtrsim 19.59\%{\rm L}_{\rm Edd}$, see Table \ref{tab:par_log}). On the contrary, during $2017-2018$ and $2018-2019$, GX 339$-$4 underwent {\it failed} outburst \cite[]{Husain-etal2022,Debnath-etal2023} yielding low luminosity possibly resulted due to the lack of matter supply from the companion. According to \cite{Wang-etal2022}, \textit{failed} and \textit{successful} outbursts follow the same initial evolutionary track, although the former class of outburst never reaches the threshold for a transition to thermally dominated accretion regimes. It is interesting to note that for $2018-2019$, $2019-2000$, $2021$ and $2023-2024$ outbursts, the rising time in $2-6$ keV energy band is longer than the same in $6-20$ keV energy band (see Table \ref{tab:lc_proper}). Fig. \ref{fig:lc-maxi-bat}a shows that the soft ($2-20$ keV) lightcurve lags behind hard ($15-50$ keV) lightcurve during $2019-2020$, $2021$ and $2023-2024$ outbursts. The fast rise and fast decay of hard photon counts unlike the soft photon counts indicates the presence of two distinct components of accretion flow comprising Keplerian and sub-Keplerian matters \cite[]{Chakrabarti-Lev1995} which accrete at two different time scales \cite[]{Aneesha-etal2019,Aneesha-Mandal2020}. Furthermore, H 1743$-$322 underwent four outbursts. Among them only $2016$a outburst was successful \cite[]{Stiele-etal2021} although all four outbursts yield similar luminosity with shorter duration. Here, we infer that during the previous outburst, a part of the disc matter may not be fully accreted by the central BH. Eventually, the left over mass can combine with the material accreted from the companion star and ultimately contributes in triggering the $2016$a outburst \cite[]{Chakrabarti-etal2019}.

\begin{figure*}
    \centering
    \includegraphics[width=0.45\textwidth]{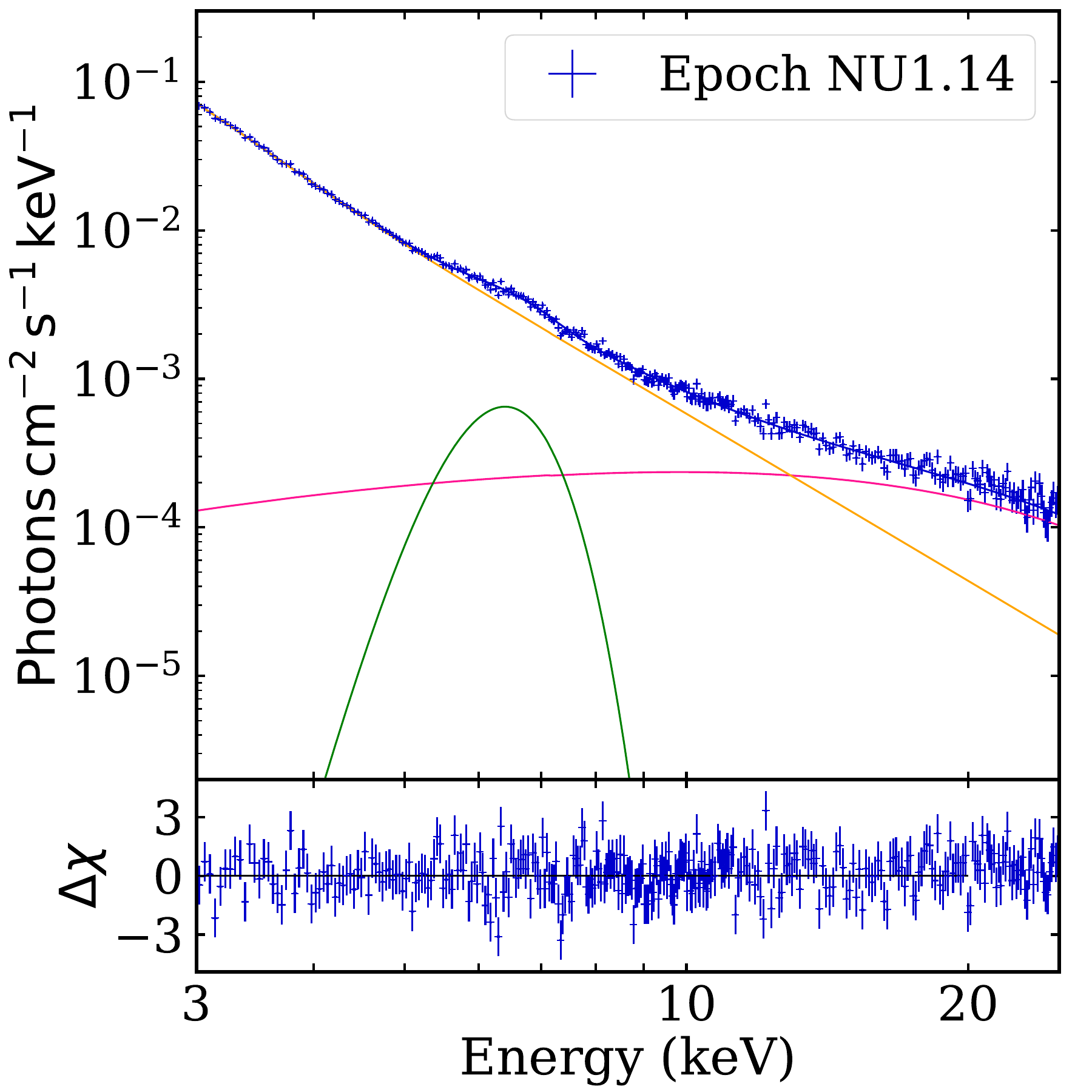}
    \hskip 0.5 cm
    \includegraphics[width=0.45\textwidth]{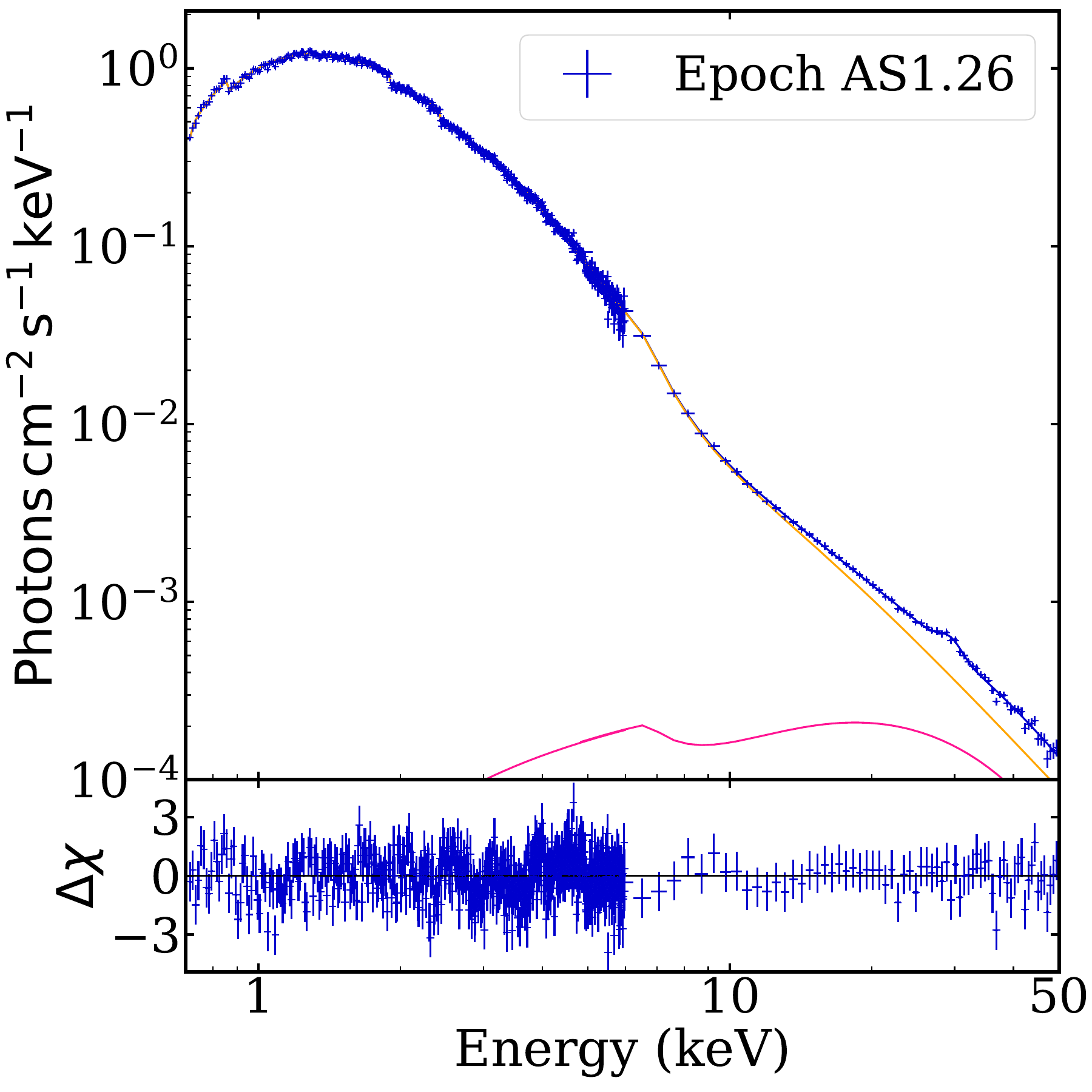}
    \caption{Energy spectra of GX 339$-$4 including annihilation lines (modelled with \texttt{bbody}) corresponding to epoch NU1.14 (left panel) and epoch AS1.26 (right panel) are modelled using \texttt{TBabs$\times$(bbody+gaussian+comptb)} with $\chi^{2}_{\rm red}=0.97 ~ (543$ dof) and \texttt{TBabs$\times$smedge$\times$(bbody+comptb)} with $\chi^{2}_{\rm red}=1.21 ~ (592$ dof),  respectively. Unfolded model components of \texttt{bbody}, \texttt{comptb} and \texttt{gaussian} are depicted by pink, orange and green solid curves, whereas plus signs represent observational data. See the text for the details.}
    \label{fig:emition-lines}  
 \end{figure*}

We observe that both GX 339$-$4 and H 1743$-$322 passed through quiescence phases multiple times during the observation period under consideration. The quiescence state spectra of GX 339$-$4 and H 1743$-$322 sources are satisfactorily described by the non-thermal Comptonization model (\texttt{Nthcomp}). We find that photon index ($\Gamma_{\rm nth}$) varies as $1.71 - 2.48$ and $1.56-2.22$ for GX 339$-$4 and H 1743$-$322, respectively, which are in accordance with the usual quiescent spectral indices of BH-XRBs \cite[]{Kong-etal2000,Corbel-etal2006,Bradley-etal2007,Plotkin-etal2013}. The luminosity of GX 339$-$4 and H 1743$-$322 are obtained as $0.03-0.06 \% ~ {\rm L}_{\rm Edd}$ and $0.10 - 0.16 \% ~ {\rm L}_{\rm Edd}$. These luminosities are generally higher compared to the quiescence state luminosity of other BH-XRBs \citep{Remillard-etal2006}. Similarly, to explain the hard state spectra of GX 339$-$4 and H 1743$-$322, we model the source spectra using \texttt{Nthcomp} model component. In the hard state of GX 339$-$4, we obtain $\Gamma=1.54-1.74$, $kT_{\rm bb} \sim 0.12-0.77$ keV and $kT_{\rm e} \sim 4.33 - 106.82$ keV, which are in good agreement with the previous findings \cite[]{Dzielak-etal2019,Wang-Ji-etal2018}. In the case of H 1743$-$322, in the hard state $\Gamma_{\rm nth} \sim 1.57-1.71$ and $kT_{\rm e} \sim 17.37-50.0$ keV which is in accordance with \cite{Dong-etal2021}. The corresponding luminosities for GX 339$-$4 and H 1743$-$322 are estimated as $0.91-11.56 \, \% \, {\rm L}_{\rm Edd}$ and $2.08-3.48 \, \% \, {\rm L}_{\rm Edd}$ which are in good agreement with the observed luminosity of other BH-XRBs in hard states \citep{Maccarone-2003}. In explaining the intermediate states of GX 339$-$4, an additional thermal disc component (\texttt{diskbb}) is required along with the \texttt{Nthcomp} component. Best fit model yields the inner disc temperature as $kT_{\rm in} \sim 0.56-0.88$ keV, $\Gamma_{\rm nth} ~\sim 1.76 - 2.66$ and $kT_{\rm e} \sim 11.43 - 50.00$ keV which agree with the results reported earlier \cite[]{Shaposhnikov-Titarchuk2009,Sridhar-etal2020}. The higher $kT_{\rm in}$ suggests that inner disc extends further towards BH. When GX 339$-$4 transits to softer state, the photon index becomes steeper $\Gamma_{\rm nth} \sim 1.46-3.26$, disc becomes hotter as $kT_{\rm in} \sim 0.82-0.88$ keV, and electron temperature decreases as $kT_{\rm e} \sim 1.63-32.36$ keV \cite[]{Plant-etal2014,Liu-etal2022}.  Moreover, we find maximum bolometric luminosity as $L_{\rm bol} \sim 30.06\%{\rm L}_{\rm Edd}$ ($2019-2020$ outburst) that possibly indicates the presence of thermally dominated disc. Additionally, we also examine the High-temperature Blackbody Bump (HBB) features in IMS spectra of $AstroSat$ and $NuSTAR$ observations. Towards this, we model $NuSTAR$ spectrum (epoch NU1.14) using model \texttt{TBabs$\times$(comptb+gaussian+bbody)} and $AstroSat$ spectrum (epoch AS1.26) using model \texttt{TBabs$\times$smedge(comptb+bbody)}. We find \texttt{bbody} colour temperature varies at $\sim 4-6$ keV in the IMS spectra. In Fig. \ref{fig:emition-lines}, we plot the energy spectra including the annihilation lines (modelled with \texttt{bbody}) for epoch NU1.14 (left panel) and AS1.26 (right panel), respectively for GX 339$-$4 source.
This possibly indicates the presence of weak signature of gravitationally redshifted annihilation line as argued by \cite{Titarchuk-Seifina2009,Seifina-Titarchuk2010,Titarchuk-Seifina2021}.

\begin{figure*}
    \centering
    \includegraphics[width=0.45\textwidth]{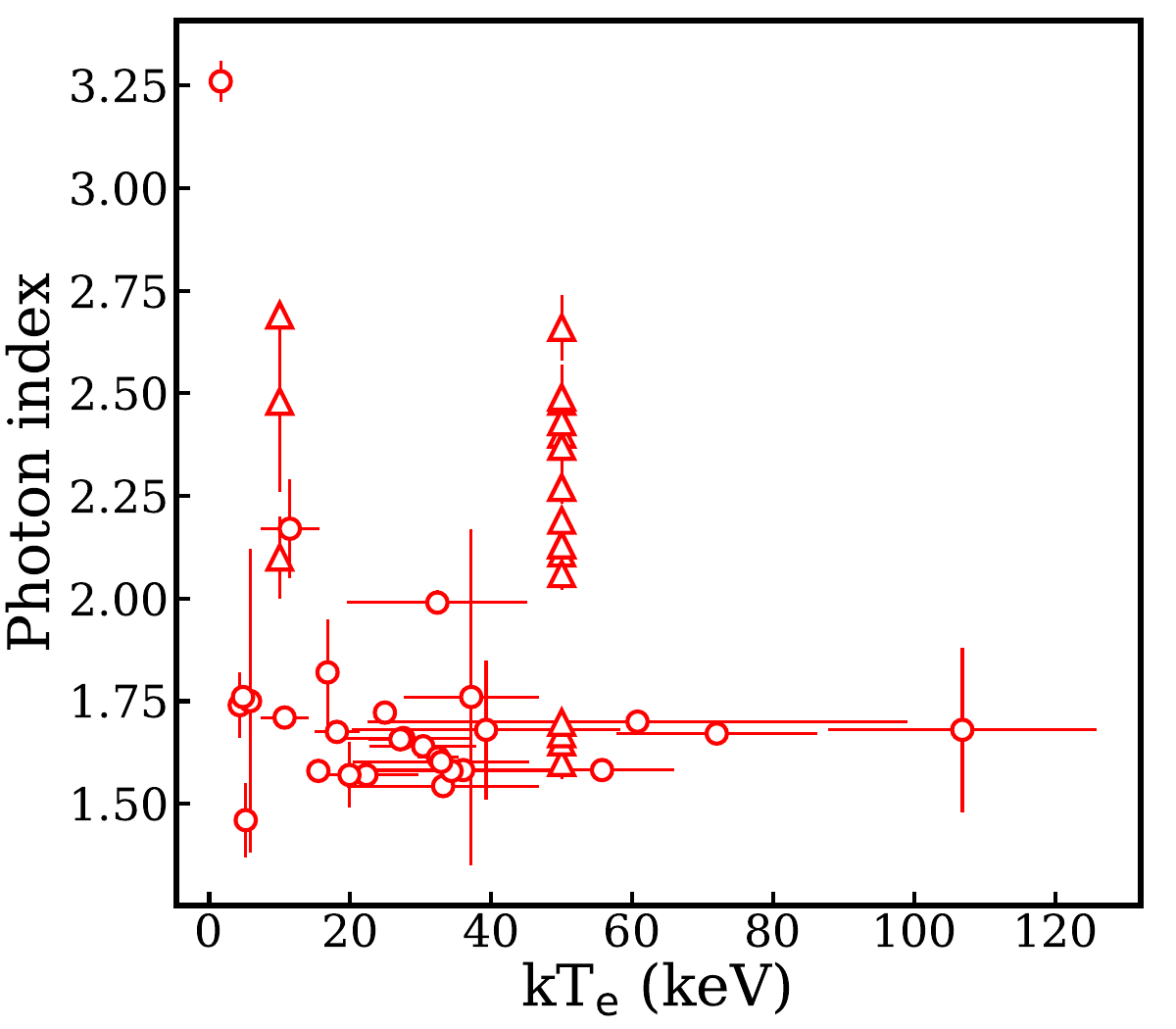}
    \hskip 0.5 cm
    \includegraphics[width=0.45\textwidth]{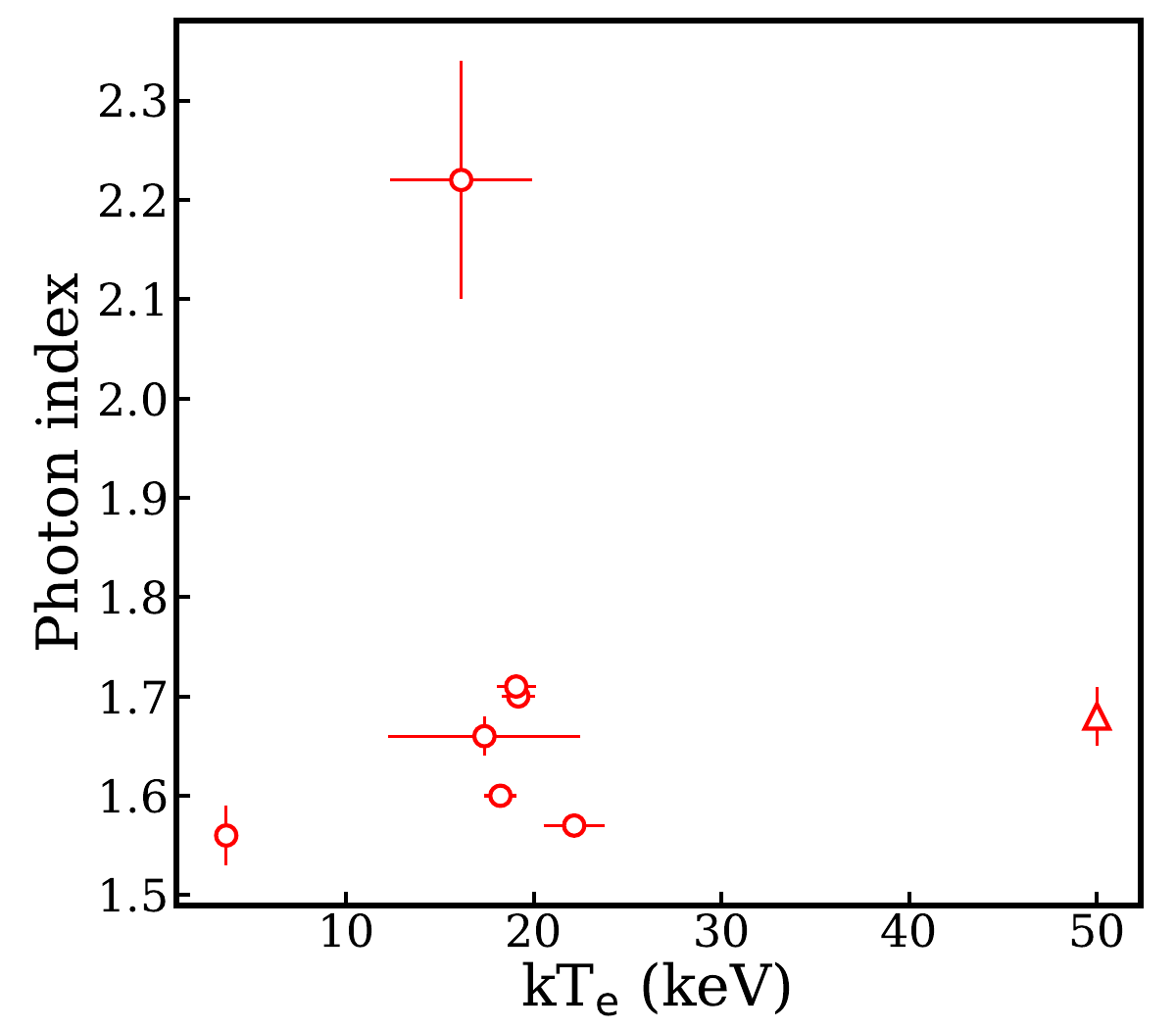}
    \caption{Variation of photon index $\Gamma_{\rm nth}$ with $kT_{\rm e}$ for GX 339$-$4 (left) and H 1743$-$322 (right). Open triangle corresponds to the results for frozen $kT_{\rm e}$, and open circles denote results when such restriction is not imposed. See the text for the details.}
    \label{fig:gamma_kT}  
 \end{figure*}

In Fig. \ref{fig:gamma_kT}, we plot the variation of photon index $\Gamma_{\rm nth}$ with electron temperature $kT_{\rm e}$ for GX 339$-$4 (left panel) and H 1743$-$322 (right panel). In both panels, open triangles represent results when $kT_{\rm e}$ is kept frozen to obtain the best fit, while open circles denote results obtained without such restriction. We observe that for GX 339$-$4, $\Gamma_{\rm nth}$ lies in the range $\sim 1.5-3.25$ for $kT_{\rm e} \sim 10$ keV. We also find that $\Gamma_{\rm nth}$ generally decreases with the increase of $kT_{\rm e}$, although few exceptions are observed. For H 1743-322, we observe $\Gamma_{\rm nth}$ ranging approximately $\sim 1.55-2.2$, with any significant variation with electron temperatures $kT_e$ below $50$ keV, as the source predominantly remains in LHS. Meanwhile, using Monte Carlo simulations, \cite{Laurent-Titarchuk2011} revealed a negative correlation between $\Gamma_{\rm nth}$ and $kT_{\rm e}$. In addition, \cite{Seifina-etal2014} reported similar variation of $\Gamma_{\rm nth}$ with $kT_{\rm e}$ for 4U 1630$-$47 source. These findings clearly indicates that as BH-XRBs transit from LHS to HSS, the corona cools down as it interacts with soft photons emitted from the accretion disc.

During the $2021$ outburst, the hard state of GX 339$-$4 was completely characterized by Comptonize emissions without any signature of disc component. As outburst progresses, the signature of disc component gradually emerges out with the increase of \texttt{Nthcomp} flux. This happens as more and more disc photons are up-scattered by the corona via Comptonization. In this spectral state, we observe type-C QPO and as time evolves, QPO frequency increases in epoch AS1.18, NU1.19 and AS1.20, respectively that evidently indicates that the overall size of the corona decreases \cite[]{Nandi-etal2012}. The evolution of the QPO frequency along the outburst is commonly observed in BH-XRB sources \cite[]{Motta-etal2011,Zhang-etal2017}. After epoch AS1.20 ($i.e.$, during epochs AS1.21, AS1.22, AS1.23), the \texttt{Nthcomp} flux decreases and we see that the QPO features also disappears. As the source moves to hard peak (epoch NU1.24: hard peak is the peak during 2021 outburst in Fig. 2 (purple curve)), the \texttt{Nthcomp} flux increases and type-C QPO reappears. This possibly indicates that QPO feature seems to be an observable provided \texttt{Nthcomp} flux is sufficiently high. After the hard peak, the disc flux contribution increases which indicates that accretion disc evolves and the source transits to intermediate state (epoch NU1.25). After epoch NU1.25, the source transits to disc dominated intermediate state ($F_{\rm diskbb} > F_{\rm nth}$) and we observe type-B QPO along with sub-harmonic and second harmonic features in the PDS during epoch AS1.26 and AS1.27. Notably, the type-C QPO is observed along with flat-top noise, while red noise accompanies the type-B QPO. Subsequently, the \texttt{Nthcomp} flux starts decreasing and ultimately, QPO disappears in epoch AS1.28. At the peak of the outburst, the disc flux increases and source transits to soft state (epoch AS1.34). We find that the source continues to stay in the soft state ($\sim {\rm MJD}~59400$) until it starts declining from peak the outburst (see Fig. \ref{fig:lc-maxi-bat}). Similarly, we observe type-C QPO during two {\it AstroSat} and four {\it NuSTAR} observations of H 1743$-$322 which belong to 2016a, 2017 and 2018 outbursts  where 2016a was successful outburst. The source was in hard state \cite[see also][]{Debnath-etal2013} and the centroid QPO frequency varies as $0.22-1.01$ Hz. As outburst progresses, QPO frequency is increased and the source spectra becomes softer which suggests the overall decrease of corona size. We see the presence of type-C QPO when F$_{\rm nth}$ value is high. It apparently suggests that the modulation of corona \cite[]{Molteni-etal1996,Das-etal2014} possibly be responsible for the origin of type-C QPO in H 1743$-$322. We notice the presence of QPO in H 1743$-$322 irrespective to \textit{successful} or \textit{failed} outburst, whereas type-B and type-C QPOs are only seen during the \textit{successful} outburst in GX 339$-$4. Recently, \cite{Lucchini-etal2023} indicated that the source variability systematically evolves few days ahead of the canonical state transition, which is eventually quantified due to the change in spectral hardness during a successful outburst. Interestingly, the evolution of such kind of variability is not seen during the failed outburst, and hence, it possibly be regarded as the predictor to recognise a given outburst as \textit{failed} or \textit{successful} one. Needless to mention that any further investigation towards this is beyond the scope of the present paper due to the lack of pointed observation during the failed as well as successful outbursts.

It may be noted that in order to explain the origin of type-C QPOs, several models were proposed in the literature, namely relativistic precession model \cite[RPM;][]{Stella-Vietri1998,Stella-etal1999}, precession of hot-inner flow model \cite[]{Ingram-etal2009}, and propagating oscillatory shock model \cite[POS;][]{Chakrabarti-etal2008,Iyer-etal2015}. However, the origin of type-B QPO is not well understood as it was indicated that type-B QPO may be resulted due to the instabilities \cite[]{Titarchuk-Osherovich1999,Tagger-Pellat1999} or possibly be associated with the disc-jet coupling mechanism \cite[]{Fender-etal2009,Radhika-nandi2014,Radhika-etal2016,Kylafis-etal2020}, and hence, it remains inconclusive till date.

Meanwhile, it is reported that QPO frequency seems to be inversely proportional to the truncation radius of the disc \cite[]{Chakrabarti-Manickam2000,Titarchuk-Osherovich2000}. As the source transits from LHS to IMS, the truncation radius moves inward resulting the increase of QPO frequency as observed during 2021 outburst of GX 339$-$4, where QPO frequency monotonically increases from $0.10$ Hz to $5.37$ Hz. Indeed, the evolution of QPO frequency during an outburst was observed earlier in GX 339$-$4 \cite[]{Nandi-etal2012} and such features are commonly observed in other BH-XRBs \cite[]{Motta-etal2011,Zhang-etal2017}. Furthermore, QPOs are not generally observed in HSS of BH-XRBs \cite[]{Belloni-2005,Nandi-etal2012,Radhika-etal2016}, and similarly, we do not find QPOs in HSS of GX 339$-$4.

Furthermore, the energy dependent PDS analyses show that for GX 339$-$4, the fundamental QPO and its harmonics (if exist) are present in $3-20$ keV. On the contrary, for H 1743$-$322, the fundamental QPO is only visible in $3-40$ keV and harmonics disappear beyond $20$ keV. We examine the energy dependent ${\rm QPO}_{\rm rms}\%$ spectra for GX 339$-$4 and H 1743$-$322. We observe that for both sources, ${\rm QPO}_{\rm rms}\%$ for type-C QPO is decreased with energy, whereas marginal changes are seen for type-B QPO (see Table \ref{tab:qpo_enedep}). We also notice that when type-B QPO is observed, the corona size is reduced due to the effect of disc thermal emission. Because of this, the inner disc temperature is increased resulting in higher rms amplitude of QPO at high energies. This finding are in agreement with the results reported by \cite{Kong_etal2020,Kara-etal2019}.

Based on the above findings, we summarize our results below.
\begin{itemize}
	\item During the observation period of $2016-2024$, H 1743$-$322 and GX 339$-$4 underwent four and six outbursts, respectively. The overall profile of the outbursts shows that GX 339$-$4 exhibits short rise slow decay, while H 1743$-$322 displays fast rise slow decay. 
	
	\item The HID shows hysteresis pattern during $2019-2020$, $2021$ and $2023-2024$ successful outbursts of GX 339$-$4, whereas the same is not observed during the successful outburst of H 1743$-$322 in $2016a$.	
	
	\item GX 339$-$4 is found to transit from quiescence ($L_{\rm bol}=0.03-0.06 \, \% {\rm L}_{\rm Edd}$) to outburst phase and vice versa. During outburst, the source exhibits hard, intermediate and soft spectral states with $L_{\rm bol}$ of $0.91-11.56 \, \% {\rm L}_{\rm Edd}$, $2.90-16.09 \, \% {\rm L}_{\rm Edd}$ and $19.59-30.06 \, \% {\rm L}_{\rm Edd}$, respectively. The quiescent state and hard state spectra are modelled with Comptonization component alone. Intermediate and soft state spectra require strong disc component along with weaker `Comptonization' component.
		
	\item H 1743$-$322 is found to transit from quiescence ($L_{\rm bol} \sim 0.10-0.16 \, \% {\rm L}_{\rm Edd}$) to hard state ($2.08-3.48 \, \% {\rm L}_{\rm Edd}$). The source does not show any thermal disc component in its energy spectrum.
	
	\item Both type-B and type-C QPO signatures in the frequency range $0.10 - 5.37$ Hz are observed in GX 339$-$4. In particular, type-C QPOs in both hard and intermediate states are observed where corona emission is dominant ($F_{\rm nth} > F_{\rm diskbb}$), whereas type-B QPO is observed for $F_{\rm diskbb} > F_{\rm nth}$. On the contrary, only type-C QPO is found in the hard state of H 1743$-$322 with frequencies in the range $0.22 - 1.01$ Hz. Note that both sources exhibit QPO harmonics.
		
	\item The energy dependent PDS study shows that in GX 339$-$4, the fundamental QPO and harmonics disappear beyond $20$ keV. In H 1743$-$322, the fundamental QPO is seen in $3-40$ keV energy band and the harmonic ceases to exist beyond $\sim 20$ keV.

\end{itemize}

\section*{Acknowledgements}
\label{sec:acknow}

Authors thank the anonymous reviewers for constructive comments and useful suggestions that help to improve the quality of the manuscript. AU and SD thank the Department of Physics, IIT Guwahati, for providing the facilities to complete this work. AU, SD, and AN acknowledge the support from ISRO sponsored project (DS\_2B-13013(2)/5/2020-Sec.2). AN also thanks GH, SAG; DD, PDMSA; Associate Director and Director, URSC for encouragement and continuous support to carry out this research. This work uses data from the {\it AstroSat} mission of the ISRO archived at the Indian Space Science Data Centre (ISSDC). This work has used the data from the Soft X-ray Telescope ({\it SXT}) developed at TIFR, Mumbai, and the {\it SXT} POC at TIFR is thanked for verifying and releasing the data and providing the necessary software tools. This work has also used the data from the {\it LAXPC} Instruments developed at TIFR, Mumbai, and the {\it LAXPC} POC at TIFR is thanked for verifying and releasing the data. This publication also made use of data from the {\it NuSTAR} mission by the National Aeronautics and Space Administration.  This work has also use data from Monitor of All-sky X-ray Image ({\it MAXI}) data provided by Institute of Physical and Chemical Research (RIKEN), Japan Aerospace Exploration Agency (JAXA), and the {\it MAXI} team.  Also this research made use of software provided by the High Energy Astrophysics Science Archive Research Center (HEASARC) and NASA’s Astrophysics Data System Bibliographic Services.

\section*{Data Availability}
Data used for this publication are currently available at the \textit{AstroSat} ISSDC website (\url{https://astrobrowse.issdc.gov.in/astro\_archive/archive}), {\it MAXI} websites (http://maxi.riken.jp/top/index.html) and  \texttt{NuSTAR} data at HEASARC (https://heasarc.
gsfc.nasa.gov/docs/archive.html).


\input{ms.bbl}
%\bibliographystyle{mnras}
%\bibliography{reference} % if your bibtex file is called example.bib

% Don't change these lines
\bsp	% typesetting comment
\label{lastpage}
\end{document}